\documentclass[useAMS,usenatbib,usegraphicx]{mn2e}
\input epsf
\newcommand{\etal}{{\it et~al.}}

\newcommand{\mstar}{$M_\odot$}

\newcommand{\microm}{$\mu$m}
\newcommand{\IRAS}{{\it IRAS}}
\newcommand{\MSX}{{\it MSX}}

\newcommand{\uchii}{{UC H{\scriptsize II}}}
\newcommand{\hii}{{H{\scriptsize II}}}

\newcommand{\mjycount}{$\mbox{mJy count}^{-1}$}
\newcommand{\jybeam}{$\mbox{Jy beam}^{-1}$}
\newcommand{\mjy}{$\mbox{mJy}$}

\newcommand{\lum}{$L_\odot$} 

\title[Millimetre continuum observations of southern massive star formation regions]{Millimetre continuum observations of southern massive star formation regions.\\
{\bf I. SIMBA observations of cold cores}}

\author[T. Hill \etal~ ]{T. Hill$^{1}$\thanks{E-mail: thill@phys.unsw.edu.au},  M. G. Burton$^{1}$, V. Minier$^{2}$, M. A. Thompson$^{3}$, A.J. Walsh$^{1}$, \and M. Hunt-Cunningham$^{1}$ and G. Garay$^{4}$\\ %, and J. L. Caswell$^{2}$\\
\\
$^{1}$ School of Physics, University of New South Wales, Sydney 2052, NSW, Australia.\\
$^{2}$ Service d'Astrophysique, DAPNIA/DSM/CEA CE de Saclay, 91191 Gif-sur-Yvette, France.\\
$^{3}$ Science \& Technology Research Institute, University of Hertfordshire, College lane, Hatfield, Herts, AL10 9AB, UK\\
$^{4}$ Departamento de Astronom\'{\i}a, Universidad de Chile, Casilla 36-D, Santiago, Chile.\\
}

\begin{document}

\date{Accepted/ Received}

\pagerange{\pageref{firstpage}--\pageref{lastpage}} \pubyear{2004}

\maketitle

\label{firstpage}

\begin{abstract}

  We report the results of a 1.2~mm continuum emission survey toward 131 star forming complexes suspected of undergoing massive star formation. These regions have previously been identified as harbouring a methanol maser and/or a radio continuum source (\uchii~ region), whose presence is in most instances indicative of massive star formation. The 1.2~mm emission was mapped using the SIMBA instrument on the 15~m Swedish ESO Submillimetre Telescope (SEST). Emission is detected toward all of the methanol maser and \uchii~ regions targeted, as well as towards 20 others lying within the fields mapped, implying that these objects are associated with cold, deeply embedded objects. Interestingly, there are also 20 methanol maser sites and 9 \uchii~ regions within the fields mapped which are devoid of millimetre continuum emission.

  In addition to the maser and \uchii~ regions detected, we have also identified 253 other sources within the SIMBA maps. All of these (253) are new sources, detected solely from their millimetre continuum emission. These  `mm-only' cores are devoid of the  traditional indicators of massive star formation, (i.e.  methanol/OH maser, \uchii~ regions, \IRAS~ point sources). At least 45\% of these mm-only cores are also without mid-infrared \MSX~ emission. The `mm-only' core may be an entirely new class of source that represents an earlier stage in the evolution of massive stars, prior to the onset of methanol maser emission. Or, they may harbour protoclusters which do not contain any high mass stars (i.e. below the \hii~ region limit).

   In total, 404 sources are detected representing  four classes of sources which are distinguished by the presence of the different combination of associated tracer/s. Their masses, estimated assuming a dust temperature of 20 K and adopting kinematic distances, range from   0.5~{$\times 10^{1}$}\mstar~ to 3.7~{$\times 10^{4}$}\mstar, with an average mass for the sample of  1.5~{$\times 10^{3}$}\mstar. The H$_{2}$ number density ($n_{H_{2}}$) of the source sample ranges from 1.4~$\times~ 10^3$~cm$^{-3}$ to 1.9~$\times~10^6$~cm$^{-3}$, with an average of 8.7~$\times~10^4$~cm$^{-3}$. The average radius of the sample is 0.5~pc. The visual extinction ranges from 10 to 500~mag with an average of 80~mag, which implies a high degree of embedding. The surface density ($\Sigma$) varies from 0.2 to 18.0~kg~m$^{-2}$ with an average of 2.8~kg~m$^{-2}$.

  Analysis of the mm-only sources shows that they are less massive ({\= M}~=~0.9~{$\times 10^{3}$}\mstar), and smaller ({\= R}~=~0.4~pc) than sources with methanol maser and/or radio continuum emission, which collectively have a mean mass of 2.5~{$\times 10^{3}$}\mstar~ and a mean radius of 0.7~pc.

\end{abstract}


\section{Introduction}

   The formation of massive stars is not a well understood process neither observationally nor physically. This may be attributed to the fact that they form deeply embedded in the cores of molecular clouds, where they are optically obscured by circumstellar dust. Additionally, massive stars form on relatively short timescales, they tend to form only in clustered mode, and occur at distances greater than the nearest examples of their low mass counterparts. Consequently, the combination of these factors hinders the study of regions undergoing massive star formation, specifically the earliest processes involved in their evolution, of which there are few clear examples. 

   Theoretical models suggest that high mass star formation could proceed either through protostellar mergers (\citealp{bonnell04}) or through the collapse of a supersonically turbulent core (\citealp{mckee03}). The first scenario requires dense clusters ($10^8$ stars pc$^{-3}$), that will trigger coalescence, whilst the second requires high accretion rates ($> 10^{-3}$ \mstar~ yr$^{-1}$) that will overcome the outward radiative pressure.

   To date our knowledge of the earliest stages of high-mass star formation has been limited, although recent work, as discussed in the following paragraphs, has brought new insights (c.f. \citealp{williams04, beuther02, szymczak00}).

   Massive stars form in dense molecular clouds, and are characterised by their high luminosity ($>10^4$ \lum~), density ($>10^4 $~cm$^{-3}$), and strong infrared dust emission (c.f. \citealp{williams04, garay03, garay04, beuther02}). Little is known of the very earliest stages in the formation of massive stars, especially the conditions inside the early protostellar core and the putative prestellar cores. These objects are expected to be massive and cold ($\sim 20$ K) (c.f. \citealp{mckee03}).

   Methanol masers have proven particularly useful as tracers of massive star formation (e.g. \citealp{batrla87, caswell95, pestalozzi04}). Methanol masers, are occasionally associated with strong radio continuum emission (i.e. \hii~ regions), \IRAS~ far-infrared colour selected sources (e.g \citet{wood89col} used log (F$_{60}$/F$_{12}$)~$\ge$~1.30 \& log~(F$_{25}$/F$_{12}$)~$\ge$ 0.57), and H$_{2}$O and OH masers \citep{caswell95}. 

   Mid infrared emission from the Midcourse Space Experiment (\MSX) satellite has also been used, in a similar way as \IRAS, in colour selecting  massive young stellar objects (MYSO). \citet{lumsden02} found that the \MSX~ colours of the youngest sources, still heavily embedded in the natal molecular clouds, are different from evolved stars which are shrouded in their own dust shells. They have developed colour-selection criteria based on \MSX~ colours, (F$_{21}$/F$_{8}$~$>$~2) designed to deliver a list of MYSO candidates.

   The coincidence of methanol masers and/or ultra compact \hii~ regions (\uchii) with massive star formation  suggests that the two (tracers and massive stars) are inextricably linked. Consequently, maser and radio continuum emission can be used as a means of tracing regions of massive star formation (MSF). High angular resolution ($\sim0.01''$ - $1''$) observations (\citealp{caswell96, phillips98, walsh98, minier01})  have shown that methanol masers are generally {\it not} directly associated with the \uchii~ regions, but rather tend to be separated from them and possibly associated with hot molecular cores (HMCs). 

    \citet{walsh98} have reported observations where $\sim25\%$ of \uchii~ regions targeted were associated with methanol masers. In this instance, the size of the \uchii~ region is generally smaller when there are masers present, suggesting such regions are possibly younger. This discovery, together with the fact that $\sim75\%$ of the masers were not associated with \uchii~ regions, led Walsh and his colleagues to propose an evolutionary sequence for massive star formation. They proposed that the methanol masers exist prior to the onset and development of the \uchii~ region, the maser emission is then destroyed by an ionising region surrounding the stars, and the \uchii~ region evolves following the destruction of the maser.

   More recently, \citet{walsh03} have shown that methanol masers are associated with sub-millimetre continuum emission, and hence trace cold, deeply embedded objects.    

   An alternative hypothesis to explain the low correlation between \uchii~ regions and methanol masers is proposed by \citet{phillips98}. They suggest instead, that maser emission can arise from intermediate mass non-ionising stars, which yield sufficient IR photons to pump the masing transition, but insufficient UV photons to produce an \uchii~ region. 

   In this paper, we present the results of a 1.2 millimetre continuum survey of massive star formation regions exhibiting signs of methanol maser and/or radio continuum emission. We will show that these tracers of massive star formation are in most instances associated with millimetre emission, and hence cold, deeply embedded sources. We will also present evidence of cores to which the only indicator to their existence is the millimetre continuum emission detected with the SEST. These millimetre sources (hereafter `mm-only cores') are devoid of methanol maser and \uchii~ emission. Other millimetre surveys have been undertaken by \citet{beuther02, faundez04, williams04}, and we draw attention to sources from these studies which overlap our own source list in Table \ref{main}.

   The main purpose of this paper is to present the results from our study of known regions of massive star formation. In subsequent papers, we will present spectral energy distribution (SED) diagrams, which we aim to use to test current theories in the evolution of massive stars (e.g. \citealp{ phillips98, walsh98, minier05}). 

  In section 2, we describe the observations and the data reduction procedure. The results of the SEST/SIMBA survey are discussed in $\S$3, derived physical quantities in $\S$4, and data including the sample images are presented in $\S$5. In $\S$6 we present the analysis of the data, and conclusions are given in $\S$7. The full set of images (including the ones presented in $\S5$ as sample images) are presented in the Appendix.

\section[]{Observations and Data Reduction}
\subsection{The sample}

   The sources chosen for this millimetre study were selected from previous studies of massive star formation regions, in particular the work of \citeauthor{walsh97}, as well as \citeauthor{thompson04}, and \citeauthor{minier01},  as discussed below.
   
The criteria for selecting the source list included:
\\
\begin{footnotesize}
%\small
\begin{enumerate}
\item sources with masers and without radio continuum emission,
\item sources with masers and with radio continuum emission,
\item radio continuum sources without methanol masers.\\
\end{enumerate}
\end{footnotesize}

   \citet{walsh97} undertook a study of \uchii~ region candidates in search of the 6.7 GHz emission line characteristic of methanol masers. The specific \uchii~ regions targeted in this survey were chosen based on their far-IR \IRAS~ colours according to the \citet{wood89col} selection criteria for identifying \uchii~ regions.

   \citet{thompson04} undertook a sub-millimetre study of \IRAS~ point sources with associated \uchii~ regions, which had previously been identified by \citet{wood89col} and \citet{kurtz94}.

    \citet{minier01} observed methanol masers at very high angular resolution (10 mas), which were devoid of radio continuum emission, and were believed to be associated with high mass protostars.

   The objective of our survey towards known methanol maser sites and \uchii~ regions was to ascertain whether these objects are associated with millimetre continuum emission, and thus, deeply embedded objects. 

\subsection{Millimetre Observations}

   The observations  were undertaken on the Swedish ESO Sub-millimetre Telescope (SEST), using the SEST IMaging Bolometer Array (SIMBA) during three separate observing periods between October 2001 and October 2002. 

   SIMBA\footnote{See http://www.ls.eso.org/lasilla/Telescopes/SEST/html/telescope-instruments/simba/index.html}  is a 37-channel hexagonal bolometer array, operating at a central frequency of 250 GHz (1.2~mm), with a main beam efficiency of about 0.50 and a bandwidth of 50~GHz. SIMBA has a HPBW of $24''$ for a single element, and the separation between elements on the sky is $44''$. Observations were taken using a fixed secondary mirror in a fast mapping observing mode, with a typical scan speed of $80''s^{-1}$. The resultant pixel size of the maps, after processing is $8''$. % Typical rms noise levels as reported by the SEST are around 40~\mjy, which is likely to improve to 30 \mjy~ in excellent weather conditions.

  The initial observations were conducted during the second commissioning period of the SIMBA instrument in October 2001. 19 regions were mapped during this period. Skydips were performed every three hours in order to correct for the atmospheric opacity. Opacities stabilised around 0.35 for the first half of the night, increasing to 1.00 toward the end of the night. Maps of Jupiter were taken for calibration purposes. Typical map integration times were 15 minutes per map, mapping regions of $600'' \times 384''$. The average residual noise in the maps for this period is $\sim150$ \mjy\footnote{This value is 4-5 times higher than typical noise values reported by the SEST in good weather conditions.}. Noise residuals in the maps may be attributed to variable sky opacities during the latter part of the observations.

   The majority of the data collection took place during seven second-half nights in June 2002 (23rd--29th inclusive). During this period, observations typically spanned 8 hours each night, allowing 115 regions to be mapped. In order to accurately monitor the sky conditions, skydips were performed following every second map ($\sim 30$ minutes). Sky opacities for this period fluctuated on a nightly basis, with average values for each night listed in Table \ref{calfactors}. No observations were taken on night three or seven due to bad weather. Maps of Uranus were taken for flux calibration purposes. 

 Due to a lack of suitable calibration data for the first night, the Uranus map from night four was adopted for calibration purposes, a process justified by the similar opacities for the two periods. Typical map integration times for this period were 15 minutes per source, mapping regions of $240'' \times 480''$. The residual noise in the maps for this period averages around 50 \mjy.

   The final set of observations were taken during four nights in October 2002 (21st--24th inclusive), over a period of $\sim5$ hours each night, mapping 38 regions in total.  The sky opacity was measured regularly by taking skydips after every second observation ($\sim 30$ minutes). Sky opacities for this period fluctuated on a nightly basis, with typical values for each night listed in Table  \ref{calfactors}. Maps of Uranus were taken each night (with the exception of the first night, where Eta Car was observed) for flux calibration purposes. Due to the unsuitability of Eta Car as a flux calibrator, Uranus data from night four were used to calibrate the data from the first night (based on similar opacities for the two nights). Typical map integration times for this period were 15 minutes per source, mapping regions of $480'' \times 480''$. The residual noise in the maps for this period averages around 60 \mjy.

 Specific calibration factors, as well as typical opacities for each night of observation are listed in Table \ref{calfactors} .
 
\begin{table} 
  \caption{Summary of the calibration factors and opacities for each night of the SEST observations. An asterisk (*) denotes calibration based on the Uranus data taken from night 4 of the same run. The dash (-) indicates that no data was taken on this occasion. Note that the opacities listed in the table are typical values for each night. \label{calfactors} }
  \begin{center}
    \footnotesize
    \begin{tabular}{|lcr|}
      \hline
      \hline
Date     & Calib. Factor             & Opacities\\
dd mm yy & \mjycount~                &  $\tau$~~~~~ \\
         &  $beam^{-1}$              & \\
\hline
26th Oct 01 & 138.8	&   $\ge0.35$ \\ %138.8
%\\
23rd Jun 02 & 65.9*	&   0.25 \\ %65.89
24th Jun 02 & 83.0	&   0.28 \\ %82.99
25th Jun 02 & -	        &    -   \\
26th Jun 02 & 66.3	&   0.18 \\ %66.32
27th Jun 02 & 66.6	&   0.24 \\ %66.56
28th Jun 02 & 84.0	&   0.28 \\ %83.99
29th Jun 02 & -         &    -   \\
%\\
21st Oct 02 & 69.7*	&   0.40 \\ %69.74
22nd Oct 02 & 76.0	&   0.25 \\ %76.01
23rd Oct 02 & 87.7	&   0.28 \\ %87.67
24th Oct 02 & 69.4	&   0.30 \\ %69.41
\hline
    \end{tabular}
  \end{center}
%\vspace{-0.4cm}
\end{table}

\subsection{Data Analysis}
 
   The data were reduced and analysed using the MOPSI reduction package\footnote{MOPSI (Mapping, On-off, Pointing, Skydip, Imaging) is a software reduction tool which was developed and is maintained by R. Zylka, IRAM, Grenoble, France. MOPSI makes use of the GreG graphics interface of the GILDAS software distribution. See http://www.mpe.mpg.de/ir/ir\_software.php} and the procedure described in the SEST manual. In summary, the data are subject to gain elevation correction, opacity correction, baseline fitting and subtraction, despiking, deconvolution,
%\footnote{Deconvolution makes use of the deconvolution algorithm developed within the SIMBA collaboration, which is designed to remove the contribution of the electronics which arises as a consequence of the fast-mapping mode of observation.},
 and sky noise reduction, prior to map creation and calibration, as described in the manual.

   The flux density values were obtained using the MOPSI photometry procedure described in the SEST manual. In brief, this procedure involves distinguishing the source from the background and subtracting the latter from the former, using Gaussian fitting. This technique uses polynomials to fit and subtract the baselines, and polygons to define the apertures. The MOPSI {\it Integrate} procedure, which employs the multiplication factor listed in Table  \ref{calfactors}, was used to determine the flux density of the source. %Due to the high density of sources in the regions that we have mapped, care had to be taken in calculating the integrated fluxes to avoid source confusion.

   The flux density of each source was also estimated using the KARMA/KVIEW package\footnote{See http://www.atnf.csiro.au/computing/software/karma} by defining a box aperture around the source and at various points in the image considered to be the background. The flux inside of each of the apertures was then measured and the background flux was subtracted. Contour levels of 5\% of the peak flux were overlaid on the sources in order to maintain source size consistency when using the box aperture.

   A comparison of the flux determination via the two methods (i.e. MOPSI and KARMA) shows that there is $\le 10\%$ difference between the two results, which arises as a consequence of the aperture definitions. Due to the greater flexibility in using the aperture applied in the MOPSI photometry procedure (i.e. polygon, which allows more accurate source definition) the fluxes determined from the MOPSI reduction procedure have been adopted as the 1.2mm flux for each of the sources, and are presented in this paper in Table \ref{main} (see also $\S5.2.1$ for further explanation).

   The free-free emission contributing to the millimetre fluxes reported in Table \ref{main} is expected to be negligible compared to the dust emission which dominates at 1.2~mm (F$_{\nu}$~$\propto$~$\nu^{-0.1}$). Assuming optically thin free-free emission for frequencies  $\ge$ 8 GHz, a typical \uchii~ region measured by \citet{walsh98} at 8~ GHz to have a peak flux of 120 mJy equates to a peak flux of 85~mJy at 250~GHz, i.e. 4\% of the 1.2 mm flux measured at the same position. Assuming that the 8~GHz flux measured by \citeauthor{walsh98} is optically thick from a hyper compact \hii~ region, with the turnover frequency at 22~GHz, the 120~mJy flux measured by \citet{walsh98} at 8~GHz equates to 700~mJy at 1.2~mm, which would still only be significant for a small fraction of our sources. In order to be optically thick at 22~GHz, a large emission measure of 2.0~$\times 10^{9}$~cm$^{-6}$~pc is required.

   Stellar winds, with $\nu^{0.6}$ (\citealp{panagia75})  are not expected to contribute more than $\sim 20\%$ to the flux density observed at 1.2 mm as derived from comparison of the SIMBA fluxes with typical radio fluxes measured by \citealp{walsh98}.
  
   On at least one occasion during the June and October 2002 observations, calibration data were taken twice in a single night, in order to determine the repeatability of the data. Analysis of these Uranus maps shows less than 3\% deviation in the calibration factor for the June data, and less than 1\% deviation in the case of the October data. Therefore, the error resulting from measurement of the calibrator is small.

\section[]{Results}

   131 known regions of massive star formation were targeted in this 1.2 mm continuum survey. The images of these regions as well as derived source properties, such as flux density and mass are presented in this paper. Of these 131 positions targeted, 69 are positions of known methanol masers, 28 have radio continuum sources, 32 are known to harbour both a methanol maser and an \uchii~ region, and the remaining two are \IRAS~ positions\footnote{\citet{walsh97} report these particular \IRAS~ sources (see Table \ref{main}) as meeting the Wood \& Churchwell \IRAS~ colour selection criteria for \uchii~ regions but having no positive methanol identification.}.  These two \IRAS~ positions are hereafter referred to as `NM-\IRAS~ positions', indicating `no maser' emission.

   Within these 131 regions, millimetre continuum emission is detected toward a total of 404 sources, 78 of which contain methanol masers, 36 have \uchii~ regions, 35 have both methanol maser and radio continuum emission, and two are the NM-\IRAS~ positions. The remaining 253 sources detected are `mm-only' cores, to which the only indicator of their existence is the millimetre-wave continuum emission detected in this survey. As discussed below, 45\% of these also have no mid-infrared emission detected by the \MSX~ satellite.

   The mm-only cores detected by SIMBA are previously unknown and are devoid of the traditional star formation identifiers, such as methanol maser and radio continuum emission. Prior to the detection of the millimetre emission of these new cores, there was no indication (i.e. tracer) that star formation was taking place in these regions. The majority of the mm-only cores are separate from and are generally offset from the targeted tracer in the same field.

  The millimetre-emitting cores are considered `sources' if they are detectable at a $3\sigma$ level above the background. Many of the images presented contain multiple sources. In such fields, it is likely that all of the millimetre sources belong to the same star forming complex as the objects targeted in the fields. We have made this assumption when assigning distances (see also $\S\ref{distances}$) to the mm-only cores,  based on maser velocities as presented in Table \ref{main}.

   Millimetre continuum emission is detected toward all 131 of the methanol maser and \uchii~ regions targeted, confirming their association with cold, deeply embedded objects. The majority of the known methanol maser and radio continuum sources are spatially encompassed within the contours of the millimetre emission, and are often directly associated with (i.e. not more than $30''$ offset from) the peak millimetre position.

   In a few cases (8), the methanol maser and/or \uchii~ region is quite offset from (i.e. $>30''$) the peak of the millimetre emission, yet their positions still fall within, or at the edge of, the  contours of the SIMBA source. The positions of these masers, presented in Table \ref{tab_offsets}, have all been determined from interferometry and are thus accurate within 1~arcsecond.

\begin{table}
\caption{Positions of methanol maser sites and \uchii~ regions within the source contours,  but offset $>30''$ from the peak millimetre continuum emission. All positions are measured by interferometry. \label{tab_offsets}}
  \begin{center}
  \footnotesize
  \begin{tabular}{|lllll|}
\hline
\hline
RA  & Dec  & Map  & Tracer & Ref\\
J2000  & 2000 &Identified & & \\
\hline
05 41 41.4 & -01 53 37 & G206.54-16.35 & radio$^\ddagger$  &$^a$\\ %ATCA 54'' in contours
05 41 45.8 & -01 54 30 & G206.54-16.35 & radio$^\ddagger$ & $^a$\\ %ATCA 49'' in contours
17 59 06.0 & -24 21 16 & G5.48-0.24 & radio$^\diamond$ & $^a$\\ % ATCA 40'' edge of contours
18 12 37.5 & -18 24 08 & G12.18-0.12 & maser$^\diamond$ & $^a$ \\ % ATCA 31''edge of contours
18 12 40.2 & -18 24 47 & G12.18-0.12 & maser$^\diamond$ & $^a$ \\ % ATCA 35'' edge of contours
18 16 59.8 & -16 14 50 & G14.60+0.01 & radio$^\ddagger$ & $^a$ \\ %ATCA 41''in contours
18 34 08.1 & -07 18 18 & G24.47+0.49 & radio$^\ddagger$ & $^a$ \\ %ATCA 37'' in contours
18 46 03.7 & -02 41 53 & G29.918-0.014 & maser$^\diamond$ & $^b$ \\ %ATCA  39'' edge of contours
\hline
\end{tabular}
\end{center}
$^a$ Positions taken from \citet{walsh98}.\\
$^b$ Positions taken from \citet{pestalozzi04}.\\ 
$^\ddagger$ Denotes sources still within contours of main source.\\
$^\diamond$ Denotes sources located at the edge of the source contours.\\
\end{table}

  Interestingly, the SIMBA maps also show that there are methanol maser sites and \uchii~ regions devoid of millimetre continuum emission. The 20 methanol masers and 9 \uchii~ regions where this occurs are listed in Table \ref{no_mm}. The positions of the majority of these tracers ($> 90\%$) have also been determined from interferometry and thus have accuracies to within 1~arcsecond. Consequently, the lack of association is not due to poor positional information. These objects are discussed further in $\S\ref{discussion}$. 

\begin{table}
  \caption{Coordinates of methanol maser sites and \uchii~ regions devoid of coincident 1.2 mm continuum emission. The rms noise in the maps is typically 50-60 mJy for all sources, except for the map of G270.25+0.84 which has a rms noise value of 150 mJy. All positions have been measured from interferometry, except those sources otherwise indicated.\label{no_mm}}
  \begin{center}
  \footnotesize
  \begin{tabular}{|lllll|}
\hline
\hline
RA  & Dec  & Map  & Tracer & Ref\\
J2000  & 2000 &Identified & & \\
\hline
05 51 06.0 & +25 45 45 & G183.34+0.59 & maser & $^c$\\ % ATCA 68''
06 08 36.1 & +21 30 28 & G189.03+0.76 & maser & $^c$\\ % ATCA 62''
06 09 13.8 & +21 53 13 & G188.79+1.02 & radio & $^a$ \\
09 16 41.4 & -47 55 46 & G270.25+0.84 & maser &   $^c$\\ % ATCA 34''
09 16 51.9 & -47 54 33 & G270.25+0.84 & radio & $^b$ \\
13 14 07.0 & -62 45 50 & G305.55+0.01 & radio & $^b$\\
13 21 27.6 & -63 00 48 & G306.33-0.30 & maser &  $^c$\\ % ATCA 44''
16 10 25.8 & -51 55 04 & G330.95-0.18 & radio &$^b$\\
18 00 50.9 & -23 21 29 & G6.53-0.10 & maser & $^b$\\
18 00 54.1 & -23 17 02 & G6.53-0.10 & maser$\dagger$ & $^c$\\
18 03 34.7 & -24 24 10 & G5.97-1.17 & radio & $^b$\\
18 03 52.4 & -24 23 48 & G5.97-1.17 & radio & $^b$\\
18 05 18.2 & -19 51 15 & G10.10+0.73 & maser$^\star$ & $^c$\\
18 10 09.9 & -19 53 22 & G10.62-0.33 & maser & $^c$\\
18 11 48.8 & -18 33 43 & G12.02-0.03 & radio$^\dagger$ & $^b$\\
18 12 41.0 & -18 26 22 & G12.18-0.12 & maser & $^b$\\
18 13 43.4 & -17 58 06 & G12.68-0.18 & maser & $^c$\\
18 27 13.5 & -11 53 16 & G19.61+0.10 & maser & $^b$\\
18 34 44.9 & -08 31 07 & G23.43-0.18 & radio & $^b$\\
18 36 24.0 & -07 04 27 & G24.84+0.08 & maser & $^c$\\
18 45 44.2 & -02 39 04 & G29.918-0.014 &  maser &$^c$\\
18 46 09.9 &  -02 36 31 & G29.918-0.014 &  maser &$^c$\\
18 47 23.8 & -01 42 39 & G30.89+0.16 & maser$^\star$ & $^c$\\
18 47 29.9 & -01 54 39 & G30.70-region &  maser & $^c$\\
18 47 37.5 & -02 08 46 & G30.59-0.04 & maser &  $^c$\\
18 51 58.9 & +00 07 27 & G33.13 -0.09 & maser &  $^c$ \\
18 54 04.2 & +02 01 36 & G35.02+0.35 & radio &  $^a$ \\
19 00 14.4 & +04 02 35 & G37.55-0.11 & maser & $^b$ \\ 
19 23 53.6 & +14 34 54 & G49.49-0.38 & maser$^\dagger$ &$^c$\\
\hline
\end{tabular}
\end{center}
$^a$ Positions taken from \citet{kurtz94}.\\
$^b$ Positions taken from \citet{walsh98}.\\
$^c$ Positions taken from \citet{pestalozzi04}.\\
$^\dagger$ Tracer located near edge of map.\\
$^\star$ Denotes positions determined from single dish measurements.\\
\end{table}

   Examination of the SIMBA sources for mid-infrared \MSX~ emission reveals that 72 sources are entirely without mid-IR emission at all wavelengths (8\microm, 12\microm, 14\microm, and 21\microm); 41 are mid-IR dark clouds\footnote{\citet{egan98} define dark clouds as small clouds seen in silhouette against the bright emission of the Galactic plane, characterised by high densities (~$>$~10$^5$~cm$^{-3}$) and low temperatures ($\sim$~10~K). The clouds are seen in absorption at the \MSX~ wave bands.}; and another 11 are potential mid-IR dark clouds, that is, they are sources devoid of emission, but it is unclear whether the lack of emission is due to absorption as is the case with the dark clouds, or simply a lack of emission. These associations are given in Table \ref{main}. Due to an excess of mid-IR emission in the fields examined, it is often not possible to distinguish individual associations due to confusion. Therefore the absence of an \MSX~ association in Table \ref{main} {\it does not} indicate a positive mid-IR identification. Consequently, the $\sim30\%$ of SIMBA sources which we report as having no mid infrared emission is a lower limit to the actual percentage. For these sources, the only indicator that star formation may be occurring is the millimetre continuum data detected in this survey.

   The SIMBA maps also reveal two types of source located at the edge of the fields. For those sources where the emission extends off the edge of the map, the flux density is not calculated due to ambiguity in source sizes. Often the peak emission can not be reported for these sources, which are denoted by a $\beta$ in column 6 of Table \ref{main}. Sources denoted by a $\gamma$ are sources which are situated close to the edge of the map, but far enough away such that we can be reasonably confident in estimating a source size and hence a flux. However due to incomplete mapping, it is possible that the size of these sources, and hence their flux has been underestimated. Therefore the flux reported for sources denoted by a $\gamma$ is likely a lower limit.

   In the case of the two NM-\IRAS~ positions targeted, no millimetre emission was detected at the targeted \IRAS~ position. These sources (G305.533+0.360 and G305.952+0.555), however, have  many other millimetre sources in the SIMBA fields. We attribute these two `non-detections' to the low spatial resolution offered by \IRAS~ in pin-pointing the peak emission of the central core.
 
   A small selection of the images taken with the SIMBA instrument are presented in $\S5$, Figure \ref{sample}, and all of the images are presented in the Appendix. Source names are derived from their Galactic coordinates. The coordinates listed in Table \ref{main}, correspond to the position of the peak millimetre emission for each source.

\section[]{Derivation of Physical Parameters} 

\subsection []{Mass \label{mass}}

  Assuming that the 1.2mm continuum emission detected toward these regions of massive star formation is from optically thin dust, the gas mass can be estimated using the following equation:

\begin{equation}
M_{gas}= \frac{S_{\nu}d^2}{{\kappa_d}B_{\nu}(T_{dust})R_d},
\label{eqmass}
\end{equation}

\noindent
where $S_{\nu}$ is the 1.2mm continuum integrated flux, $d$ is the distance to the source, ${\kappa}_d$ is the %{\kappa}_0({\nu}/{\nu}_0)^{\beta}$ the 
mass absorption coefficient per unit mass of dust, $B_{\nu}$ is the Planck function for a blackbody of temperature $T_{dust}$, and $R_{d}$ is the dust to gas mass ratio.
   
   While values of ${\kappa}_d$ and $R_d$ may vary between sources, we adopt a value of 0.1~m$^2$~kg$^{-1}$ for the mass absorption coefficient ($\kappa_d$) as per the opacity models of \citet{ossenkopf94} (c.f. \citealp{minier05}). This gives mass estimates four times smaller than those derived using the opacity models of \citet{hildebrand83}. We assume a dust to gas ratio ($R_d$) of 1:100 (i.e. 1\%). $B_{\nu}(T_{dust})$ has been evaluated for a dust temperature of 20~K (see $\S\ref{temperature}$).

  The masses determined from equation \ref{eqmass} are listed in column 10 of Table \ref{main}. A histogram of the distribution of masses is included in Fig. \ref{hist}, with a cumulative distribution shown in Fig. \ref{cumul}.

\subsection{Temperature \label{temperature}}

 The derivation of the mass of a source depends on temperature as per equation \ref{eqmass}. Assuming that these sources are cold cores, as indicated by the presence (or lack-there-of in the instance of the mm-only cores) of methanol maser or radio continuum emission, we have assumed a temperature of 20 K for the purposes of mass derivation, which is consistent with temperatures reported in the literature. \citet{faundez04} found that 1.2 mm cores associated with \IRAS~ sources have typical dust temperatures of 32 K, while \citet{garay04} found that 1.2 mm massive dust cores without emission at far infrared wavelengths have temperatures of 17 K. \citet{minier05} derive core temperatures $< 50$ K and as low as 16 K for sources with no mid-IR emission, and \citet{motte03} adopt temperatures of 20-30 K for their sample of sub-millimetre cores.

   We refrain from categorising the sources based on the presence of their tracer, and thus do not apply different temperatures to those cores suspected of being at different stages of evolution. We aim to constrain the temperature of each of the individual cores in forthcoming papers by compiling spectral energy distributions (SEDs). The aim of this paper is to present the images and the mass estimates for each millimetre core of the SIMBA survey.

   In assuming a temperature, there is an additional uncertainty in the final mass estimate derived. We present in Table \ref{scalef}, the scaling factor for deriving the mass for different temperatures, from the value determined for a temperature of 20 K. For example, if the temperature were 40 K instead of 20 K, then the mass that we report will be 2.3 times larger than the actual value.

%which can be extrapolated from the 20 K mass estimate. For example, if the temperature of a source were 40 K instead of the 20 K, then the mass that we report will be 2.3 times larger then the true mass of the object.

% {\bf mass Prop 1/temp therefore for 20 K mass prop 1/20 to get 40k prop 1/temp*2.5, where 1/2.5 =0.4. therefore need to do mass * 0.4 to get real mass.}

\begin{table} 
  \caption{Scaling factor for mass estimates using different temperatures to the 20~K taken here. For other temperatures, divide the value reported in Table \ref{main} by the scaling factor in this table. \label{scalef}}
  \begin{center}
%    \footnotesize
    \begin{tabular}{|cc|}
      \hline
      \hline
Temp. & Scaling \\
 (K)      &  Factor\\
\hline
10 & 0.4\\%0.354
20 & 1.0\\%1.00
30 & 1.7\\ %1.671
40 & 2.3\\ %2.349
50 & 3.0\\ %3.031
60 & 3.7\\ %3.713
70 & 4.4\\ %4.396 
\hline
    \end{tabular} 
  \end{center}
\end{table}

\subsection[]{Distances \label{distances}}
  
   The distance for each of the methanol masers and \uchii~ regions targeted in this survey is taken from the literature, and listed in column 9 of Table \ref{main}, together with the literature reference\footnote{In some instances, this involved correcting the data for a Galactic Centre distance of 8.5~kpc rather than 10~kpc.}.

  For the mm-only cores, we have assumed that they are at the same distance as the targeted source in the field, as indicated in column 5 of Table \ref{main}. According to \citet{blitz93} the mean diameter of a giant molecular cloud (GMC) is 45~pc in diameter. Projecting the SIMBA maps on the sky, gives a spatial size of 0.4 $\times$ 0.7~pc and 20 $\times$ 40~pc for the distance extremes of 0.3 and 16.7~kpc respectively for a map size of $240'' \times 480''$. Therefore all of the cores in a single map are likely to be located within the same GMC and hence we can make the assumption that the mm-only cores lie at the same heliocentric distance as the maser and/or the \uchii~ region cores within the same fields.

  Many of the sources have a near-far distance ambiguity. 195 sources have a clear kinematic distance, whilst for 197, the near-far distance ambiguity exists. The two NM-\IRAS~ positions targeted have no distance estimate. The near and far distances are listed in column 9 of Table \ref{main}, with the near distances preceding the far and separated by a `/'. For 12 sources in total, there is no known distance, which is indicated by the `Ind' (Indeterminate) in the same column.

   For analytical purposes, we have assumed the near distance for the 197 sources with a distance ambiguity (c.f. \citealp{williams04}). We do not expect the results and conclusions to be significantly affected by this assumption. As confirmation, we also analyse the sources with no distance ambiguity separately and present the results (c.f. $\S\ref{histograms}$).

   G10.62-0.33 is reported by \citet{walsh97} to have a kinematic distance of 21.6 kpc, which has likely been overestimated. Due to the uncertainty in this distance estimate, and the close proximity of this source to G10.62-0.38 (c.f. Fig. A1, map: G10.62-0.33), we have adopted the distance of the latter to the former and its mm-only companion. According to \citet{walsh03} G10.62-0.38 has a kinematic distance of 6 kpc.

\subsection[]{Source Size and H$_{2}$ Number Density ($n_{H_{2}}$)}

  The spatial size (FWHM) of each of the sources listed in Table \ref{main}, was determined using the Graphical Astronomy and Image Analysis Tool (GAIA)\footnote{See http://www.starlink.rl.ac.uk/gaia}. GAIA determines the FWHM (size) of the source in arcseconds, which can then be transformed into parsecs (pc) when the distance is known. %measures using the relation: $W_{(pc)}= d~tan~{\theta}$, where $W_{(pc)}$ is the size of the source in parsecs, $d$ is the distance to the source in pc, and ${\theta}$ is the angle that the source projects on the sky. 
Column 8 of Table \ref{main}, lists the FWHM of the sources in arcseconds, and column 4 of Table \ref{main2}, lists the radius of each source in parsecs.
%Column 2 of Table \ref{main2}, lists the size of the sources in arcseconds. Column 3 indicates the method used in determining the source size (either `A' denoting automatic or `M' indicating a manual determination), with a pixel size of $8''$ used for manual determinations.

  The H$_{2}$ number density ($n_{H_{2}}$) of each source was derived from its mass and volume estimates, assuming a spherical geometry and a mean mass per particle of $\mu~=~2.29~m_{H}$, which accounts for a 10\% contribution of Helium (c.f. \citealp{faundez04}). Using the parameters determined in Tables \ref{main} and \ref{main2} (i.e. mass, radius and $n_{H_{2}}$),  it is possible to determine the column density  ($N_{H_{2}}$) and the surface density ($\Sigma$) of the sources in our sample. The visual extinction ($A_v$) of the sources can also be estimated from the column density ($N_{H_{2}}$), where,  $A_v=N_{\rm H_2}/0.94\times10^{21}$~mag (\citealp{frerking82}). The parameters of $N_{H_{2}}$, $\Sigma$ and $A_v$ for each of the individual sources in the sample, have not been included here, however, average values of $\Sigma$ and $A_v$ are reported in $\S\ref{discussion}$.

% nh2= M/[4/3 pi r^3] / Mass Hydrogen (cm^-3)
% Nh2= M/pi r^2  / Mass Hydrogen (cm^-2)
% Surface density (\Sigma) = Mass/ pi r^2 (g cm-2)
% Av = $A_v=N_{\rm H_2}/0.94\times10^{21}

 The H$_{2}$ number density of each source is listed in column 5 of Table \ref{main2}. For all sources with a reported distance ambiguity in the literature, the H$_{2}$ number density ($n_{H_{2}}$) for both the near and far distances are reported, with the near preceding the far, and separated by a `/'. Histogram plots of the H$_{2}$ number density are given in Fig. \ref{hist}, with cumulative distributions given in Fig. \ref{cumul}.

\subsection[]{G10.10+0.73: an exception}

   G10.10+0.73 has an associated radio continuum source with a methanol maser source offset ($\sim80''$) from it. Millimetre continuum emission is detected toward the radio continuum source, but not toward the methanol maser.
   
   Interestingly, this maser source is also the only maser source toward which \citet{walsh03} did not detect submillimetre continuum emission in their SCUBA survey of methanol masers. Taking the far distance of 16.3 kpc and the 3$\sigma$ sensitivity limit of 150 mJy yields an upper limit for the mass of any continuum source at 600 \mstar.

   However the radio continuum source is associated with a planetary nebula, which would suggest that G10.10+0.73 is located at the near distance (0.3 kpc), rather than the far.

   Given the Galactic latitude of G10.10+0.73, if it were located at the far distance of 16.3 kpc, then it would lie $\sim$~200 pc above the Galactic plane. The probability of a massive star forming region existing this far above the Galactic plane seems small (\citealp{reed00}). Taking the near distance for the maser source, however,  in accordance with this argument yields a mass of 0.2\mstar~ at a 3$\sigma$ sensitivity limit of 150 mJy at 1.2~mm. With an upper limit of 0.2\mstar, it is unlikely that any star will form let alone develop a maser source. This therefore contradicts the previous argument and suggests that the maser source is more likely located at the far distance.
   
   Due to the uncertainty in assigning a distance to this source, we exclude it from the following analysis.
   
\section[]{Presentation of the Data}

\subsection{Sample Images \label{sampleimages}}

   Each of the sources detected in this survey, has been visually classified according to its morphology. The classification scheme used is as follows:

  An `S' denotes {\it singular sources}. A `D' indicates a single millimetre source at a contour level of 10\% of the peak emission of the SIMBA  map, within which there is a {\it double} core, each with their own peak of emission. An `A' is assigned to those cores quite clearly differentiated as separate millimetre sources at a contour level of 10\%, which are arranged spatially {\it adjacent} to each other in the SIMBA maps. Double cores differ from adjacent cores in that the emission is continuous between cores at a contour level of 10\% of the peak emission, whilst for adjacent cores, there is no continuous emission between cores, indicating that they are separate cores. `L' indicates those sources arranged spatially in a {\it linear} association, while `I' represents those sources closely associated, and spatially arranged in an {\it irregular} arrangement. A `U' is assigned to those sources whose morphology is {\it unknown}, which is usually given to those sources coincident with the edge of the map.
 
   The sources have been further categorised according to the strength of the clustering of the core. This classification was made according to visual analysis of each of the regions, based on their angular separations. Each source has been assigned either an `L', an `M', or a `T', indicating a {\it low}, {\it medium} or {\it tight} clustering association respectively. A {\it low} clustering association has been assigned to sources that appear in the same map, but which are distributed over a wider field than the {\it medium} and {\it tight} counterparts.

\begin{figure*}
  \begin{minipage}{1.0\textwidth}
         \includegraphics[width=7.0cm, height=7.0cm]{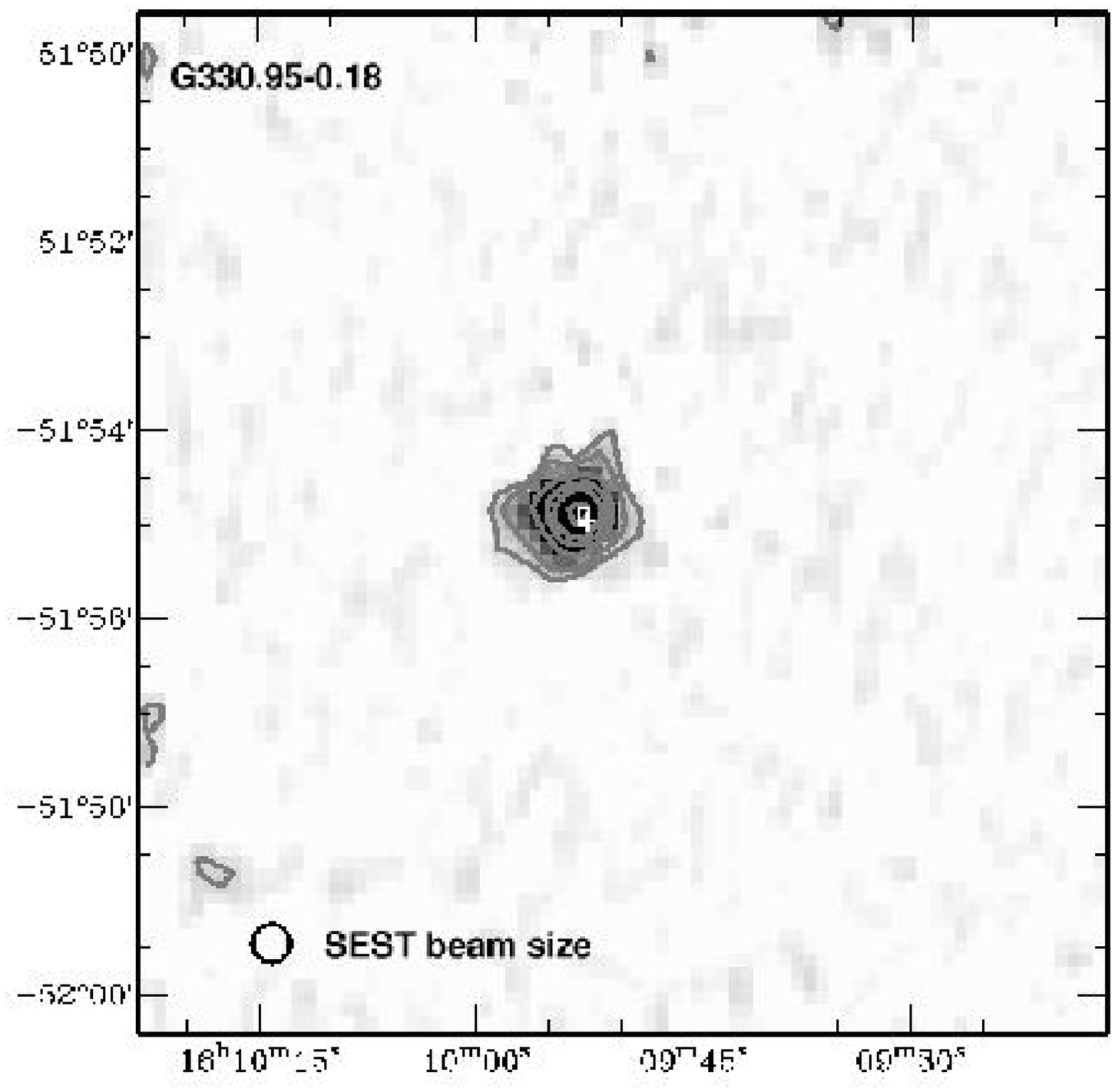}
	 \includegraphics[width=7.0cm, height=7.0cm]{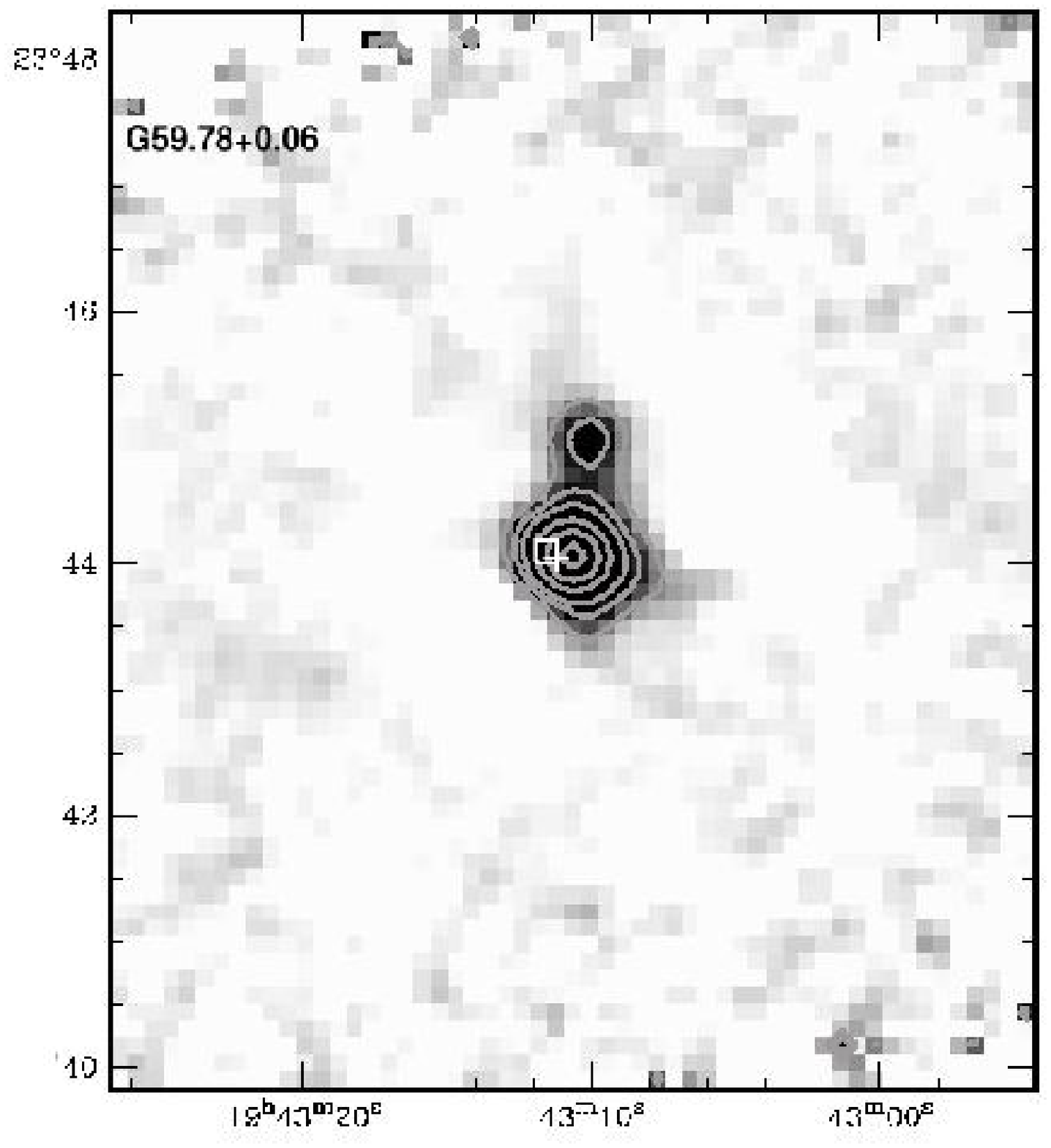}
  \end{minipage}
  \begin{minipage}{1.0\textwidth}
         \includegraphics[width=7.0cm, height=7.0cm]{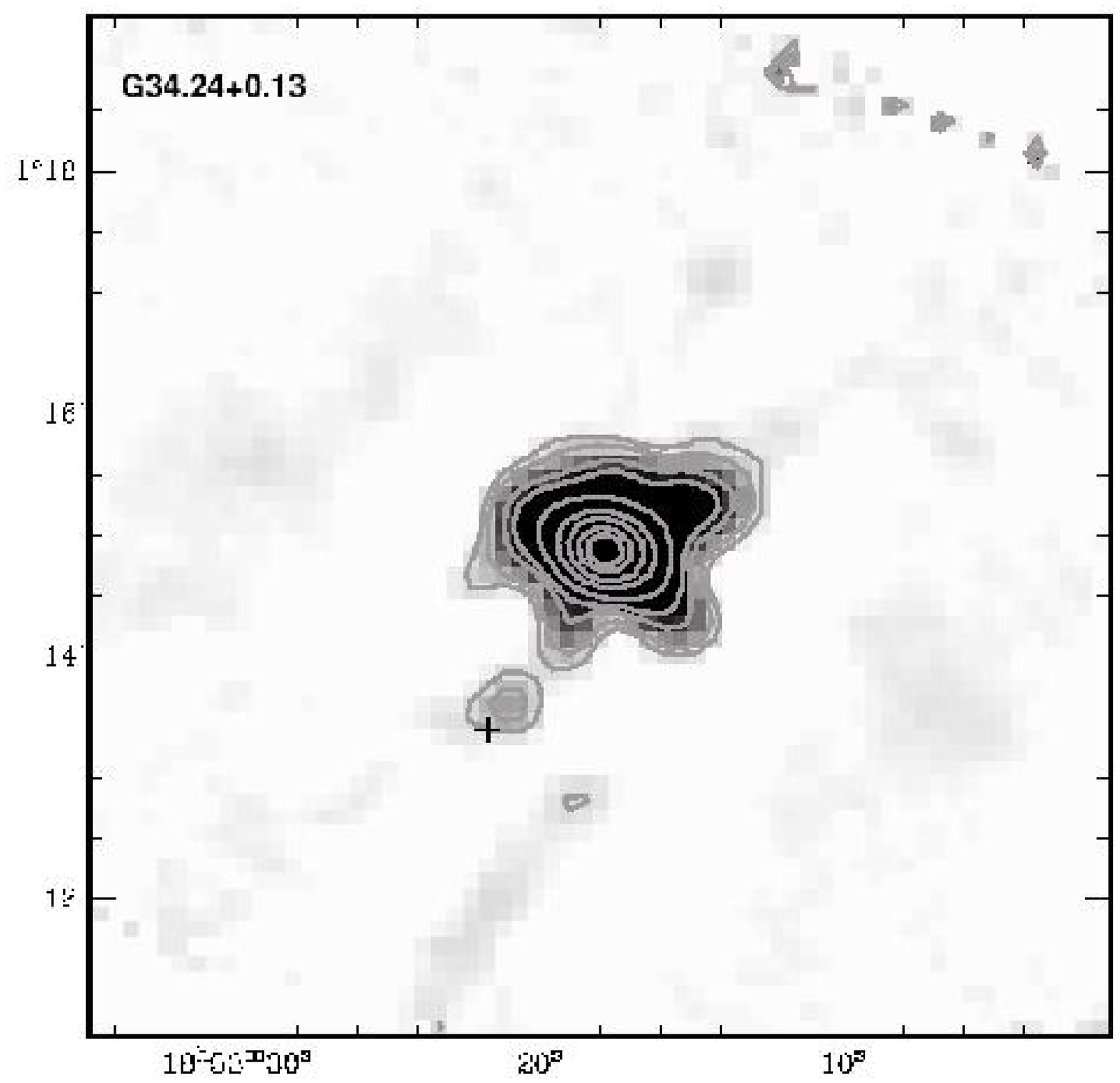}
	 \includegraphics[width=7.0cm, height=7.0cm]{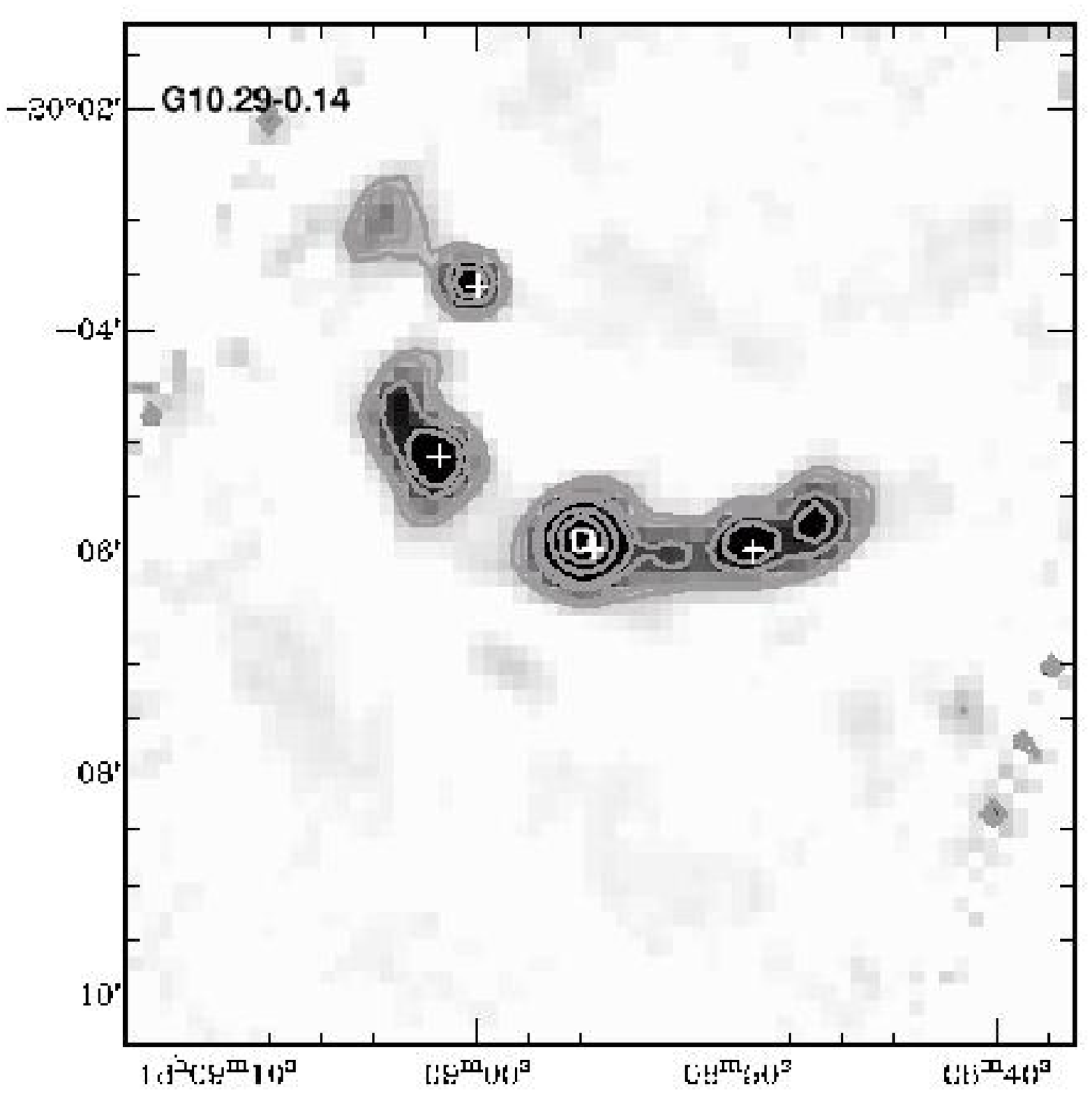}
  \end{minipage}
  \begin{minipage}{1.0\textwidth}
         \includegraphics[width=7.0cm, height=7.0cm]{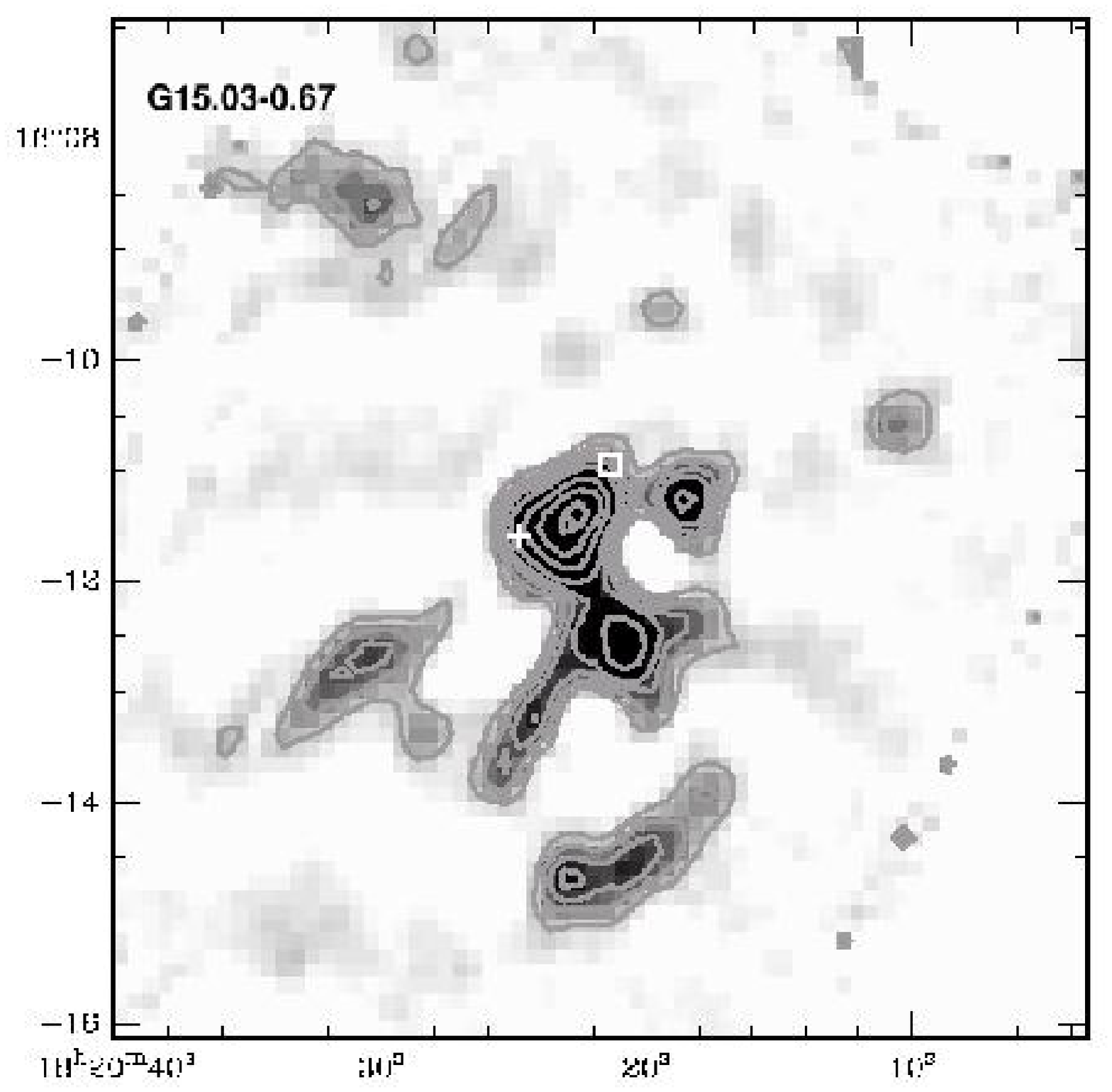}
         \includegraphics[width=7.0cm, height=7.0cm]{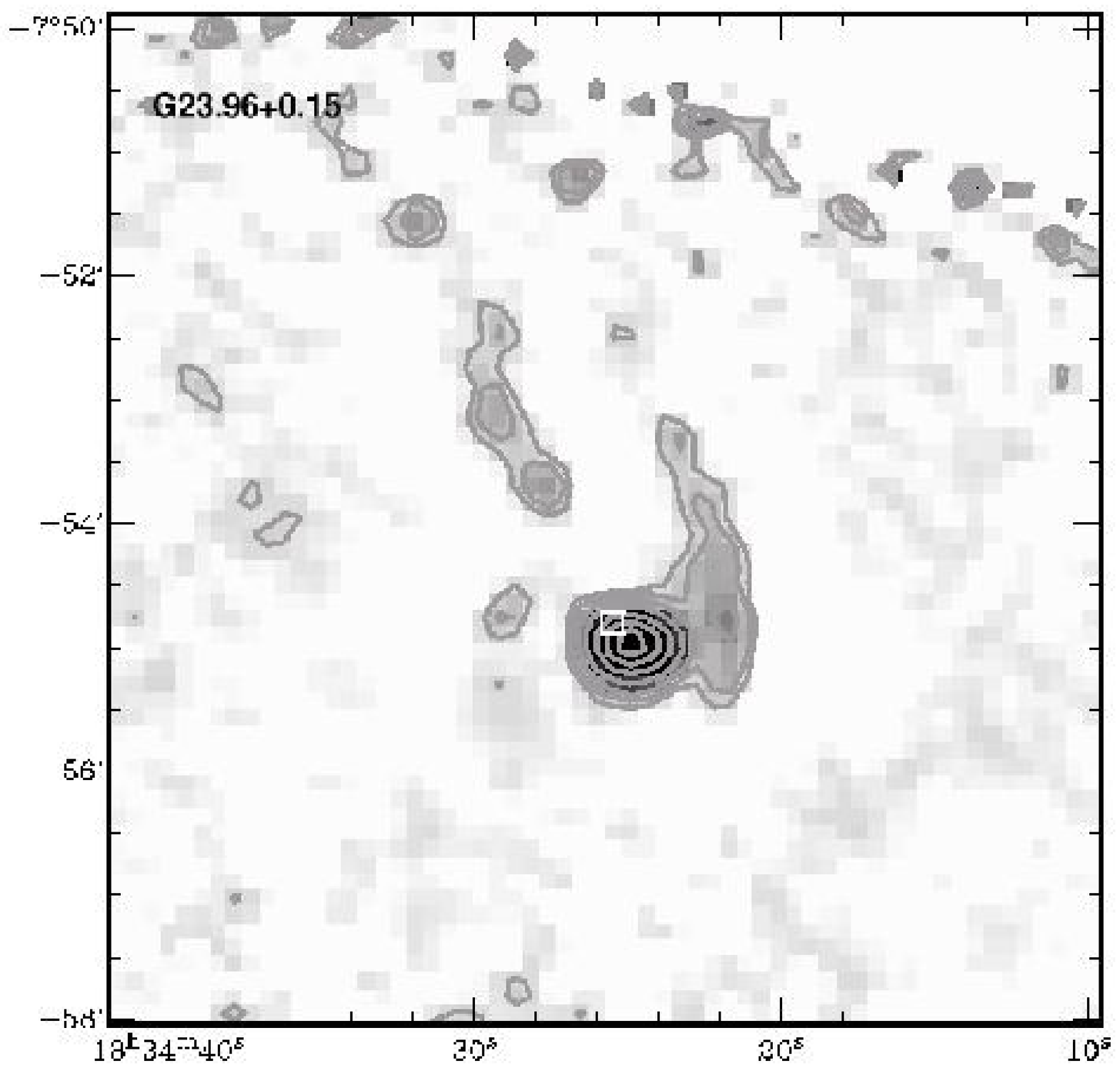} 
  \end{minipage}
  \vspace{+0.1cm}
  \caption{Sample Images from the SIMBA survey. The grey contours represent the strength of the 1.2mm continuum emission, with levels set at 10, 30, 50, 70, 90\% of the peak flux. The `plus' symbol represents the position of a methanol maser source, and the `box' a radio continuum source (\uchii~ region), which were the objects targeted in this survey. {\it Top Left:} G330.95-0.18, a {\it singular} source with the beam size of SEST indicated on the image. {\it Top Right:} G59.78+0.06 with a {\it double}  core  belonging to the {\it double}  morphology class. {\it Middle Left:} G34.24+0.13 is representative of {\it adjacent} sources. {\it Middle Right:}  G10.29-0.14, a {\it linearly} aligned region. {\it Bottom Left:} G15.03-0.67 is an example of an {\it irregularly} distributed, {\it tightly} clustered region. {\it Bottom Right:} G23.96+0.15 is an {\it irregularly} arranged region with a {\it low} degree of clustering. The entire image set from this survey can be found in the Appendix.}
\label{sample}
\end{figure*}

   Figure \ref{sample} presents a small selection of images from this survey, chosen to highlight each of the morphological and clustering classes mentioned above. These assignments are also listed in Table \ref{main2}.

   Also of interest in these images presented in Fig. \ref{sample}, is the indication of the source tracer. The methanol maser is depicted as a `plus' symbol, while the radio continuum source is indicated by a `box'.% The beam size of the SEST has also been added to the top-left image in this figure.

   All of the images from the SIMBA survey, including those in this sample set, can be found in the Appendix. Those sources found near the centre of the frame provided the tracer used to target the position of each field.

\subsection{Physical Parameters of the Star Forming Regions}

\subsubsection{Main Parameters}

We present the results of this survey in Table \ref{main}. The table lists all of the sources in right ascension order (RA). Columns 1 and 2 give the coordinates of the source in right ascension and declination, using J2000 epoch. Column 3 lists the source name in G-name nomenclature, derived from the Galactic coordinates of each object. Galactic names to two (or less) decimal places are consistent with those reported by \citeauthor{walsh98, thompson04, minier01} from which they were targeted. Source names given to three decimal places, denote those sources identified in this survey, with the extended Galactic names intended to distinguish closely associated sources. Column 4 indicates the identifier of the source, where `$m$', `$r$', \& `$mm$' represent the presence of a maser, a radio continuum source, or a `mm-only' source, respectively. The two NM-\IRAS~ sources are identified with `\IRAS' in the same column. Column 5 lists the millimetre map in which the source was identified (for the `mm-only' sources only). Columns 6 and 7 give the integrated flux (Jy) and the peak flux (\jybeam) for each source respectively. Column 8 gives the FWHM of each of the sources in arcseconds. Column 9 lists the distance to the source in kpc, and column 10 lists the mass of the source estimated from equation \ref{eqmass}. For those sources for which a distance ambiguity exists, the near distance precedes the far distance as is the case with the mass estimates. Column 11 indicates the lack of an 8$\mu$m \MSX~ correlation, those sources devoid of mid infrared emission are denoted by an `N', whilst those sources associated with a mid-IR dark cloud are denoted with a `DC'. An absence in this column does not indicate an \MSX~ correlation, since the infrared images are often too confused to distinguish whether a specific source has mid-IR emission or not.
   We have also drawn attention to sources previously studied in other sub-millimetre surveys such as \citet{beuther02, williams04, faundez04} in column 3.

\begin{table*}
  \begin{center}
    \caption{Parameters for the 404 1.2mm continuum sources found in the survey. \label{main}}
\vspace{-0.5cm}
   \end{center}
     \begin{tabular}{@{}lllclcccccc@{}}
       \hline
       \hline
       \multicolumn{2} {c} {Peak Position~~~~~~} &  & \multicolumn{2} {c} {Identifier~~~~~~~~~~~} & \multicolumn{2} {c} {1.2mm Flux}&&&\\

RA  & Dec & Source Name & Tracer & mm map & Integ. & Peak  &  FWHM & Distance & Mass & \MSX~\\
 (J2000)&(J2000)&   $^a$    &        &        &  Jy$^b$& Jy/bm  &  arcsec$^c$     &  kpc$^d$ &  \mstar & corr$^e$\\
(1)&(2)&(3)&(4)&(5)&(6)&(7)&(8)&(9)&(10)& (11)\\

\hline
05 41 42.7 & -01 54 23& G206.535-16.356 & mm & G206.54-16.35 & 10.0 & 2.3	& 86&	0.5&		 	4.3E+01	&\\
05 41 45.4 & -01 55 51	& G206.54-16.35	&  mr & 	& 19.2	& 4.5	&  	 47 &	0.5$^\epsilon$	&	8.2E+01 &\\
05 51 06.0 & +25 45 46 & G183.34+0.59	&  m & 		& 5.3 & 1.3 &  	 45 &	Ind$^\zeta$	& 	 	Ind	&\\
%	&		&		&		&	&	&	&		&			&&\\
06 07 47.9  & -06 22 57 & G213.61-12.6 &  mr &  & 24.3$^\alpha$ & 11.1 &  	 77&	2.1$^\zeta$ &  	 	1.8E+03	&\\
06 08 34.5 & +20 38 51 & G189.78+0.34 &  m & 	& 2.8	& 0.1	&  	 36 &	1.8$^\zeta$	&	 	1.5E+02 &\\
06 08 40.1	&	+21 31 00 & G189.03+0.76 &  r &  & 6.0 & 1.5&  59&	 	0.7$^\epsilon$ &  	5.7E+01 &\\
06 08 45.3 & +21 31 48 & G189.028+0.805 &  mm & G189.03+0.76 & 2.0 & 0.6 &  47&	 	0.7 &  	 	1.7E+01	        &\\
06 09 06.5 & +21 50 27 & G188.79+1.02 &  r &  & 1.9 & 0.6 &  	 40&	4.1$^\eta$ &  	 	5.5E+02	                &\\
06 12 52.9 & +18 00 35 & G192.581-0.042 &  r & G192.60-0.05 & 4.4 & 2.1 & 42& 	 	2.6 &  	 	5.0E+02	        &\\
06 12 54.0 & +17 59 23 & G192.60-0.05 &  m &  & 4.0 & 2.6 &  	 	35&2.6$^\xi$ &  	 	4.6E+02         &\\
06 12 54.0  & +17 59 47 & G192.594-0.045 &  mm & G192.60-0.05 & 2.7 & 1.2 &  	 	18& 2.6 &  	 	3.1E+02 &\\
% &  &  &  &  &  &  & 		 & 			&\\
08 35 31.5 & -40 38 28 & G259.94-0.04 &  m &  & 1.9 & 0.8 &  	 31&	Ind$^\zeta$ &  	 	Ind	                &\\
% &  &  &  &  &  &  & 		 & 			&\\
09 03 13.5 & -48 55 22 & G269.45-1.47 &  mr &  & 1.3 & 0.3 &  	 55&	7.9$^\zeta$ &  	 	1.4E+03	                &\\
09 03 32.3 & -48 28 00 & G269.15-1.13$^{\ast}$ &  m &  & 5.2 & 2.8 &  	31& 	3.6$^\zeta$ &  	 	1.1E+03 &\\
09 16 41.4 & -47 55 46 & G270.25+0.84 &  m &  & 6.8 & 3.6 &  	 30&	2.1$^\zeta$ &  	 	5.1E+02	                &\\
% &  &  &  &  &  &  & 		 & 			&\\
10 23 47.0 & -57 48 38 & G284.271-0.391 &  mm & G284.35-0.42 & 0.4 & 0.1 & 20& 	4.9 &  	 	1.6E+02	                &\\
10 24 03.0 & -57 47 58 & G284.295-0.362 &  mm & G284.35-0.42 & -$^\beta$ & - &  	- &	4.9 &  	 	Ind	        &\\
10 24 04.0 & -57 49 02 & G284.307-0.376 &  mm & G284.35-0.42 & 0.6 & 0.2 &  	 46&	4.9 &  	 	2.5E+02	        &\\
10 24 06.0 & -57 52 06 & G284.338-0.417 &  mm & G284.35-0.42 & 0.1 & 0.1 &  	 15&    4.9 &  	 	5.7E+01	        &\\
10 24 10.0 & -57 52 39 & G284.35-0.42 &  m  &  & 1.3 & 0.5 & 24& 	 	4.9$^\zeta$ &  	 	5.2E+02	        &\\
10 24 12.0 & -57 51 42 & G284.345-0.404 &  mm & G284.35-0.42 & 1.4 & 0.4 &  24&	 	4.9 &  	 	5.6E+02	        &\\
10 24 14.0 & -57 50 46 & G284.341-0.389 &  mm & G284.35-0.42 & 2.5 & 0.6 &  	 58&	4.9 &  	 	1.0E+03	        &\\
10 24 15.0 & -57 49 10 & G284.328-0.365 &  mm & G284.35-0.42 & 0.2 & 0.2 &  	 14&	4.9 &  	 	9.0E+01	        &\\
10 24 18.0 & -57 54 46  & G284.384-0.441 &  mm & G284.35-0.42 & 0.2 & 0.1 &  	 23&	4.9 &  	 	7.4E+01	        &N\\
10 24 21.0 & -57 49 42 & G284.344-0.366 &  mm & G284.35-0.42 & 1.9 & 0.4 &  	 24&	4.9 &  	 	7.9E+02	        &\\
10 24 27.0 & -57 49 18 & G284.352-0.353 &  mm & G284.35-0.42 & 2.9$^\gamma$ & 0.7 & 48& 	 4.9  & 1.2E+03	        &\\
10 48 05.2 & -58 26 40 & G287.37+0.65 &  m  &  & 0.8 & 0.4&  30&	 	5$^\zeta$ &   	3.3E+02	                        &\\
10 57 33.0 & -62 58 54 & G290.40-2.91 &  m  &  & 1.5 & 0.6 & 34& 3$^\zeta$ &  		2.3E+02	                        &\\
% &  &  &  &  &  &  & 		 & 			&\\
11 11 33.9 & -61 21 22 & G291.256-0.769 &  mm & G291.27-0.70 & 2.9 & 0.9 & 42& 3.1  &  	 	4.7E+02	                &N\\
11 11 38.3 & -61 19 54 & G291.256-0.743 &  mm & G291.27-0.70 & 5.0 & 1.2 & 51& 3.1 &  	 	8.1E+02	                &DC\\
11 11 54.8 & -61 18 26 & G291.27-0.70 &  mr &  & 63.0 & 15.5 & 122& 3.1$^\zeta$ &  	 	1.0E+04	                &\\
11 12 00.6 & -61 18 34 & G291.288-0.706 &  mm & G291.27-0.70 & 1.2 & 0.7 & 40&3.1 &  	 	1.9E+02	                &\\
11 12 09.4 & -61 18 10 & G291.302-0.693 &  mm & G291.27-0.70 & 2.4 & 0.6 & 41&3.1 &  	 	3.9E+02	                &\\
11 12 15.0 & -61 17 38 & G291.309-0.681 &  mm & G291.27-0.70 & 6.8 & 1.4 & 61&3.1 &  	 	1.1E+03	                &DC\\
11 12 16.1 & -58 46 19 & G290.37+1.66 &  m &  & 1.5$^\alpha$ & 0.4 & 34& 3$^\zeta$ &  	 	2.3E+02	                &\\
11 14 54.5 & -61 13 32 & G291.587-0.499 &  mm & G291.59-0.4 & 6.1 & 3.1 &  74&	 	7.5 &  	 	5.8E+03	        &\\
11 14 57.8 & -61 11 40 & G291.576-0.468 &  mm & G291.59-0.4 & 1.8 & 0.7 &  	40& 	7.5 &  	 	1.7E+03	        &\\
11 14 58.9 & -61 10 36 & G291.572-0.450 &  mm & G291.59-0.4 & 0.3 & 0.2 &  	20& 	7.5 &  	 	3.3E+02	        &N\\
11 15 01.1 & -61 15 56 & G291.608-0.532 &  mm & G291.59-0.4 & 4.1 & 1.2 &  	 60&	7.5 &  	 	3.9E+03	        &\\
11 15 02.2 & -61 13 40 & G291.597-0.496 &  mm & G291.59-0.4 & 3.0 & 1.7 &  	 71&	7.5 &  	 	2.8E+03	        &\\
11 15 06.4 & -61 09 38 & G291.58-0.53 &  m  &  & 7.5 & 2.3 &  	38& 	7.5$^\zeta$ &  	 	7.2E+03	                &N\\
11 15 08.9 & -61 17 08 & G291.630-0.545 &  mm & G291.59-0.4 & 11.9 & 2.0 & 82& 	 	7.5 &  	 	1.1E+04	        &\\
11 15 19.9 & -61 11 08 & G291.614-0.443 &  mm & G291.59-0.4 & 0.4 & 0.1 &  	10& 	7.5 &  	 	3.4E+02	        &\\
11 32 01.4 & -62 13 18 & G293.824-0.762 &  mm & G293.82-0.74 & 0.5 & 0.2 &  	 23&	10.6 &  	9.4E+02         &\\
11 32 06.1 & -62 12 22 & G293.82-0.74 &  mr &  & 1.8 & 0.8 &  	 27&	10.6$^\zeta$ &  	 	3.5E+03	        &\\
11 32 32.4 & -62 15 42 & G293.892-0.782 &  mm & G293.82-0.74 & 0.2 & 0.1&  13&	 	10.6 &  3.8E+02                 &\\
11 32 42.0 & -62 22 35 & G293.95-0.8 &  mr &  & 1.2 & 0.6 &  	30& 	11$^\zeta$ &  	 	2.4E+03	                &\\
11 32 42.0 & -62 21 55 & G293.942-0.876 &  mm & G293.95-0.8 & 1.2 & 0.5 &  29&	 	11 &  	 	2.4E+03	        &\\
11 32 55.8 & -62 26 11 & G293.989-0.936 &  mm & G293.95-0.8 & 1.2 & 0.5 &  	34& 	11 &  	 	2.5E+03	        &\\
11 35 31.0 & -63 14 36 & G294.52-1.60$^\ast$ &  m  &  & 2.2 & 1.0 & 30& 1/6.1$^\zeta$ & 3.7E+01 / 1.4E+03	                &\\
11 38 57.1 & -63 28 46 & G294.945-1.737 &  mm & G294.97-1.7 & 0.3 & 0.2 & 29& 1.3/5.8 & 8.4E+00 / 1.7E+02	        &\\
11 39 09.0 & -63 28 38 & G294.97-1.7 &  r &  & 4.7 & 1.2 & 49& 1.3/5.8$^\zeta$ & 1.4E+02 / 2.7E+03	                &\\
11 39 22.1 & -63 28 30 & G294.989-1.720 &  m  & G294.97-1.7 & 1.6 & 0.8 & 30&1.3/5.8 & 4.7E+01 / 9.4E+02          	&\\
% &  &  &  &  &  &  & 		 & 			&\\
12 11 45.4 & -61 45 42 & G298.26+0.7 &  m  &  & 1.5 & 0.6 & 36& 4$^\zeta$ &  	 	4.1E+02	                        &\\
12 17 18.6 & -62 28 40 & G299.02+0.1 &  m  &  & 0.9 & 0.4 &  	 	37& 10$^\zeta$ &  	 	1.4E+03	        &\\
12 17 30.2 & -62 29 04 & G299.024+0.130 &  mm & G299.02+0.1 & 0.3 & 0.2 & 24& 	 	10 &  	 	4.6E+02	        &\\
12 29 37.3 & -62 57 23 & G300.455-0.190 &  mm & G300.51-0.1 & 0.4 & 0.2 &  	35& 	9.4 &  	 	5.4E+02	        &\\
12 30 02.0 & -62 56 35 & G300.51-0.1 &  m &  & 1.9 & 0.9 &  	 29&	9.4$^\zeta$ &  	 	2.8E+03	                &\\
12 35 29.1 & -63 01 32 & G301.14-0.2 &  mr &  & 16.1 & 6.4 & 70& 4.4$^\zeta$ &  	 	5.3E+03	                &\\
12 43 32.1 & -62 55 06 & G302.03-0.06 &  mr &  & 2.6 & 0.9 & 35&4.5$^\zeta$ &  	 	8.8E+02	                        &\\
% &  &  &  &  &  &  & 		 & 			\\
13 08 13.5 & -62 10 20 & G304.890+0.636 &  mm & G305.952+0.555 & -$^\beta$ & - & -& 	 	Ind &  	 	Ind	&N\\
\end{tabular}
\end{table*}

\begin{table*}
\begin{center}
\contcaption{}
\vspace{-0.3cm}
    \begin{tabular}{@{}lllclcccccc@{}}
       \hline
       \hline
 \multicolumn{2} {c} {Peak Position~~~~~~} &  & \multicolumn{2} {c} {Identifier~~~~~~~~~~~} & \multicolumn{2} {c} {1.2mm Flux}&&&\\

RA  & Dec & Source Name & Tracer & mm map & Integ. & Peak  &  FWHM & Distance & Mass & \MSX~\\
 (J2000)&(J2000)& $^a$      &        &        &  Jy$^b$& Jy/bm   &  arcsec$^c$     &  kpc$^d$ &  \mstar & corr$^e$\\
(1)&(2)&(3)&(4)&(5)&(6)&(7)&(8)&(9)&(10)& (11)\\
\hline
13 08 23.8 & -62 13 56 & G304.906+0.574 &  mm & G305.952+0.555 & 0.3 & 0.1 & 29&  	 	Ind &  	 	Ind	&N\\
13 08 31.9 & -62 15 48 & G304.919+0.542 &  mm & G305.952+0.555 & 0.4 & 0.2 &  	39& 	Ind &  	 	Ind	        &\\
13 08 34.2 & -62 15 00 & G305.952+0.555 & IRAS &  & -$^\delta$ & - &  	 -&	Ind &  	 	Ind	                &\\
13 08 35.4 & -62 17 00 & G304.952+0.522 &  mm & G305.952+0.555 & 0.6 & 0.2 & 39& 	 	Ind &  	 	Ind	&DC\\
13 08 38.8 & -62 15 32 & G304.933+0.546 &  mm & G305.952+0.555 & 0.7 & 0.3 &  36&	 	Ind &  	 	Ind	&\\
13 08 43.3 & -62 15 16 & G304.942+0.550 &  mm & G305.952+0.555 & 0.2 & 0.2 &  	 16&	Ind &  	 	Ind	        &\\
13 10 40.5 & -62 34 53 & G305.145+0.208 &  mm & G305.21+0.21 & 0.4 & 0.1 & 17& 3.9/5.9 & 9.6E+01 / 2.2E+02	        &\\
13 10 42.0 & -62 43 13 & G305.137+0.069 &  mm & G305.20+0.02 & 1.6 & 0.6 & 34& 3/6.8 & 2.4E+02 / 1.2E+03	        &DC\\
13 11 08.3 & -62 32 37 & G305.201+0.241 &  mm & G305.21+0.21 & 0.2 & 0.4 & 8& 3.9/5.9 & 4.1E+01 / 9.5E+01	        &N\\
13 11 09.4 & -62 33 17 & G305.202+0.230 &  mm & G305.21+0.21 & 1.4 & 0.4 & 38& 3.9/5.9 & 3.7E+02 / 8.4E+02	        &\\
13 11 12.3 & -62 44 57 & G305.20+0.02 &  r &  & 6.2 & 1.9 & 71& 3/6.8$^\zeta$ & 9.4E+02 / 4.8E+03	                &\\
13 11 13.6 & -62 47 29 & G305.192-0.006 &  m  & G305.20+0.02 & 2.3 & 0.8 & 31& 3/6.8 & 3.5E+02 / 1.8E+03	        &\\
13 11 14.1 & -62 34 45 & G305.21+0.21 &  m &  & 10.6 & 3.1 & 87& 3.9/5.9$^\iota$ & 2.7E+03 / 6.3E+03	                &\\
13 11 15.9 & -62 46 41 & G305.197+0.007 &  mm & G305.20+0.02 & 2.3 & 0.7 & 42& 3/6.8 & 3.6E+02 / 1.8E+03	        &\\
13 11 17.0 & -62 45 53 & G305.200+0.02 &  m  & G305.20+0.02 & 0.7 & 0.4 & 21& 3/6.8 & 1.1E+02 / 5.8E+02	        &\\
13 11 19.8 & -62 30 29 & G305.226+0.275 &  mm & G305.21+0.21 & 0.8 & 0.3 & 28& 3.9/5.9 & 2.1E+02 / 4.7E+02	        &N\\
13 11 20.9 & -62 29 48 & G305.228+0.286 &  mm & G305.21+0.21 & 0.2 & 0.1 & 14& 3.9/5.9 & 3.9E+01 / 8.9E+01	        &N\\
13 11 26.7 & -62 31 16 & G305.238+0.261 &  mm & G305.21+0.21 & 1.4$^\alpha$ & 0.3 & 47 & 3.9/5.9 & 3.6E+02 / 8.2E+02	&N\\
13 11 30.3 & -62 33 24 & G305.242+0.225 &  mm & G305.21+0.21 & 0.2 & 0.1 &14&  3.9/5.9 & 6.2E+01 / 1.4E+02	        &\\
13 11 32.5 & -62 32 12 & G305.248+0.245 &  m  & G305.21+0.21 & 1.2 & 0.5 & 32& 3.9/5.9 & 3.1E+02 / 7.1E+02            &\\
13 11 35.8 & -62 48 16 & G305.233-0.023 &  mm & G305.20+0.02 & 0.7 & 0.3 & 38& 3/6.8 & 1.0E+02 / 5.1E+02	        &N\\
13 11 54.4 & -62 47 19 & G305.269-0.010 &  mm & G305.20+0.02 & 3.4$^\gamma$ & 1.3 & 65& 3/6.8 & 5.2E+02 / 2.7E+03	&\\
13 12 30.5 & -62 34 43 & G305.355+0.194 &  mm & G305.37+0.21 & -$^\alpha$ & - & -& 3/6.8 &  	 	Ind	        &\\
13 12 31.6 & -62 34 11 & G305.358+0.203 &  mm & G305.37+0.21 & 17.9$^\alpha$ & 3.2 & 98& 3/6.8 & 2.7E+03 / 1.4E+04    &\\
13 12 34.7 & -62 35 15 & G305.362+0.185 &  m  & G305.37+0.21 & 3.5 & 2.3 & 56& 3/6.8 & 5.3E+02 / 2.7E+03	        &\\
13 12 35.2 & -62 37 15 & G305.361+0.151 &  m  & G305.37+0.21 & 5.3 & 1.4 & 33&3/6.8 & 8.2E+02 / 4.2E+03	        &\\
13 12 36.3 & -62 33 39 & G305.37+0.21 &  r &  & 5.8 & 1.6 & 56&3/6.8$^\iota$ & 8.8E+02 / 4.5E+03	                &\\
13 12 38.1 & -62 36 42 & G305.340-0.172 &  mm & G305.37+0.21 & 0.5 & 0.2 & 35&3/6.8 & 8.1E+01 / 4.2E+02	        &\\
13 13 46.0 & -62 25 37 & G305.513+0.333 &  mm & G305.533+0.360 & 0.1 & 0.3 & 13& 	 	Ind &  	 	Ind	&N\\
13 13 55.2 & -62 23 53 & G305.533+0.360 & IRAS &  & -$^\delta$ & - &  	 -&	Ind &  	 	Ind	                &\\
13 13 58.7 & -62 25 05 & G305.538+0.340 &  mm & G305.533+0.360 & 1.1 & 0.6 & 41& 	 	Ind &  	 	Ind	&\\
13 14 06.1 & -62 47 53 & G305.519-0.040 &  mm & G305.55+0.01 & -$^\beta$ & - & -& 3.6/6.3 &  	 	Ind	        &\\
13 14 06.1 & -62 46 41 & G305.520-0.020 &  mm & G305.55+0.01 & 0.2 & 0.1 & 22& 3.6/6.3 & 4.2E+01/ 1.3E+02	        &\\
13 14 20.0 & -62 45 13 & G305.549+0.002 &  mm & G305.55+0.01 & 1.0 & 0.3 & 41& 3.6/6.3 & 2.2E+02 / 6.8E+02	        &\\
13 14 21.2 & -62 44 33 & G305.55+0.01 &  m &  & 0.9 & 0.5 & 35& 3.6/6.3$^\zeta$ & 2.1E+02 / 6.4E+02	                &\\
13 14 22.4 & -62 46 01 & G305.552+0.012 &  mm & G305.55+0.01 & 2.0 & 0.7 & 61& 3.6/6.3 & 4.5E+02 / 1.4E+03	        &\\
13 14 25.8 & -62 44 32 & G305.561+0.012$^\ast$ &  r & G305.55+0.01 & 3.2 & 1.1 & 76& 3.6/6.3 & 7.1E+02/ 2.2E+03	        &\\
13 14 35.1 & -62 43 02 & G305.581+0.033 &  mm & G305.55+0.01 & 0.5$^\alpha$ & 0.1 & 78&3.6/6.3 & 1.2E+02 / 3.7E+02	&\\
13 14 49.1 & -62 44 24 & G305.605+0.010 &  mm & G305.55+0.01 & 0.3 & 0.1 & 35&3.6/6.3 & 6.6E+01 / 2.0E+02	        &\\
13 16 31.5 & -62 59 02 & G305.776-0.251 &  mm & G305.81-0.25 & 0.4 & 0.2 & 32&3.9/6 & 1.1E+02/ 2.7E+02	                &\\
13 16 39.6 & -62 57 49 & G305.81-0.25 &  mr &  & 6.1 & 3.0 & 52& 3.9/6$^\zeta$ & 1.6E+03 / 3.8E+03	                &\\
13 16 58.3 & -62 55 25 & G305.833-0.196 &  mm & G305.81-0.25 & 0.5 & 0.2 & 34& 3.9/6 & 1.3E+02 / 3.2E+02	        &N\\
13 21 18.2 & -63 00 43 & G306.33-0.3 &  m &  & 0.5 & 0.2 & 32& 2/8.1$^\zeta$ & 3.2E+01 / 5.3E+02	                &\\
13 21 18.2 & -63 01 07 & G306.319-0.343 &  mm & G306.33-0.3 & 0.2 & 0.1 & 12& 2/8.1 & 1.0E+01 / 1.7E+02	        &\\
13 21 32.3 & -62 58 26 & G306.343-0.302 &  mm & G306.33-0.3 & 0.7 & 0.1 & 14& 2/8.1 & 4.5E+01 / 7.4E+02	        &N\\
13 21 34.6 & -62 59 54 & G306.345-0.345 &  mm & G306.33-0.3 & 0.4 & 0.1 & 14 & 2/8.1 & 2.9E+01 / 4.8E+02	        &\\
13 50 38.2 & -61 34 20 & G309.917+0.494 &  mm & G309.92+0.4 & 0.2 & 0.1 &  15 &		5.5 & 	 	9.8E+01	        &N\\
13 50 41.6 & -61 35 15 & G309.92+0.4$^\ast$ &  m  &  & 5.5 & 2.2 &  	 64&	 5.5$^\mu$ &  	 	2.8E+03	                &\\
% &  &  &  &  &  &  & 		 & 			&\\
15 00 33.6 & -58 58 05 & G318.913-0.162 &  r & G318.92-0.68 & 3.5 & 1.6 & 35& 	 	2 &  	 	2.4E+02	        &\\
15 00 55.3 & -58 58 54 & G318.92-0.68 &  m  &  & 4.8 & 1.9 &  	 36&	2$^\mu$ &  	 	3.3E+02	                &\\
15 31 41.6 & -56 30 11 & G323.74-0.3 &  m  &  & 5.9 & 2.6 & 36& 3.4/10.3$^\iota$ & 1.2E+03 / 1.1E+04	                &\\
% &  &  &  &  &  &  & 		 & 		&	\\
16 09 48.8 & -51 53 51 & G330.95-0.18 &  mr &  & 32.8 & 15.3 & 90&5.3/9.6$^\iota$ & 1.6E+04 / 5.2E+04	                &\\
16 11 26.9 & -51 41 57 & G331.28-0.19$^\ast$ &  m  &  & 6.9 & 1.9 & -$\ddagger$ & 4.7/10.1$^\iota$ & 2.6E+03 / 1.2E+04	&\\
16 19 30.5 & -51 02 40 & G332.640-0.586 &  mm & G332.73-0.62 & 0.7 & 0.3 & 16&3.3/11.8 & 1.3E+02 / 1.7E+03	        &\\
16 19 38.1 & -51 03 12 & G332.648-0.606$^\ast$ &  m  & G332.73-0.62 & 15.7$^\alpha$ & 1.8 & 134&3.3/11.8 & 2.9E+03 / 3.7E+04	&\\
16 19 47.4 & -51 00 08 & G332.701-0.587 &  mm & G332.73-0.62 & 0.8 & 0.3 & 30&3.3/11.8 & 1.5E+02 / 1.9E+03	        &\\
16 19 48.3 & -51 02 00 & G332.646-0.647 &  mm & G332.73-0.62 & 2.0 & 0.8 & 30&3.3/11.8 & 3.7E+02 / 4.8E+03	        &\\
16 19 51.7 & -51 01 20 & G332.695-0.609 &  mm & G332.73-0.62 & 3.9 & 1.3 & 43& 3.3/11.8 & 7.3E+02 / 9.3E+03           &\\
16 20 02.7 & -51 00 32 & G332.73-0.62 &  m  &  & 0.6 & 0.3 & 11&3.3/11.8$^\zeta$ & 1.1E+02 / 1.4E+03	                &N\\
16 20 07.0 & -50 56 48 & G332.777-0.584 &  mm & G332.73-0.62 & 0.6 & 0.3 &15& 3.3/11.8 & 1.2E+02 / 1.5E+03	        &N\\
16 20 07.0 & -51 00 00  & G332.627-0.511 &  mm & G332.73-0.62 & 0.5 & 0.3 & 13&3.3/11.8 & 9.1E+01 / 1.2E+03	        &DC\\

\end{tabular}
\end{center}
\end{table*}

\begin{table*}
\begin{center}
\contcaption{}
   \begin{tabular}{@{}lllclcccccc@{}}
       \hline
       \hline
 \multicolumn{2} {c} {Peak Position~~~~~~} &  & \multicolumn{2} {c} {Identifier~~~~~~~~~~~} & \multicolumn{2} {c} {1.2mm Flux}&&& \\

RA  & Dec & Source Name & Tracer & mm map & Integ. & Peak  & FWHM & Distance & Mass & \MSX~\\
 (J2000)&(J2000)& $^a$  &        &        &  Jy$^b$& Jy/bm   &  arcsec$^c$     &  kpc$^d$ &  \mstar & corr$^e$\\
(1)&(2)&(3)&(4)&(5)&(6)&(7)&(8)&(9)&(10)& (11)\\

\hline
16 20 12.0  & -50 53 20 & G332.827-0.552$^\ast$ &  mm & G332.73-0.62 & -$^\beta$ & - & -&3.3/11.8 &  	 	Ind     &\\
16 20 15.0 & -50 56 40 & G332.794-0.598 &  mm & G332.73-0.62 & 0.9 & 0.3 & 25&3.3/11.8 & 1.7E+02 / 2.2E+03	        &\\
% &  &  &  &  &  &  & 		 & 			&\\
17 45 54.3 & -28 44 08 & G0.204+0.051 &  mm & G0.21-0.00 & 0.6 &0.3 & 37&8.4/8.6 & 7.5E+02 / 7.8E+02	                &DC\\
17 46 03.9 & -28 24 58 & G0.49+0.19 &  m  &  & 1.2 & 0.5 & 30&2.6/14.5$^\zeta$ & 1.4E+02 / 4.4E+03	                &\\
17 46 07.1 & -28 41 28 & G0.266-0.034 &  mm & G0.21-0.00 & 1.0 & 0.4 & 37&8.4/8.6 & 1.2E+03 / 1.2E+03	                &DC\\
17 46 07.7 & -28 45 20 & G0.21-0.00 &  mr &  & 1.2 & 0.4 & 42&8.4/8.6$^\zeta$ & 1.5E+03 / 1.6E+03	                &\\
17 46 08.2 & -28 25 23 & G0.497+0.170 &  mm & G0.49+0.19 & 0.9 & 0.2 & 26&2.6/14.5 & 1.0E+02 / 3.2E+03	                &\\
17 46 09.5 & -28 43 36 & G0.24+0.01 &  mm & G0.21-0.00 & 7.2 & 1.2 & 74&8.4/8.6 & 8.6E+03 / 9.0E+03	                &DC\\
17 46 10.1 & -28 23 31 & G0.527+0.181 &  r & G0.49+0.19 & 2.4 & 0.8 & 38& 2.6/14.5 & 2.8E+02 / 8.7E+03	        &\\
17 46 10.7 & -28 41 36 & G0.271+0.022 &  mm & G0.21-0.00 & 0.6$^\alpha$ & 0.3 & 31&8.4 / 8.6 & 6.9E+02 / 7.2E+02	&DC\\
17 46 11.4  & -28 42 40 & G0.26+0.01 &  mm & G0.21-0.00 & 6.7 & 1.2 & 119&8.4/8.6& 8.0E+03 / 8.4E+03	                &DC\\
17 46 52.8 & -28 07 35 & G0.83+0.18 &  m  &  & 1.2 & 0.5 & 29&4.5/12.5$^\zeta$ & 4.2E+02 / 3.3E+03	                &\\
17 47 00.0 & -28 45 20 & G0.331-0.164 &  mm & G0.32-0.20 & 0.8 & 0.2 & 28&8/9 & 8.6E+02 / 1.1E+03	                &\\
17 47 01.2 & -28 45 36 & G0.310-0.170 &  mm & G0.32-0.20 & 0.2 & 0.1 & 12& 8/9 & 1.9E+02 / 2.3E+02	                &\\
17 47 09.1 & -28 46 16 & G0.32-0.20 &  mr &  & 5.9 & 1.2 & 126&8/9$^\zeta$ & 6.4E+03 / 8.1E+03	                        &\\
17 47 20.1 & -28 47 04 & G0.325-0.242 &  mm & G0.32-0.20 & 0.3 & 0.1 & 11& 8/9 & 3.5E+02 / 4.4E+02	                &DC\\
17 48 31.6 & -28 00 31 & G1.124-0.065 &  mm & G1.13-0.11 & 0.5 & 0.2 &  28&	 	8.5 &  	 	5.9E+02         &\\
17 48 34.7 & -28 00 16 & G1.134-0.073 &  mm & G1.13-0.11 & 0.2 & 0.1 &  18&	 	8.5 &  	 	2.5E+02         &\\
17 48 36.4 & -28 02 31 & G1.105-0.098 &  mm & G1.14-0.12 & 1.8 & 0.4 & 19& 8.5 & 2.2E+03	                        &\\
17 48 41.9 & -28 01 44 & G1.13-0.11$^\ast$ &  r &  & 7.9 & 1.7 &  	115& 	8.5$^\eta$ &  	 	9.7E+03	                &\\
17 48 48.5 & -28 01 13  & G1.14-0.12 &  m  &  & 0.2 & 0.2 & 47& 8.5$^\iota$ &  3.0E+02	                                &\\
17 50 14.5 & -28 54 31 & G0.55-0.85 &  mr &  & 15.8 & 5.7 &  	94& 	2$^\chi$ &  	 	1.1E+03	                &\\
17 50 18.8 & -28 53 14 & G0.549-0.868 &  mm & G0.55-0.85 & 4.0 & 0.2 &  70& 	 	2 &  	 	2.7E+02	        &\\
17 50 24.9 & -28 50 15 & G0.627-0.848 &  mm & G0.55-0.85 & 0.3 & 0.2 &  	26& 	2 &  	 	2.2E+01	        &N\\
17 50 26.7 & -28 52 23 & G0.600-0.871 &  mm & G0.55-0.85 & 0.8 & 0.3 &  	 39&	2 &  	 	5.7E+01	        &N\\
17 50 46.5 & -26 39 45 & G2.54+0.20 &  m  &  & 2.1 & 0.4 & 63& 2.7/14.2$^\zeta$ & 2.6E+02 / 7.2E+03	                &DC\\
17 59 04.6 & -24 20 55 & G5.48-0.24 &  r &  & 1.1 & 0.3 &  39&	 	12.5$^\nu$ &  	 	3.0E+03	                &\\
17 59 07.5 & -24 19 19  & G5.504-0.246 &  mm & G5.48-0.24 & 0.7 & 0.1 & 31& 	 	12.5 &  	 	1.9E+03	&N\\
% &  &  &  &  &  &  & 		 & 		&	\\
18 00 31.0 & -24 03 59 & G5.89-0.39 &  r &  & 23.2 & 9.5 &  	90& 	2.6$^\nu$ &  	 	2.7E+03	                &\\
18 00 40.9 & -24 04 21 & G5.90-0.42 &  m  &  & 8.1 & 2.9 & 62&2.7/14.2$^\zeta$ & 1.0E+03 / 2.8E+04	                &\\
18 00 43.9 & -24 04 47 & G5.90-0.44 &  mm & G5.90-0.42 & 5.0 & 1.5 & 69&2.7/14.2 & 6.2E+02 / 1.7E+04	                &\\
18 00 50.9 & -23 21 29 & G6.53-0.10 &  r &  & 3.0 & 0.9 &  	 41&	14$^\nu$ &  	 	1.0E+04	                &\\
18 00 54.1 & -23 17 02 & G6.60-0.08$^\star$ &  m  &  & 0.3 & 0.1 & 15 &0.3/16.6$^\zeta$ &  	 5.4E-01 / 1.6E+03	        &\\
18 00 59.2 & -23 17 02 & G6.620-0.100$^\star$ & mm & G6.60-0.08 & 0.2 &0.1 & 21 & 0.3/16.6& 3.7E-01 / 1.1E+03                  &N\\
18 02 52.8 & -21 47 54 & G8.127+0.255 &  mm & G8.13+0.22 & 0.9 & 0.2 &  	31& 	3.4 &  	 	1.7E+02	        &N\\
18 02 56.2 & -21 47 38 & G8.138+0.246 &  mm & G8.13+0.22 & 1.9 & 0.5 &  	 66&	3.4 &  	 	3.6E+02	        &\\
18 03 00.8 & -21 48 10 & G8.13+0.22$^\ast$ &  mr &  & 8.0 & 2.0 &  	69& 	3.4$^\nu$ &  	 	1.6E+03                 &\\
18 03 26.3 & -24 22 29 & G5.948-1.125 &  mm & G5.97-1.17 & 0.3 & 0.2 & 39& 	 	1.9 &  	 	2.0E+01	        &N\\
18 03 34.5 & -24 21 41 & G5.975-1.146 &  mm & G5.97-1.17 & 0.5 & 0.2 &  	29& 	1.9 &  	 	2.8E+01	        &N\\
18 03 36.8 & -24 22 13 & G5.971-1.158 &  mm & G5.97-1.17 & 0.8 & 0.5 &  	48& 	1.9 &  	 	5.0E+01	        &\\
18 03 40.9 & -24 22 37 & G5.97-1.17 &  r &  & 9.2 & 2.2 &  88&	 	1.9$^\nu$ &  	 	5.7E+02	                &\\
18 05 15.6 & -19 50 55 & G10.10+0.73 &  r &  & 0.6 & 0.4 & 30& 0.4/16.3$^\iota$ & 0.2E+01 / 2.9E+03	                &\\
18 06 14.8 & -20 31 37 & G9.62+0.19$^\ast$ &  mr &  & 8.5 & 3.3 &  	94& 	 2$^\mu$ &  	 	5.8E+02	                &\\
18 06 18.9 & -21 37 21 & G8.67-0.36 &  mr &  & 13.9 & 5.0 &  	81& 	4.7$^\nu$ &  	 	5.2E+03	                &\\
18 06 23.5 & -21 36 57 & G8.686-0.366 &  m & G8.67-0.36 & 3.4 & 1.4 &  55&	 	4.7 &  	 	1.3E+03	        &DC\\
18 06 24.6 & -21 40 01 & G8.644-0.395 &  mm & G8.67-0.36 & 0.3 & 0.2 &  26&	 	4.7 &  	 	1.1E+02         &DC\\
18 06 26.4 & -21 35 29 & G8.713-0.364 &  mm & G8.67-0.36 & 0.9 & 0.3 &  66&	 	4.7 &  	 	3.2E+02	        &DC\\
18 06 28.7 & -21 34 17 & G8.735-0.362 &  mm & G8.67-0.36 & 1.5$^\gamma$ & 0.3 & 66& 	 4.7 &   	5.7E+02	        &\\
18 06 36.1 & -21 36 01 & G8.724-0.401a &  mm & G8.67-0.36 & 0.2 & 0.2 &  	 21&	4.7 &  	 	8.3E+01	        &DC\\
18 06 36.7 & -21 37 05 & G8.724-0.401b &  mm & G8.67-0.36 & 0.6 & 0.3 &  	 35&	4.7 &  	 	2.1E+02	        &DC\\
18 06 37.3 & -21 36 33 & G8.718-0.410 &  mm & G8.67-0.36 & 0.1 & 0.1 &  	 8&	4.7 &  	 	5.3E+01	        &DC\\
18 07 45.8 & -20 19 47  & G9.966-0.020 &  mm & G9.99-0.03 & 0.2 & 0.1 & 10& 5.0 & 8.1E+01	                        &N\\
18 07 50.4 & -20 18 51 & G9.99-0.03 &  m  &  & 1.2 & 0.6 & 45& 5.0$^\chi$ & 5.2E+02 	                                &N\\
18 07 53.2 & -20 18 19 & G10.001-0.033 &  r & G9.99-0.03 & 0.4 & 0.2 & 22& 5.0 & 1.5E+02 	                        &\\
18 08 37.9 & -19 51 41 & G10.47+0.02$^\ast$ &  mr &  & 24.0$^\alpha$ & 10.1 & 122& 6/10.8$^\zeta$ & 1.5E+04 / 4.8E+04	&\\
18 08 44.9 & -19 54 38 & G10.44-0.01 &  m  &  & 1.6 & 0.6 & 51& 5.9/10.8$^\zeta$ & 9.4E+02 / 3.2E+03	                &DC\\
18 08 45.9 & -20 05 34 & G10.287-0.110 &  mm & G10.32-0.15 & 2.8 & 1.0 & 54& 1.8/15 & 1.6E+02 / 1.1E+04               &N\\
18 08 49.4 & -20 05 58 & G10.284-0.126 &  m  &  & 2.6 & 1.1 &  	 46&	1.9$^\nu$ &  	 	1.6E+02	                &\\
18 08 52.4 & -20 05 58 & G10.288-0.127 &  mm & G10.32-0.15 & 0.9 & 0.6 & 40& 1.8/15 & 5.2E+01 / 3.6E+03	        &\\
18 08 55.5 & -20 05 58 & G10.29-0.14 &  mr &  & 7.8 & 2.5 &  	73& 1.9$^\nu$ &  	 	4.8E+02	                &\\
18 09 00.0 & -20 05 34 & G10.343-0.142 &  m  & G10.32-0.15 & 1.7 & 0.9 & 26&1.8/15 & 9.6E+01 / 6.6E+03	                &\\
\end{tabular}
\end{center}
\end{table*}

\begin{table*}
\begin{center}
\contcaption{}
  \begin{tabular}{@{}lllclcccccc@{}}
       \hline
       \hline
\multicolumn{2} {c} {Peak Position~~~~~~} &  & \multicolumn{2} {c} {Identifier~~~~~~~~~~~} & \multicolumn{2} {c} {1.2mm Flux}&&& \\

RA  & Dec & Source Name & Tracer & mm map & Integ. & Peak  &  FWHM& Distance & Mass& \MSX~ \\
 (J2000)&(J2000)& $^a$  &        &        &  Jy$^b$& Jy/bm   &  arcsec$^c$    &  kpc$^d$ &  \mstar & corr$^e$\\
(1)&(2)&(3)&(4)&(5)&(6)&(7)&(8)&(9)&(10)& (11)\\

\hline
18 09 01.5 & -20 05 08 & G10.32-0.15 &  m  &  & 5.5 & 1.2 & 108& 1.8/15$^\zeta$ & 3.0E+02 / 2.1E+04	              &\\
18 09 03.5 & -20 02 54 & G10.359-0.149 &  mm & G10.32-0.15 & 1.4 & 0.4 & 47&1.8/15 & 8.0E+01 / 5.5E+03	              &N\\
18 09 14.2 & -20 18 53 & G10.146-0.314 &  mm & G10.15-0.34 & 0.5 & 0.2 &  43&	 	6 &  	 	3.0E+02	      &\\
18 09 18.2 & -20 16 21 & G10.191-0.307 &  mm & G10.15-0.34 & 0.2 & 0.7 &  	11& 	6 &  	 	9.8E+01	      &\\
18 09 18.2 & -20 19 17 & G10.148-0.331 &  mm & G10.15-0.34 & -$^\alpha$ & - &  	 37&	6 &  	 	Ind	      &\\
18 09 20.5 & -20 15 01 & G10.214-0.305 &  mm & G10.15-0.34 & 0.2 & 0.1 &  	 16&	6 &  	 	1.0E+02	      &DC\\
18 09 21.6 & -20 16 21 & G10.191-0.308 &  mm & G10.15-0.34 & 0.5 & 0.2 &  	 29&	6 &  	 	3.0E+02	      &DC\\
18 09 21.6 & -20 19 25 & G10.15-0.34 &  r &  & 6.5 & 1.4 &  	70& 	6$^\nu$ &  	 	        4.0E+03	      &\\
18 09 25.0 & -20 15 41 & G10.213-0.326 &  mm & G10.15-0.34 & 3.9 & 1.1 &  62&		6 &  	 	2.4E+03	      &DC\\
18 09 26.2 & -20 17 33 & G10.188-0.344 &  mm & G10.15-0.34 & 1.0 & 0.4 &  	58& 	6 &  	 	5.9E+02	      &DC?\\
18 09 26.7 & -20 21 25 & G10.133-0.378 &  mm & G10.15-0.34 & 0.4 & 0.2 &  	 30& 6 &  	 	2.3E+02	      &\\
18 09 26.7 & -20 19 17 & G10.164-0.360 &  mm & G10.15-0.34 & 7.7$^\alpha$ & 1.2 &  109& 	6 &  	4.7E+03	      &\\
18 09 28.4 & -20 14 29 & G10.237-0.328 &  mm & G10.15-0.34 & 0.1 & 0.1 &  	15& 	6 &  	 	7.4E+01	      &DC\\
18 09 29.6 & -20 16 45 & G10.206-0.350 &  mm & G10.15-0.34 & 1.0 & 0.6 &  	 56&    6 &  	 	6.4E+02	      &N\\
18 09 31.3 & -20 18 29 & G10.184-0.370 &  mm & G10.15-0.34 & 0.2 & 0.1 &  	 17& 6 &  	 	1.4E+02	      &\\
18 09 33.5 & -20 17 49 & G10.198-0.372 &  mm & G10.15-0.34 & 0.3 & 0.2 &  	 30&	 6 &  	 	2.0E+02	      &N\\
18 09 34.6 & -20 22 21 & G10.138-0.419 &  mm & G10.15-0.34 & 0.2 & 0.1 &  	 20&	6 &  	 	1.2E+02	      &DC\\
18 09 35.3 & -20 21 25 & G10.149-0.407 &  mm & G10.15-0.34 & 0.3 & 0.2 &  	 29&	6 &  	 	2.1E+02	      &N\\
18 09 36.4 & -20 20 29 & G10.165-0.403 &  mm & G10.15-0.34 & 0.4 & 0.2 &  	 29&     6 &  	 	2.1E+02	      &N\\
18 09 36.4 & -20 18 29 & G10.194-0.387 &  mm & G10.15-0.34 & 0.5 & 0.2 &  	 30&	 6 &  	 	3.0E+02	      &DC\\
18 09 39.2 & -20 19 25 & G10.186-0.404 &  mm & G10.15-0.34 & 0.2$^\gamma$ & 0.1 & 12& 		6 &  	1.0E+02	      &\\
18 10 14.5 & -19 57 17 & G10.575-0.347 &  mm & G10.62-0.33 & -$^\beta$ & 0.2 &  	-& 	6.0 &  	 Ind	      &N\\
18 10 15.7 & -19 54 45 & G10.63-0.33B &  mm & G10.62-0.33 & 1.4 & 0.5 &  	 56&	6.0 &  	 	8.7E+02	      &\\
18 10 18.0 & -19 54 05 & G10.62-0.33 &  m  &  & 3.7 & 0.7 &  	63& 	6.0$^\iota$ &  	 	        2.2E+03	      &\\
18 10 19.0 & -20 45 33 & G9.88-0.75 &  r &  & 5.5 & 0.9 &  	 117& 	3.9$^\epsilon$ &  	 	1.4E+03	      &\\
18 10 24.1 & -20 43 09 & G9.924-0.749 &  mm & G9.88-0.75 & 0.7 & 0.3 & 42& 		3.9 &  	 	1.7E+02	      &\\
18 10 29.4 & -19 55 41 & G10.62-0.38 &  mr &  & 27.9 & 10.2 &  	 111&	6.0$^\nu$ &  	 	1.7E+04	              &\\
18 10 41.1 & -19 57 41 & G10.620-0.441 &  mm & G10.62-0.33 & 0.4 & 0.2 &  	19& 	6.0 &  	 	2.3E+02	      &N\\
18 11 24.4 & -19 32 04 & G11.075-0.384 &  mm & G11.11-0.34 & 0.9 & 0.3 &  	 38&	5.2 &  	 	4.0E+02	      &\\
18 11 31.8 & -19 30 44 & G11.11-0.34 &  r &  & 3.3 & 0.9 &  	91& 	5.2$^\epsilon$ &  	 	1.5E+03	      &\\
18 11 35.8 & -19 30 44 & G11.117-0.413 &  mm & G11.11-0.34 & 0.5 & 0.3 &  	31& 	5.2 &  	 	2.5E+02	      &N\\
18 11 51.4 & -17 31 30 & G12.88+0.48$^{\diamond \dagger \ast}$ &  m  &  & 6.9 & 2.4 & 87& 4/12.5$^\zeta$ & 1.9E+03 / 1.8E+04	 &\\
18 11 52.9 & -18 36 03 & G11.948-0.003 &  mm & G12.02-0.03 & 1.2$^\gamma$ & 0.5 & 42& 6.6/10 & 9.0E+02 / 2.1E+03     &\\
18 11 53.6 & -17 30 02 & G12.914+0.493$^\dagger$ &  mm & G12.88+0.48 & 0.7 & 0.4 & 29& 4/12.5 & 2.0E+02 / 2.0E+03	      &\\
18 12 01.9 & -18 31 56 & G12.02-0.03$^\diamond$ &  m  &  & 0.6 & 0.2 & 28& 6.6/10$^\zeta$ & 4.1E+02 / 9.4E+02	              &\\
18 12 02.1 & -18 40 26 & G11.902-0.100 &  mm & G11.93-0.14 & 0.2 & 0.1 & 16& 4.6/12.1 & 6.5E+01 / 4.5E+02	      &N\\
18 12 11.1 & -18 41 27 & G11.903-0.140 &  mr & G11.93-0.14 & 2.2 & 0.7 & 77& 4.6/12.1 & 7.9E+02 / 5.5E+03	      &N\\
18 12 15.6 & -18 44 58 & G11.861-0.183 &  mm & G11.93-0.14 & 0.1 & 0.1 & 15& 4.6/12.1 & 4.3E+01 / 3.0E+02	      &N\\
18 12 17.3 & -18 40 03 & G11.93-0.14 &  m  &  & 0.6 & 0.3 & 31& 4.6/12.1$^\iota$ & 2.3E+02 / 1.6E+03	              &\\
18 12 19.6 & -18 39 54 & G11.942-0.157 &  mm & G11.93-0.14 & 0.7 & 0.3 & 34& 4.6/12.1 & 2.3E+02 / 1.6E+03	      &\\
18 12 23.5 & -18 22 49 & G12.200-0.003 &  mm & G12.20-0.09 & 0.7 & 0.5 &  	28& 	14 &  	 	2.4E+03	      &\\
18 12 25.7 & -18 39 46 & G11.956-0.177 &  mm & G11.93-0.14 & 0.1 & 0.1& 13& 4.6/12.1 & 4.0E+01 / 2.7E+02	      &N\\
18 12 33.1 & -18 30 05 & G12.112-0.125 &  mm & G12.18-0.12 & 0.6 & 0.3 &  	50& 	14 &  	 	1.9E+03	      &N\\
18 12 39.2 & -18 24 17 & G12.20-0.09 &  mr &  & 4.3 & 2.3 &  	73& 	14$^\nu$ &  	 	1.4E+04               &\\
18 12 41.6 & -18 24 47 & G11.942-0.256 &  mm & G11.99-0.27 & 0.6 & 0.3 &26& 5.2/11.4 & 2.9E+02 / 1.4E+03	      &N\\
18 12 42.7 & -18 25 08 & G12.18-0.12 &  m  &  & 0.6 & 0.1 &  	42&  	14$^\nu$ &  	 	2.1E+03	              &\\
18 12 44.4 & -18 24 25 & G12.216-0.119 &  mm & G12.18-0.12 & 1.2 & 0.5 & 25& 	 	14 &  	 	3.9E+03	      &\\
18 12 51.2 & -18 40 40 & G11.99-0.27 &  m  &  & 0.3 & 0.2 & 28& 5.2/11.4$^\zeta$ & 1.5E+02 / 7.3E+02	              &\\
18 12 56.4 & -18 11 04  & G12.43-0.05 &  r &  & 0.9 & 0.2 &  	 42& 	16.7$^\nu$ &  	 	4.1E+03	              &\\
18 13 54.7 & -18 01 41 & G12.68-0.18 &  m  &  & 5.6 & 1.3 & 102& 4.7/11.8$^\zeta$ & 2.1E+03 / 1.3E+04	              &\\
18 13 58.5 & -18 54 21 & G11.94-0.62B &  mm & G11.93-0.61 & 4.5 & 1.3 &  	65& 	3.6 &  	 	9.8E+02       &N\\
18 14 00.9 & -18 53 27 & G11.93-0.61$^\ast$ &  mr &  & 5.9 & 2.2 &  	67& 	3.6$^\nu$ &  	 	1.3E+03	              &\\
18 14 07.6 & -18 00 37 & G12.722-0.218 &  mm & G12.68-0.18 & 1.9 & 0.8 & 50& 4.7/11.8 & 7.2E+02 / 4.5E+03	      &\\
18 14 25.5 & -17 53 52 & G12.855-0.226 &  mm & G12.90-0.26 & -$^\beta$ & - & -& 3.9/12.6 &  	 	Ind	      &\\
18 14 28.3 & -17 52 08 & G12.885-0.222 &  mm & G12.90-0.26 & 0.4 & 0.2 & 18& 3.9/12.6 & 1.0E+02 / 1.1E+03	      &\\
18 14 30.0 & -17 51 52 & G12.892-0.226 &  mm & G12.90-0.26 & 0.3 & 0.2 & 21& 3.9/12.6 & 6.5E+01 / 6.8E+02	      &\\
18 14 34.3 & -17 51 56 & G12.90-0.25B &  mm & G12.90-0.26 & 1.6 & 0.8 & 62& 3.9/12.6 & 4.1E+02 / 4.3E+02	      &DC?\\
18 14 36.1 & -17 54 56 & G12.859-0.272 &  mm & G12.90-0.26 & 2.3 & 0.7 & 50&3.9/12.6 & 5.9E+02 / 6.2E+03	      &\\
18 14 36.1 & -16 45 44 & G13.87+0.28 &  m &  & 6.0 & 1.8 &  79&	 	4.5$^\eta$ &  	 	2.1E+03	              &\\
18 14 39.5 & -17 52 00 & G12.90-0.26$^\ast$ &  m  &  & 8.6 & 2.4 &90& 3.9/12.6$^\iota$ & 2.2E+03 / 2.3E+04	              &\\
18 14 41.7 & -17 54 24 & G12.878-0.226 &  mm & G12.90-0.26 & 0.3 & 0.2 & 24& 3.9/12.6 & 7.0E+01 / 7.3E+02	      &\\
\end{tabular}
\end{center}
\end{table*}

\begin{table*}
\begin{center}
\contcaption{}
\begin{tabular}{@{}lllclcccccc@{}}
       \hline
       \hline
 \multicolumn{2} {c} {Peak Position~~~~~~} &  & \multicolumn{2} {c} {Identifier~~~~~~~~~~~} & \multicolumn{2} {c} {1.2mm Flux}&&& \\

RA  & Dec & Source Name & Tracer & mm map & Integ. & Peak  &  FWHM & Distance & Mass& \MSX~ \\
 (J2000)&(J2000)& $^a$  &        &        &  Jy$^b$& Jy/bm   &  arcsec$^c$     &  kpc$^d$ &  \mstar& corr$^e$ \\
(1)&(2)&(3)&(4)&(5)&(6)&(7)&(8)&(9)&(10)&(11)\\

\hline
18 14 42.9 & -17 53 12 & G12.897-0.281 &  mm & G12.90-0.26 & 0.3 & 0.1 & 12& 3.9/12.6 & 6.7E+01 / 7.0E+02	      &N\\
18 14 44.5 & -17 52 16 & G12.914-0.280 &  mm & G12.90-0.26 & 0.2 & 0.2 & 14& 3.9/12.6  & 6.2E+01 / 6.5E+02	      &\\
18 14 45.7 & -17 50 48 & G12.938-0.272 &  mm & G12.90-0.26 & 0.2 & 0.1 & 11 & 3.9/12.6 & 5.4E+01 / 5.7E+02	      &\\
18 16 22.1 & -19 41 27 & G11.49-1.48$^\ast$ &  m  &  & 4.9 & 1.5 & 97& 1.1/15.6$^\zeta$ & 1.0E+02 / 2.0E+04	              &\\
18 17 00.5 & -16 14 44 & G14.60+0.01 &  mr &  & 2.2 & 0.8 & 78& 2.8/13.7$^\zeta$ & 3.0E+02 / 7.2E+03	              &\\
18 19 12.6 & -20 47 31  & G10.84-2.59 &  r &  & 5.0 & 2.0 &  74 &		1.9$^\epsilon$ &  	 3.1E+02      &\\
18 20 10.3 & -16 10 35 & G15.022-0.618 &  mm & G15.03-0.37 & 1.7 & 0.8 & 30& 2.4/14 & 1.7E+02 / 5.7E+03	      &DC\\
18 20 17.6 & -16 13 55 & G14.987-0.670 &  mm & G15.03-0.37 & 1.5 & 0.8 & 24& 2.4/14 & 1.4E+02 / 4.9E+03	      &DC?\\
18 20 18.1 & -16 11 15 & G15.027-0.651 &  mm & G15.03-0.37 & 5.5 & 2.4 & 51& 2.4/14 & 5.4E+02 / 1.8E+04	      &N\\
18 20 19.2 & -16 09 31 & G15.054-0.641 &  mm & G15.03-0.37 & 0.7 & 0.5 & 18& 2.4/14 & 6.5E+01 / 2.2E+03	      &DC\\
18 20 20.9 & -16 14 35 & G14.983-0.687 &  mm & G15.03-0.37 & 4.5 & 1.5 & 57& 2.4/14 & 4.4E+02 / 1.5E+04	      &DC?\\
18 20 20.9 & -16 12 35 & G15.012-0.671 &  mm & G15.03-0.37 & 14.4 & 3.0 & 71& 2.4/14 & 1.4E+03 / 4.8E+04	      &N\\
18 20 23.1 & -16 11 16 & G15.03-0.67 &  mr &  & 30.0 & 7.6 & 98& 2.4/14$^\zeta$ & 2.9E+03 / 1.0E+05	              &\\
18 20 23.1 & -16 14 43 & G14.99-0.70 &  mm & G15.03-0.37 & 3.6 & 1.6 & 43& 2.4/14 & 3.5E+02 / 1.2E+04	              &N\\
18 20 24.2 & -16 13 15 & G15.009-0.688 &  mm & G15.03-0.37 & 3.1 & 1.4 & 25& 2.4/14 & 3.1E+02 / 1.0E+04	      &\\
18 20 25.3 & -16 13 39 & G15.005-0.695 &  mm & G15.03-0.37 & 1.1 & 1.1 & 24& 2.4/14 & 1.0E+02 / 3.5E+03	      &\\
18 20 27.0 & -16 08 51 & G15.079-0.663 &  mm & G15.03-0.37 & 1.5$^\alpha$ & 0.5 & 43& 2.4/14 & 1.4E+02 / 4.9E+03    &\\
18 20 28.1 & -16 13 15 & G15.016-0.702 &  mm & G15.03-0.37 & 1.2 & 0.6 & 22& 2.4/14 & 1.2E+02 / 4.1E+03	      &\\
18 20 29.8 & -16 12 35 & G15.029-0.703 &  mm & G15.03-0.37 & 6.4$^\alpha$ & 1.2 & 104 & 2.4/14 & 6.3E+02 / 2.1E+04  &\\
18 20 30.3 & -16 08 35 & G15.089-0.673 &  mm & G15.03-0.37 & 3.8 & 1.1 & 44& 2.4/14 & 3.8E+02 / 1.3E+04	      &\\
18 20 31.4 & -16 12 51 & G15.028-0.710 &  mm & G15.03-0.37 & -$^\alpha$ & 1.3 & -& 2.4/14 &  	 	Ind	      &\\
18 20 33.1 & -16 08 19 & G15.098-0.681 &  mm & G15.03-0.37 & 1.0 & 0.7 & 31& 2.4/14 & 1.0E+02 / 3.4E+03	      &\\
18 21 14.6 & -14 32 52 & G16.580-0.079 &  mm & G16.58-0.05 & 0.5 & 0.2 &45&  4.5/11.8 & 1.7E+02 / 1.2E+03	      &DC?\\
18 21 09.1 & -14 31 49 & G16.58-0.05$^{\dagger \ast }$ &  m  &  & 3.0 & 1.5 & 58& 4.5/11.8$^\iota$ & 1.0E+03 / 7.1E+03	              &\\
18 24 56.0 & -13 19 03 & G18.087-0.292 &  mm & G18.15-0.28 & 0.4 & 0.2 &  20&	 	2.6 &  	 	4.7E+01	      &DC\\
18 24 58.6 & -13 18 47 & G18.095-0.299 &  mm & G18.15-0.28 & 0.4 & 0.2 &  	33& 	2.6 &  	 	5.1E+01	      &DC\\
18 25 00.8 & -13 18 23 & G18.105-0.304 &  mm & G18.15-0.28 & 0.5 & 0.2 &  	 31&	2.6 &  	 	5.3E+01       &DC\\
18 25 01.3 & -13 15 35 & G18.15-0.28 &  r &  & 2.5 & 0.9 &  	 60&	2.6$^\epsilon$ &  	 	2.8E+02	      &\\
18 25 03.5 & -13 16 15 & G18.142-0.297 &  mm & G18.15-0.28 & 1.1 & 0.3 &  	 56&	2.6 &  	 	1.2E+02	      &\\
18 25 05.1 & -13 14 55 & G18.165-0.293 &  mm & G18.15-0.28 & 0.3 & 0.2 &  	 22&	2.6 &  	 	3.2E+01	      &\\
18 25 05.1 & -13 18 31 & G18.112-0.321 &  mm & G18.15-0.28 & 0.2 & 0.2 &  	 20&	2.6 &  	 	2.5E+01	      &N\\
18 25 07.3 & -13 14 23 & G18.177-0.296 &  mm & G18.15-0.28 & 0.9 & 0.5 &  	 29&	2.6 &  	 	1.1E+02	      &N\\
18 25 42.2 & -13 10 32 & G18.30-0.39 &  r &  & 5.6 & 1.6 &  	 75&	2.9$^\epsilon$ &  	 	8.1E+02	      &\\
18 27 16.3 & -11 53 51 & G19.61-0.1 &  m &  & 1.3 & 0.7 & 43 & 4/12$^\zeta$ & 3.5E+02 / 3.2E+03	              &\\
18 27 38.2 & -11 56 38 & G19.607-0.234 &  mr & G19.70-0.27A & 13.4 & 5.6 & 57&3.5/12.5 & 2.8E+03 / 3.6E+04	      &\\
18 27 55.5 & -11 52 39 & G19.70-0.27A &  m &  & 1.1 & 0.5 & 35& 3.5/12.5$^\zeta$ & 2.3E+02 / 2.9E+03	              &\\
18 29 24.2 & -15 15 34 & G16.871-2.154 &  mm & G16.86-2.15 & -$^\alpha$ & - & -& 1.7/14.6 &  	 	Ind   &\\
18 29 24.4 & -15 16 04 & G16.86-2.15 &  m &  & 16.9$^\alpha$ & 2.7 & 109 & 1.7/14.6$^\zeta$ & 8.3E+02 / 6.2E+04      &\\
18 29 33.1 & -15 15 50 & G16.883-2.188 &  mm & G16.86-2.15 & 0.5 & 0.2 & 21& 1.7/14.6 & 2.3E+01 / 1.7E+03	      &N\\
18 31 02.1 & -09 49 14 & G21.87+0.01 &  mr &  & 1.0 & 0.6 & 30& 1.9/13.9$^\zeta$ & 6.4E+01 / 3.4E+03	              &\\
18 31 43.2 & -09 22 25 & G22.36+0.07B$^{\dagger \ast}$ &  m &  & 2.5 & 0.6 & 43& 5.1/10.7$^\zeta$ & 1.1E+03 / 4.8E+03	              &\\
18 31 44.1 & -09 22 12 & G22.35+0.06$^{\diamond \dagger}$ &  m  &  & 2.1 & 0.6 & 39& 5/10.8$^\iota$ & 8.9E+02 / 4.2E+03	              &\\
18 33 53.6 & -08 07 15  & G23.71+0.17$^\ast$ &  r &  & 3.0 & 1.0 &  	38& 	6.5$^\nu$ &  	 	2.2E+03	              &\\
18 33 53.6 & -08 08 43 & G23.689+0.159 &  mm & G23.71+0.17 & 0.4 & 0.1 &  15&	 	6.5 &  	 	2.5E+02	      &N\\
18 34 07.6 & -07 19 05 & G24.450+0.489 &  mm & G24.47+0.49 & 0.2 & 0.1 & 10& 5.7/9.8 & 1.0E+02 / 2.9E+02	      &\\
18 34 10.3 & -07 17 45  & G24.47+0.49$^\ast$ &  r &  & 3.8 &0.8 & 53& 5.7/9.8$^\eta$ & 2.1E+03 / 6.2E+03	              &\\
18 34 20.9 & -05 59 40 & G25.65+1.04$^\ast$ &  mr &  & 6.5 & 2.5 &  	63&  	3.2$^\chi$ &  	 	1.1E+03               &\\
18 34 21.7 & -07 54 45 & G23.949+0.163 &  mm & G23.96+0.15 & 0.8 & 0.2 & 36& 	 	5 &  	 	3.2E+02	      &\\
18 34 24.9 & -07 54 53  & G23.96+0.15$^\ast$ &  r &  & 2.2 & 0.9 &  	 55&	5$^\nu$ &  	 	9.2E+02	              &\\
18 34 25.4 & -08 40 23 & G23.281-0.201 &  mm & G23.25-0.24 & 0.3 & 0.2 & 23& 4.3/11.3 & 1.1E+02 / 7.4E+02	      &DC\\
18 34 25.9 & -08 41 19 & G23.268-0.210 &  mm & G23.25-0.24 & 0.6 & 0.2 & 36& 4.3/11.3 & 1.7E+02 / 1.2E+03	      &DC?\\
18 34 26.5 & -07 51 09 & G24.012+0.173 &  mm & G23.96+0.15 & 0.1 & 0.1 &  	13& 	5 &  	 	4.7E+01	      &N\\
18 34 27.6 & -07 53 41 & G23.976+0.150 &  mm & G23.96+0.15 & 0.2 & 0.1 &  	 16&	5 &  	 	7.2E+01	      &DC?\\
18 34 28.7 & -07 54 53 & G23.960+0.137 &  mm & G23.96+0.15 & 0.1 & 0.1 &  	 4&	5 &  	 	4.7E+01	      &\\
18 34 29.2 & -07 53 09 & G23.987+0.148 &  mm & G23.96+0.15 & 0.2 & 0.1 &  	 19&	5 &  	 	8.1E+01	      &DC\\
18 34 31.3 & -08 42 47 & G23.25-0.24 &  m &  & 0.4 & 0.2 & 20& 4.3/11.3$^\zeta$ & 1.1E+02 / 7.8E+02	              &\\
18 34 31.9 & -07 51 33 & G24.016+0.150 &  mm & G23.96+0.15 & 0.1 & 0.1 & 7& 	 	5 &  	 	5.1E+01	      &DC\\
18 34 36.2 & -08 42 39 & G23.268-0.257 &  mm & G23.25-0.24 & 3.7 & 0.6 & 53& 4.3/11.3 & 1.2E+03 / 8.0E+03	      &\\
18 34 39.4 & -08 31 33 & G23.43-0.18 &  m &  & 4.0 & 1.7 & 55&6/9.6$^\zeta$ & 2.5E+03 / 6.3E+03	              &\\
18 34 45.6 & -08 34 21 & G23.409-0.228 &  mm & G23.43-0.18 & 0.9 & 0.5 & 22& 6/9.6 & 5.7E+02 / 1.5E+03	      &\\
18 34 48.4 & -08 33 57 & G23.420-0.235 &  mm & G23.43-0.18 & 1.0 & 0.4 & 25& 6/9.6 & 6.0E+02 / 1.5E+03	      &\\
\end{tabular}
\end{center}
\end{table*}

\begin{table*}
\begin{center}
\contcaption{}
\begin{tabular}{@{}lllclcccccc@{}}
       \hline
       \hline
\multicolumn{2} {c} {Peak Position~~~~~~} &  & \multicolumn{2} {c} {Identifier~~~~~~~~~~~} & \multicolumn{2} {c} {1.2mm Flux}&&& \\

RA  & Dec & Source Name & tracer & mm map & Integ. & Peak  &  FWHM & Distance & Mass& \MSX~ \\
 (J2000)&(J2000)& $^a$  &        &        &  Jy$^b$& Jy/bm   &  arcsec$^c$    &  kpc$^d$ &  \mstar & corr$^e$ \\
(1)&(2)&(3)&(4)&(5)&(6)&(7)&(8)&(9)&(10)& (11)\\
\hline
18 34 50.7 & -08 41 03 & G23.319-0.298 &  mm & G23.25-0.24 & 0.7 & 0.2 & 38& 4.3/11.3 & 2.3E+02 / 1.6E+03	   &\\
18 36 06.1 & -07 13 47 & G23.754+0.095 &  mm & G24.78+0.08 & 1.1 & 0.5 & 38& 6.6/8.8 & 8.2E+02 / 1.5E+03	   &N\\
18 36 09.4 & -07 11 39 & G24.792+0.099 &  mm & G24.78+0.08 & 4.1 & 1.1 & 45& 6.6/8.8 & 3.1E+03 / 5.4E+03          &\\
18 36 12.6 & -07 12 11 & G24.78+0.08 &  m &  & 13.6 & 4.1 & 73& 6.6/8.8$^\zeta$ & 1.0E+04 / 1.8E+04	           &\\
18 36 18.4 & -07 08 52 & G24.84+0.08 &  m &  & 1.0 & 0.4 & 23& 8/10.1$^\zeta$ & 1.1E+03 / 1.8E+03	           &\\
18 36 19.5 & -07 09 00 & G24.850+0.082 &  mm & G24.84+0.08 & 0.5 & 0.3 & 13& 8/10.1 & 5.3E+02 / 8.5E+02          &\\
18 36 25.9 & -07 05 08 & G24.919+0.088 &  mm & G24.84+0.08 & 2.6 & 0.7 & 47& 8/10.1 & 2.8E+03 / 4.4E+03	   &\\
18 38 03.0 & -06 24 01 & G25.70+0.04 &  mr &  & 1.9 & 0.5 &  	46& 	11.9$^\chi$ &  	 	4.5E+03	           &\\
18 38 57.0 & -06 24 53 & G25.802-0.159 &  r & G25.83-0.18 & 1.1 & 0.5 & 34& 5.5/9.8 & 5.5E+02 / 1.8E+03	   &\\
18 39 03.6 & -06 24 10 & G25.83-0.18 &  m &  & 5.4 & 2.3 & 60& 5.5/9.8$^\zeta$ & 2.8E+03 / 8.8E+03	           &DC\\
18 42 42.6 & -04 15 32 & G28.14-0.00 &  m &  & 0.8 & 0.4 & 40& 6.2/8.8$^\zeta$ & 5.2E+02/ 1.1E+03	           &\\
18 42 43.1 & -04 09 56 & G28.231+0.367 &  mm & G28.20-0.04 & 0.5 & 0.2 & 31& 6/8.9 & 2.8E+02 / 6.1E+02	   &\\
18 42 54.9 & -04 07 40 & G28.287+0.010 &  mm & G28.20-0.04 & 0.5$^\alpha$ & 0.2 & 44& 6/8.9 & 3.1E+02 / 6.9E+02  &\\
18 42 58.1 & -04 13 56 & G28.20-0.04 &  mr &  & 7.0 & 2.6 & 70& 6/8.9$^\zeta$ & 4.3E+03 / 9.5E+03	           &\\
18 43 00.8 & -04 14 28 & G28.198-0.063 &  mm & G28.20-0.04 & 0.3 & 0.2 & 21& 6/8.9 & 1.5E+02 / 3.4E+02	           &DC\\
18 43 02.4 & -04 14 59 & G29.193-0.073 &  mm & G28.20-0.04 & 0.3 & 0.2 & 19& 6/8.9 & 2.0E+02 / 4.5E+02	           &DC\\
18 44 14.2 & -04 17 59 & G28.28-0.35 &  mr &  & 5.1 & 1.2 &  	 64& 	2.8$^\chi$ &  	 	6.5E+02	           &\\
18 44 22.0 & -04 17 38 & G28.31-0.38 &  m &  & 1.1& 0.3 & 39& 5/9.9$^\zeta$ & 4.6E+02 / 1.8E+03	           &\\
18 45 52.8 & -02 42 29 & G29.888+0.001 &  mm & G29.918-0.014 & 0.5 & 0.3 & 9& 6/8.7 & 2.8E+02 / 5.9E+02	   &N\\
18 45 54.4 & -02 42 37 & G29.889-0.006 &  mm & G29.918-0.014 & 1.0 & 0.3 & 25& 6/8.7 & 6.4E+02 / 1.3E+03	   &\\
18 45 59.7 & -02 41 17 & G29.918-0.014 &  mm & target Oct02 & 0.3 & 0.2 & 11& 6/8.7$^\zeta$ & 1.8E+02 / 3.7E+02  &N\\
18 46 00.2 & -02 45 09 & G29.86-0.04 &  m &  & 1.0 & 0.5 & 32& 6.4/8.3$^\zeta$ & 7.0E+02 / 1.2E+03	           &\\
18 46 01.3 & -02 45 25 & G29.861-0.053 &  mm & G29.918-0.014 & 0.7 & 0.4 & 26& 6/8.7 & 4.1E+02 / 8.6E+02	   &\\
18 46 02.4 & -02 45 57 & G29.853-0.062 &  mm & G29.918-0.014 & 0.8 & 0.4 & 28& 6/8.7 & 5.1E+02 / 1.1E+03	   &\\
18 46 04.0 & -02 39 25 & G29.96-0.02B$^\ast$ &  mr &  & 9.3 & 4.0 & 80& 6/8.7$^\zeta$ & 5.7E+03 / 1.2E+04	           &\\
18 46 05.0 & -02 42 29 & G29.912-0.045 &  mm & G29.918-0.014 & 3.4 & 0.7 & 73& 6/8.7 & 2.1E+03 / 4.3E+03	   &\\
18 46 06.1 & -02 41 25 & G29.930-0.040 &  mm & G29.918-0.014 & 0.4 & 0.3 & 10& 6/8.7 & 2.6E+02 / 5.5E+02	   &\\
18 46 08.8 & -02 39 09 & G29.969-0.033 &  mm & G29.918-0.014 & 0.5 & 0.3 & 27& 6/8.7 & 3.2E+02 / 6.7E+02	   &\\
18 46 09.8 & -02 41 25 & G29.937-0.054 &  mm & G29.918-0.014 & 1.2 & 0.5& 23& 6/8.7 & 7.2E+02 / 1.5E+03	   &\\
18 46 11.5 & -02 42 05 & G29.945-0.059 &  mm & G29.918-0.014 & 2.3 & 0.7 & 62& 6/8.7 & 1.4E+03 / 2.9E+03	   &\\
18 46 12.5 & -02 39 09 & G29.978-0.050 &  m  & G29.918-0.014 & 1.9 & 0.8 & 55& 6/8.7 & 1.2E+03 / 2.5E+03	   &DC\\
18 46 58.6 & -02 07 27 & G30.533-0.023 &  mm & G30.59-0.04 & 0.6$^\gamma$ & 0.4 & 22& 3/11.6 & 8.6E+01 / 1.3E+03 &\\
18 47 07.0 & -01 46 50 & G30.855+0.149 &  mm & G30.89+0.16 & 1.4$^\alpha$ & 0.3 & 38& 6.9/7.7 & 1.1E+03 / 1.4E+03 &\\
18 47 08.6 & -01 44 02 & G30.89+0.16 &  m  &  & 0.8 & 0.5 & 28& 6.9/7.7$^\zeta$ & 6.2E+02 / 7.8E+02	           &\\
18 47 13.4 & -01 44 58 & G30.894+0.140 &  mm & G30.89+0.16 & 1.1 & 0.4 & 31& 6.9/7.7 & 8.8E+02 / 1.1E+03	   &DC?\\
18 47 15.5 & -01 44 18 & G30.908+0.137 &  m  & G30.89+0.16 & 0.2 & 0.1 & 5& 6.9/7.7 & 1.2E+02 / 1.5E+02	   &DC?\\
18 47 15.5 & -01 47 06 & G30.869+0.116 &  r & G30.89+0.16 & 2.7 & 1.2 & 43& 6.9/7.7 & 2.2E+03 / 2.7E+03	   &\\
18 47 18.9 & -02 06 07 & G30.59-0.04 &  m  &  & 3.2 & 1.2 & 51& 3/11.6$^\zeta$ & 4.9E+02 / 7.3E+03	           &\\
18 47 26.7 & -01 44 42 & G30.924+0.092 &  mm & G30.89+0.16 & 0.5 & 0.2 & 28& 6.9/7.7 & 4.3E+02 / 5.4E+02         &DC?\\
18 47 34.2 & -01 56 41 & G30.760-0.027 &  mm & G30.76-0.05 & 1.6 & 0.7 & 34& 5.8/8.8 & 9.3E+02 / 2.2E+03	   &\\
18 47 34.3 & -01 12 47 & G31.41+0.30 &  mr &  & 15.2 & 6.8 &  	 67&	7.9$^\chi$ &  	 	1.6E+04	           &\\
18 47 35.4 & -02 02 07 & G30.682-0.072 &  mm & G30.76-0.05 & 2.5 & 0.6 & 43& 5.8/8.8 & 1.1E+03 / 3.3E+03	   &\\
18 47 35.8 & -01 55 13 & G30.78-0.02 &  m  &  & 6.7 & 1.1 &  	91&  	5.5$^\eta$ &  	 	3.5E+03	           &\\
18 47 36.0 & -02 01 05 & G30.705-0.065 &  m & G30.76-0.05 & 8.4 & 2.5 & 70& 5.8/8.8 & 4.8E+03 / 1.1E+04	   &DC\\
18 47 37.9 & -01 57 45 & G30.76-0.05 &  mm &  & 2.3 & 1.3 & 64& 5.8/8.8$^\zeta$ & 1.3E+03 / 3.0E+03	           &\\
18 47 38.5 & -01 56 57 & G30.769-0.048 &  mm &  & 1.4 & 0.6 & 30& 5.8/8.8 & 7.7E+02 / 1.8E+03	                   &\\
18 47 39.2 & -01 58 41 & G30.740-0.060 &  mm & G30.76-0.05 & 3.4 & 1.1 & 52& 5.8/8.8 & 2.0E+03 / 4.5E+03         &N\\
18 47 39.6 & -01 53 53 & G30.749-0.048 &  mm & G30.76-0.05 & 2.2 & 0.6 & 30& 5.8/8.8 & 1.3E+03 / 2.9E+03	   &DC?\\
18 47 41.3 & -01 35 05 & G31.094+0.111 &  mm & G31.06+0.09 & 0.2 & 0.1 & 9& 1.2/13.3 & 0.5E+01 / 6.6E+02	   &\\
18 47 41.3 & -02 00 33 & G30.716-0.082 &  mm & G30.76-0.05 & 4.1 & 1.5 & 41& 5.8/8.8 & 2.4E+03 / 5.5E+03	   &\\
18 47 41.3 & -01 59 45 & G30.729-0.078 &  mm & G30.76-0.05 & 1.4 & 0.6 & 35& 5.8/8.8 & 7.8E+02 / 1.8E+03	   &DC\\
18 47 41.3 & -01 37 21 & G31.06+0.09 &  m  &  & 0.3 & 0.2 & 25& 1.2/13.3$^\zeta$ & 0.8E+01 / 9.9E+02	           &\\
18 47 46.5 & -01 54 16 & G30.81-0.05 &  m  &  & 16.9 & 5.7 & 75& 6.7/7.9$^\zeta$ & 1.3E+04 / 1.8E+04	           &DC\\
18 47 50.9 & -01 38 17 & G31.065+0.051 &  r & G31.06+0.09 & 0.4$^\alpha$ & 0.1 & 23& 1.2/13.3 & 1.0E+01 / 1.2E+03&\\
18 48 01.6 & -01 36 01 & G31.119+0.029 &  mm & G31.06+0.09 & 0.2$^\alpha$ & 0.1 & 20& 1.2/13.3 & 0.6E+01 / 7.2E+02 &\\
18 48 12.4 & -01 26 23 & G31.28+0.06 &  mr &  & 5.3 & 1.5 &  	91&  	6.1$^\chi$ &  	 	3.3E+03	           &\\
18 48 09.7 & -01 27 50 & G31.256+0.061 &  mm & G31.28+0.06 & 0.8 & 0.2&  	31&  	6.1 &  	 	5.3E+02	   &DC\\
18 49 33.1 & -01 29 04  & G31.40-0.26$^\ast$ &  r &  & 3.1 & 1.3 &  	 52&	7.3$^\rho$ &   	 	2.8E+03	           &\\
18 49 34.2 & -01 29 44 & G31.388-0.266 &  mm & G31.40-0.26 & 0.1& 0.1 &  	 15&	7.3 &  	 	1.3E+02	   &\\
18 50 30.7 & -00 02 00 & G32.80+0.19$^\ast$ &  r &  & 9.2 & 3.8 &  	56& 	13$^\epsilon$ &  	 	2.6E+04	   &\\
18 52 08.0 & +00 08 10 & G33.13-0.09 &  mr &  & 2.7 & 1.2 &  	 56& 	6.0$^\epsilon$ &  	 	1.7E+03	   &\\
\end{tabular}
\end{center} 
\end{table*}

\begin{table*}
\begin{center}
\contcaption{}
\begin{tabular}{@{}lllclcccccc@{}}
       \hline
       \hline
\multicolumn{2} {c} {Peak Position~~~~~~} &  & \multicolumn{2} {c} {Identifier~~~~~~~~~~~} & \multicolumn{2} {c} {1.2mm Flux}&&& \\

RA  & Dec & Source Name & Tracer & mm map & Integ. & Peak  &  FWHM & Distance & Mass & \MSX~\\
 (J2000)&(J2000)&  $^a$ &        &        &  Jy$^b$& Jy/bm   &  arcsec$^c$     &  kpc$^d$ &  \mstar & corr$^e$ \\
(1)&(2)&(3)&(4)&(5)&(6)&(7)&(8)&(9)&(10)&(11)\\

\hline
18 52 50.2 & +00 55 28 & G33.92+0.11 &  r &  & 6.2 & 1.7 &  	68&  	8.2$^\nu$ &  	 	7.1E+03	              &\\
18 53 18.2 & +01 14 57 & G34.256+0.155 &  mm & G34.24+0.13  & 51.8 & 13.5 &   130& 	 	3.5$^\nu$ &  1.1E+04  &\\
18 53 21.7 & +01 13 37 & G34.24+0.13$^\ast$ &  m &  & 0.7 & 0.4 &  	30& 	3.5 &  	 	1.4E+02	                      &\\
18 54 00.5 & +02 01 16 & G35.02+0.35 &  mr &  & 2.5 & 1.0 & 55& 3/10.9$^\zeta$ & 3.9E+02/ 5.1E+03	              &\\
18 56 01.2 & +02 22 59 & G35.57+0.07 &  r &  & 1.5 & 0.6 &  	 46& 	10.2$^\eta$ &  	 	2.6E+03	              &\\
18 56 03.9 & +02 23 23 & G35.586+0.061 &  mm & G35.57+0.07 & 0.5 & 0.2 &  15&	 	10.2 &  	8.3E+02	      &N\\
18 56 05.5 & +02 22 27 & G35.575+0.048 &  mm & G35.57+0.07 & 0.3 & 0.2 &  	16& 	10.2 &  	5.0E+02	      &\\
18 56 13.5 & +02 21 23 & G35.575+0.010 &  mm & G35.57+0.07 & 0.5 & 0.3 &  	 28&	10.2 &  	9.4E+02	      &N\\
18 56 14.0 & +02 21 15 & G35.574+0.007 &  mm & G35.57+0.07 & 0.4 & 0.3&  	 21&	10.2 &  	7.8E+02	      &N\\
18 56 22.4 & +02 20 26 & G35.58-0.03 &  r &  & 3.5 & 1.5 &  	 65&	10.1$^\eta$ &  	 	6.1E+03	              &\\
18 56 22.4 & +02 20 58 & G35.585-0.026 &  mm & G35.58-0.03 & 0.2 & 0.1 & 18& 	 	10.1 &  	 3.7E+02      &\\
18 56 22.4 & +02 19 30 & G35.564-0.037 &  mm & G35.58-0.03 & 0.2 & 0.1 &  	8& 	10.1 &  	 2.8E+02      &N\\
18 56 22.4 & +02 22 02 & G35.601-0.018 &  mm & G35.58-0.03 & 0.2 & 0.1&  	 29&	10.1 &  	 4.0E+02      &\\
18 57 09.0 & +01 39 05 & G35.05-0.52 &  r &  & 0.7 & 0.2 &  	36& 	11.5$^\nu$ &  	 	1.5E+03	              &\\
% &  &  &  &  &  &  & 		 & 		&	\\
19 00 06.9 & +03 59 39 & G37.475-0.106 &  m  & G37.55-0.11 & 0.2 & 0.1 &  	22& 	9.8 &  	 	3.3E+02	      &\\
19 00 16.0 & +04 03 15  & G37.55-0.11 &  r &  & 1.7 & 0.7 &  	 48&	9.8$^\nu$ &  	 	2.8E+03	&\\
19 23 24.9 & +14 30 56 & G49.459-0.317 &  mm & G49.49-0.37 & 0.3 & 0.2 & 7&5.5  &  	 	1.3E+02	&N\\
19 23 27.7 & +14 30 33 & G49.458-0.330 &  mm & G49.49-0.37 & 1.0 & 0.3 & 38& 5.5&  	 	5.1E+02	&N\\
19 23 27.1 & +14 31 04 & G49.465-0.324 &  mm & G49.49-0.37 & 0.6 & 0.3 & 22& 5.5 &  	 	3.3E+02 &N\\
19 23 32.6 & +14 29 52 & G49.456-0.354 &  mm & G49.49-0.37 & 3.6 & 1.2 & 44& 5.5 &  	 	1.8E+03	&\\
19 23 35.9 & +14 31 04 & G49.482-0.355 &  mm & G49.49-0.37 & 1.2 & 0.7 & 31& 5.5 &  	 	6.0E+02	&\\
19 23 37.1 & +14 30 16 & G49.472-0.366 &  m  & G49.49-0.37 & 3.8 & 1.7 & 34& 5.5 &  	 	2.0E+03	&N\\
19 23 38.2 & +14 34 16 & G49.533-0.338 &  mm & G49.49-0.37 & 0.2 & 0.1 & 7& 5.5 &  	 	1.1E+02	&N\\
19 23 39.8 & +14 32 40 & G49.513-0.356 &  mm & G49.49-0.37 & 0.8 & 0.3 & 16& 5.5 &  	 	4.0E+02	&N\\
19 23 39.8 & +14 33 44 & G49.528-0.348 &  mm & G49.49-0.37 & 1.1 & 0.4 & 21& 5.5 &  	 	5.6E+02	&N\\
19 23 39.8 & +14 31 05 & G49.49-0.37 &  m  &  & 31.5 & 10.7 & 64&  5.5$^\zeta$ &  	 	1.6E+04	&\\
19 23 43.1 & +14 30 32 & G49.488-0.385 &  mm & G49.49-0.37 & 71.5 & 17.1 & 87 &5.5 &  	 	3.7E+04	&\\
19 23 45.9 & +14 29 44 & G49.481-0.401 &  mm & G49.49-0.37 & 6.2 & 2.8 &  44 & 5.5 &  	 	3.2E+03	&\\
19 23 47.5 & +14 28 24 & G49.465-0.417 &  mm & G49.49-0.37 & 0.2 & 0.1 & 7& 5.5 &  	 	1.1E+02	&N\\
19 23 48.6 & +14 28 08 & G49.462-0.423 &  mm & G49.49-0.37 & 0.6 & 0.3& 23& 5.5 &  	 	3.0E+02	&N\\
19 23 49.2 & +14 28 48 & G49.474-0.420 &  mm & G49.49-0.37 & 0.9 & 0.2 & 26& 5.5 &  	 	4.4E+02	&N\\
19 23 50.3 & +14 29 28  & G49.48-0.419 &  mm & G49.49-0.37 & 0.2 & 0.1 & 7 & 5.5 &  	 	1.0E+02	&\\
19 23 50.3 & +14 32 48 & G49.536-0.395 &  mm & G49.49-0.37 & 0.3 & 0.1 & 11 & 5.5 &  	 	1.8E+02	&\\
19 23 50.8 & +14 29 52 & G49.494-0.420 &  mm & G49.49-0.37 & 0.6 & 0.3 & 14& 5.5  &  	 	3.0E+02	&\\
19 23 50.8 & +14 30 56 & G49.508-0.409 &  mm & G49.49-0.37 & 0.4 & 0.2 & 19& 5.5  &  	 	2.3E+02	&N\\
19 23 53.0 & +14 30 24 & G49.505-0.421 &  mm & G49.49-0.37 & 1.2 & 0.3 & 36 & 5.5 &  	 	6.3E+02	&N\\
19 43 10.0 & +23 44 59 & G59.794+0.076 &  mm	& G59.78+0.06 &	0.8 & 0.2 & 30& 2.6/6 & 9.5E+01 / 5.0E+02	     &N\\
19 43 11.2 & +23 44 03 & G59.78+0.06$^\diamond$	&  r & & 4.7	& 1.3	& 65&2.6/6$^\zeta$ & 5.4E+02 / 2.9E+03	     &\\
\hline
\end{tabular}
\end{center}

\begin{flushleft}
$^a$ Source names given to two (or less) decimal places are consistent with those reported by \citeauthor{walsh98, thompson04, minier01} from which they were targeted. Source names given to three decimal places, denote those source identified in this survey, with the extended Galactic names intended to distinguish closely associated sources. Footnotes in this column indicate sources which have been targeted in the millimetre and submillimetre studies of \citet{beuther02, williams04, faundez04} denoted by $^\diamond$, $^\dagger$ and  $^\ast$ respectively. Sources denoted by a $^\star$ are faint millimetre sources, which produce masses uncharacteristic of massive star formation regions if located at the near distance.

 $^b$ 1.2mm Fluxes in Jy. $^{\alpha}$denotes sources that have more than one millimetre peak encompassed, for which it was not possible to clearly distinguish the individual cores. The flux quoted here is for all the sources. $^{\beta}$denotes sources located too close to the edge of the map, for their fluxes to be calculated. In the majority of cases it is not possible to determine the peak of the millimetre emission either. $^{\gamma}$denotes sources situated quite close to the edge of the map, with some uncertainty in source size. The fluxes quoted here are a lower limit. $^{\delta}$denotes the two NM-\IRAS~ positions where no millimetre emission is detected at the reported \IRAS~ coordinates.\\

$^c$ Sources denoted with a $^\ddagger$ indicate those for which a radius could not be determined using the GAIA program.\\

$^d$ Distances given in kpc, with the footnote indicating the literature reference. Only those sources that were targeted have the distance reference indicated. The `mm-only' sources have distances adopted from the nearby tracer, and accordingly their distance reference is the same as the nearby tracer. $^{\epsilon}$\citet{kurtz94}, $^{\zeta}$\citet{pestalozzi04}, $^{\eta}$\citet{thompson04}, $^{\xi}$\citet{minier05}, $^{\iota}$\citet{walsh97}, $^{\mu}$\citet{norris93}, $^{\nu}$\citet{wood89morph},  $^{\rho}$\citet{palagi93}, as quoted in \citet{thompson04}, $^{\chi}$\citet{walsh03}. `Ind' indicates those sources for which a distance is {\it indeterminate}.\\

$^e$ Indicates the lack of mid-IR \MSX~ association (`N') or an association with a mid-IR dark cloud (`DC'). A `DC?' in this column indicates that there is no mid-IR emission, however it is not clear whether the lack of emission is due to absorption (i.e. a dark cloud) or simply an absence of emission.  Note that an absence in this column {\it does not} indicate that an \MSX~ source is present.

\end{flushleft}
\end{table*}

\subsubsection{Derived Parameters}

   We present the morphological class, the clustering type, the radius of the source and the H$_{2}$ number density ($n_{H_{2}}$) for each of the sources listed in Table \ref{main}. Column 1 lists the source names in G-name nomenclature, listed in right ascension order as per Table \ref{main}. Column 2 gives the morphological class of the source, while column 3 indicates the strength of the clustering of the sources in the maps (as discussed in $\S\ref{sampleimages}$). Column 4 gives the radius of the source in parsecs (pc) and column 5 lists the H$_{2}$ number density ($n_{H_{2}}$) for each source. Two values are listed in columns 4 and 5  if there is an ambiguity in the distance to the source, with the result for the near distance preceding that of the far distance.

\clearpage
\begin{table*}
\begin{center}
\caption{Derived Parameters of the 1.2mm sources from Table \ref{main}. \label{main2} }
\begin{tabular}{@{}lcccclccccc@{}}
       \hline
       \hline

Source Name~$^a$& M~$^b$ & C~$^c$  & Radius~$^d$  & $n_{H_2}$~$^d$ &Source Name~$^a$ & M~$^b$ & C~$^c$ & Radius~$^d$ & $n_{H_2}$~$^d$  \\
       & & & parsecs~ &cm$^{-3}$~ & & &   & parsecs~ &cm$^{-3}$~\\
(1)&(2)&(3)&(4)&(5) &(1)&(2)&(3)&(4)&(5)\\
\hline
G206.535-16.356 & D & T 	&  0.10	&  1.6E+05 &\vline~	G304.906+0.574 & S & L	&  Ind	&  Ind	\\
G206.54-16.35 & D & T 	&  0.06	&  1.9E+06&	\vline~	G304.919+0.542 & I  & L	&  Ind	&  Ind	\\
G183.34+0.59 & S & -	&  Ind	&  Ind 	   &    \vline~	G305.952+0.555 & I  & L	&  Ind	&  Ind	\\
G213.61-12.6 & S & -	&  0.39	&  1.3E+05&     \vline~	G304.952+0.522 & S & L	&  Ind	&  Ind	\\
G189.78+0.34 & S & -	&  0.16	&  1.7E+05&     \vline~	G304.933+0.546 & I  & L	&  Ind	&  Ind	\\
G189.03+0.76 & D & M 	&  0.10	&  2.1E+05&	\vline~	G304.942+0.550 & I & L	&  Ind	&  Ind	\\
G189.028+0.805 & D & M 	&  0.08	&  1.5E+05&     \vline~	G305.145+0.208 & S & - & 0.16/0.25 & 9.4E+04/6.2E+04	\\
G188.79+1.02 & S & -	&  0.40	&  3.7E+04&	\vline~	G305.137+0.069 & S & - & 0.25/0.57 & 6.5E+04/2.9E+04	\\
G192.581-0.042 & L & M 	&  0.27	&  1.1E+05&	\vline~ G305.201+0.241 & I  & M  & 0.08/0.12 & 3.7E+05/2.5E+05	\\
G192.60-0.05 & L & M 	&  0.22	&  1.8E+05&	\vline~	G305.202+0.230 &  I & M  & 0.36/0.54 & 3.5E+04/3.2E+04	\\
G192.594-0.045 & L & M 	&  0.12	&  8.5E+05&	\vline~	G305.20+0.02 & L & M  & 0.52/1.17 & 2.9E+04/2.2E+03	\\
G259.94-0.04 & S & -	&  Ind	&  Ind	  &     \vline~	G305.192-0.006 & L & M  & 0.23/0.51 & 1.3E+05/5.6E+04	\\
G269.45-1.47 & S & -	&  1.06	&  5.1E+03&	\vline~	G305.21+0.21 & I  & M  & 0.83/1.25 & 2.1E+04/1.4E+04	\\
G269.15-1.13 & S & -	&  0.27	&  2.4E+05&	\vline~	G305.197+0.007 & L & M  & 0.31/0.70 & 5.2E+04/2.3E+04	\\
G270.25+0.84 & S & -	&  0.15	&  6.3E+05&	\vline~	G305.200+0.02 & L & M  & 0.15/0.34 & 1.5E+05/6.4E+04	\\
G284.271-0.391 & S & L 	&  0.23	&  5.3E+04&	\vline~	G305.226+0.275 & I  & M  & 0.26/0.40 & 4.8E+04/3.2E+04	\\
G284.295-0.362 & D & L 	&  Ind	&  Ind   &	\vline~ G305.228+0.286 & I  & M  & 0.13/0.20 & 7.1E+04/4.7E+04	\\
G284.307-0.376 & D & L 	&  0.54	&  6.7E+03&	\vline~ G305.238+0.261 & I  & M  & 0.44/0.67 & 1.7E+04/1.0E+04	\\
G284.338-0.417 & S & L 	&  0.18	&  4.5E+04&	\vline~	G305.242+0.225 & I  & M  & 0.14/0.21 & 1.0E+05/6.9E+04	\\
G284.35-0.42 & I  & L 	&  0.28	&  9.9E+04&	\vline~	G305.248+0.245 & I  & M  & 0.30/0.46 & 4.7E+04/3.1E+04	\\
G284.345-0.404 & I  & L 	&  0.29	&  1.0E+05&	\vline~	G305.233-0.023 & S & - & 0.28/0.63 & 1.9E+04/8.6E+03	\\
G284.341-0.389 & I & L 	&  0.69	&  1.3E+04&	\vline~	G305.269-0.010 & S & - & 0.48/1.08 & 2.1E+04/9.0E+03	\\
G284.328-0.365 & I  & L 	&  0.17	&  8.1E+04&	\vline~	 G305.355+0.194 & I  & T 	&  Ind	&  Ind	\\ 
G284.384-0.441 & I  & L 	&  0.28	&  1.5E+04&	\vline~ G305.358+0.203 & I  & T  & 0.71/1.62 & 3.2E+04/1.4E+04	\\
G284.344-0.366 & I  & L 	&  	0.29	&  	1.4E+05&\vline~	G305.362+0.185 & I  & T  & 0.41/0.93 & 3.3E+04/1.5E+04	\\
G284.352-0.353 & I  & L 	&  	0.57	&  	2.8E+04&\vline~	G305.361+0.151 & D & M  & 0.24/0.54 & 3.1E+05/1.1E+05	\\
G287.37+0.65 & S & -	&  	0.37	&  	2.8E+04&	\vline~ G305.37+0.21 & I  & T  & 0.40/0.92 & 5.7E+04/2.5E+04	\\
G290.40-2.91 & S & -	&  	0.24	&  	6.6E+04&	\vline~ G305.340-0.172 & D & M  & 0.25/0.57 & 2.1E+04/9.4E+03	\\
G291.256-0.769 & S & T 	&  	0.31	&  	6.5E+04&	\vline~	G305.513+0.333 & S & -	&  	Ind	&  	Ind	\\
G291.256-0.743 & I  & T 	&  	0.39	&  	6.1E+04&\vline~	G305.533+0.360 & S & -	&  	Ind	&  	Ind	\\
G291.27-0.70 & I  & T 	&  	0.92	&  	5.7E+04&  \vline~	G305.538+0.340 & S & -	&  	Ind	&  	Ind	\\
G291.288-0.706 & I  & T 	&  	0.30	&  	3.0E+04&\vline~ G305.519-0.040 & S & L	&  	Ind	&  	Ind	\\
G291.302-0.693 & I  & T 	&  	0.31	&  	5.5E+04&\vline~	G305.520-0.020 & S & L & 0.19/0.33 & 2.7E+04/1.5E+04	\\
G291.309-0.681 & I  & T 	&  	0.46	&  	5.0E+04&\vline~	G305.549+0.002 & I  & L & 0.36/0.63 & 2.1E+04/1.2E+04	\\
G290.37+1.66 & S & -	&  	0.24	&  	6.7E+04&	\vline~	G305.552+0.013 & I  & L & 0.30/0.53 & 3.2E+04/1.8E+04	\\ 
G291.587-0.499 & D & L 	&  	1.34	&  	1.0E+04&	\vline~	G305.552+0.012 & I  & L & 0.53/0.93 & 1.3E+04/7.2E+03	\\
G291.576-0.468 & D & L 	&  	0.73	&  	1.9E+04	&       \vline~ G305.561+0.012 & I  & L & 0.67/1.16 & 1.0E+04/5.9E+03	\\	
G291.572-0.450 & D & L 	&  	0.37	&  	2.8E+04&	\vline~	G305.581+0.033 & S & L & 0.68/1.19 & 1.6E+03/9.3E+02	\\
G291.608-0.532 & D & L 	&  	1.09	&  	1.3E+04&	\vline~	G305.605+0.010 & S & L & 0.31/0.54 & 9.8E+03/5.6E+03	\\
G291.597-0.496 & D & L 	&  	1.29	&  	5.6E+03&	\vline~	G305.776-0.251 & S & - & 0.30/0.47 & 1.8E+04/1.1E+04	\\
G291.58-0.53 & S & L 	&  	0.68	&  	9.6E+04&	\vline~	G305.81-0.25 & S & - & 0.50/0.76 & 5.6E+04/3.6E+04	\\
G291.630-0.545 & D & L 	&  	1.48	&  	1.5E+04&	\vline~	G305.833-0.196 & S & - & 0.32/0.49 & 1.8E+04/1.2E+04	\\
G291.614-0.443 & S & L 	&  	0.19	&  	2.2E+05&	\vline~	G306.33-0.3 & S & L & 0.16/0.63 & 3.7E+04/9.0E+03	\\
G293.824-0.762 & A & M 	&  	0.60	&  	1.9E+04&	\vline~	G306.319-0.343 & S & L & 0.06/0.24 & 2.2E+05/5.5E+04	\\
G293.82-0.74 & A & M 	&  	0.70	&  	4.4E+04&	\vline~	G306.343-0.302 & S & L & 0.07/0.28 & 6.0E+05/1.5E+05	\\
G293.892-0.782 & S & -	&  	0.34	&  	4.3E+04&	\vline~	G306.345-0.345 & S & L & 0.07/0.28 & 4.0E+05/9.8E+04	\\
G293.95-0.8 & S & -	&  	0.81	&  	1.9E+04&	\vline~	G309.917+0.494 & A & L	&  	0.20	&  	5.5E+04	\\
G293.942-0.876 & S & -	&  	0.77	&  	2.2E+04&	\vline~	G309.92+0.4 & A & L	&  	0.85	&  	1.9E+04	\\
G293.989-0.936 & S & -	&  	0.90	&  	1.5E+04&	\vline~	G318.913-0.162 & S & -	&  	0.17	&  	2.0E+05	\\
G294.52-1.60 & S & - & 0.07/0.44 & 4.3E+05/7.0E+04&	\vline~	G318.92-0.68 & S & -	&  	0.18	&  	2.6E+05	\\
G294.945-1.737 & S & L & 0.09/0.41 & 4.5E+04/1.0E+04&	\vline~	G323.74-0.3 & S & - & 0.30/0.90 & 1.9E+05/6.2E+04	\\
G294.97-1.7 & D & L & 0.15/0.69 & 1.6E+05/3.6E+04&	\vline~	G330.95-0.18 & S & - & 1.15/2.09 & 4.4E+04/2.4E+04	\\
G294.989-1.720 & D & L & 0.10/0.43 & 2.3E+05/5.1E+04&	\vline~	G331.28-0.19 & S & -	&  	Ind	&  	Ind	\\
G298.26+0.7 & S & -	&  	0.35	&  	4.1E+04&	\vline~	G332.640-0.586 & D & L & 0.13/0.45 & 2.8E+05/7.7E+04	\\
G299.02+0.1 & D & M 	&  	0.89	&  	8.7E+03&	\vline~	G332.648-0.606 & D & L & 1.07/3.84 & 1.0E+04/2.8E+03	\\
G299.024+0.130 & D & M 	&  	0.59	&  	9.4E+03&	\vline~	G332.701-0.587 & S & L & 0.24/0.87 & 4.4E+04/1.2E+04	\\
G300.455-0.190 & S & L	&  	0.79	&  	4.6E+03&	\vline~	G332.646-0.647 & D & L & 0.24/0.85 & 1.2E+05/3.4E+04	\\
G300.51-0.1 & S & L	&  	0.66	&  	4.3E+04&	\vline~	G332.695-0.609 & D & L & 0.35/1.24 & 7.5E+04/2.1E+04	\\
G301.14-0.2 & S & L	&  	0.75	&  	5.4E+04&	\vline~	G332.73-0.62 & S & L & 0.09/0.33 & 6.0E+05/1.7E+05	\\
G302.03-0.06 & S & L	&  	0.38	&  	6.6E+04&	\vline~	G332.777-0.584 & S & L & 0.12/0.43 & 2.8E+05/7.9E+04\\
G304.890+0.636 & U & -	&  	Ind	&  	Ind&	\vline~	G332.627-0.511 & S & L & 0.10/0.36 & 3.9E+05/1.1E+05\\
\hline
\end{tabular}
\end{center}
\end{table*}

\clearpage
\begin{table*}
\begin{center}
\contcaption{}
\begin{tabular}{@{}lcccclccccc@{}}
       \hline
       \hline

Source Name~$^a$& M~$^b$ & C~$^c$  & Radius~$^d$  & $n_{H_2}$~$^d$ &Source Name~$^a$ & M~$^b$ & C~$^c$ & Radius~$^d$ & $n_{H_2}$~$^d$  \\
       & & & parsecs~ &cm$^{-3}$~ & & &   & parsecs~ &cm$^{-3}$~\\
(1)&(2)&(3)&(4)&(5) &(1)&(2)&(3)&(4)&(5)\\
\hline										
G332.827-0.552 & U & L &  	 Ind &  	Ind & \vline~	G10.32-0.15 & D & M  & 0.47/3.93 & 1.2E+04/1.5E+03 & \\
G332.794-0.598 & S & L & 0.20/0.72 & 8.8E+04/2.5E+04 & \vline~	G10.359-0.149 & D & M  & 0.21/1.72 & 3.9E+04/4.7E+03	\\
G0.204+0.051 & S & M  & 0.74/0.76 & 7.7E+03/7.5E+03 & \vline~	G10.146-0.314 & I & T  & 	0.63 &  	5.2E+03	\\
G0.49+0.19 & A & L & 0.19/1.04 & 9.3E+04/1.7E+04 & \vline~	G10.191-0.307 & S & T  &  	 0.16 &	 	1.0E+05	\\
G0.266-0.034 & I & M  & 0.74/0.76 & 1.2E+04/1.2E+04 & \vline~	G10.148-0.331 & I & T  &  	 0.53 &  	 	Ind	\\
G0.21-0.00 & S & M  & 0.85/0.87 & 1.0E+04/1.0E+04 & \vline~	G10.214-0.305 & S & T  &  	0.24 & 	 	3.3E+04	\\
G0.497+0.170 & A & L & 0.16/0.90 & 1.0E+05/1.9E+04 & \vline~	G10.191-0.308 & S & T  &  	 0.41 & 	 	1.8E+04	\\
G0.24+0.01 & I & M  & 1.51/1.54 & 1.1E+04/1.0E+04 & \vline~	G10.15-0.34 & I & T  &  	1.01 & 	 	1.6E+04	\\
G0.527+0.181 & S & L & 0.24/1.32 & 9.0E+04/1.6E+04 & \vline~	G10.213-0.326 & I & T  &  	 	0.91 & 	 	1.4E+04	\\
G0.271+0.022 & I & M  & 0.63/0.64 & 1.2E+04/1.1E+04 & \vline~	G10.188-0.344 & S & T  &  	0.85 & 	 	4.1E+03	\\
G0.26+0.01 & I & M  & 2.43/2.48 & 2.4E+03/2.3E+03 & \vline~	G10.133-0.378 & I & T  &  	 0.44 &	 	1.2E+04	\\
G0.83+0.18 & S & - & 0.31/0.87 & 5.8E+04/2.1E+04 & \vline~	G10.164-0.360 & I & T  &  	 	1.58 & 	 	5.1E+03	\\
G0.331-0.164 & L & L & 0.55/0.62 & 2.2E+04/1.9E+04 & \vline~	G10.237-0.328 & S & T  &  	 	0.22 & 	 	3.1E+04	\\
G0.310-0.170 & L & L & 0.24/0.27 & 5.8E+04/5.1E+04 & \vline~	G10.206-0.350 & I & T  &  	 0.81 &  	5.1E+03	\\
G0.32-0.20 & S & L & 2.44/2.75 & 1.9E+03/1.7E+03 & \vline~	G10.184-0.370 & S & T  &  	 0.25 &  	3.7E+04	\\
G0.325-0.242 & S & L & 0.22/0.25 & 1.4E+05/1.2E+05 & \vline~	G10.198-0.372 & S & T  &  	 	0.44 &   	1.0E+04	\\
G1.124-0.065 & D & L &  	0.58 &  	1.3E+04 & \vline~G10.138-0.419 & S & T  &  	 0.29 &  	2.1E+04	\\
G1.134-0.073 & D & L &  	 0.37 &   	2.0E+04 & \vline~ G10.149-0.407 & S & T  & 	0.42 &   	1.2E+04	\\	
G1.105-0.098 & D & L & 0.40 & 1.5E+05 & \vline~	G10.165-0.403 & S & T  &	0.41 & 	 	1.3E+04	\\
G1.13-0.11 & D & L &  	 2.36 &  		3.1E+03 & \vline~ G10.194-0.387 & S & T  &  	 0.44 &  	1.5E+04	\\
G1.14-0.12 & S & L & 0.98 & 1.3E+03 & \vline~	G10.186-0.404 & S & T  &  	0.18 &	 	8.0E+04	\\
G0.55-0.85 & D & L &  	 0.46 &  	 	4.8E+04 & \vline~ G10.575-0.347 & S & L &  	 	Ind &  	 	Ind	\\	
G0.549-0.868 & D & L &  	0.34 & 	 	2.9E+04 & \vline~ G10.63-0.33B & D & L &  	0.81 & 	 	7.0E+03	\\
G0.627-0.848 & S & L &  	0.13 & 	 	4.9E+04 & \vline~ G10.62-0.33 & D & L &  	 	0.91 & 	 	1.2E+04	\\
G0.600-0.871 & S & L &  	0.19 & 	 	3.5E+04 & \vline~ G9.88-0.75 & S & L &  	 	1.11 & 	 	4.5E+03	\\
G2.54+0.20 & S & - & 0.41/2.18 & 1.6E+04/3.0E+03 & \vline~	  G9.924-0.749 & S & L &  	 0.4 & 	 	1.1E+04	\\
G5.48-0.24 & S & L &  	 	1.19 & 	 	7.7E+03 & \vline~ G10.62-0.38 & D & L &  	 	1.61 &  	1.7E+04	\\	
G5.504-0.246 & S & L &  	0.92 & 	 	1.0E+04 & \vline~ G10.620-0.441 & D & L &  	0.28 & 	 	4.3E+04	\\
G5.89-0.39 & S & L &  	 	0.57 & 	 	6.2E+04 & \vline~ G11.075-0.384 & S & L &  	 0.47 &  	1.6E+04	\\
G5.90-0.42 & A & L & 0.41/2.15 & 6.3E+04/1.2E+04 & \vline~	  G11.11-0.34 & D & L &  	 	1.14 &   	4.4E+03	\\
G5.90-0.44 & A & L & 0.45/2.37 & 2.9E+04/5.5E+03 & \vline~        G11.117-0.413 & D & L &  	 	0.39 &   	1.8E+04	\\
G6.53-0.10 & S & L &  	 	1.38 &  	1.6E+04 & \vline~ G12.88+0.48 & S & L & 0.85/2.65 & 1.3E+04/4.2E+03	\\
G6.60-0.08 & L & L & 0.01/0.60 & 9.3E+06/1.7E+05 & \vline~	  G11.948-0.003 & S & - & 0.67/1.01 & 1.3E+04/8.5E+03	\\	
G6.620-0.10 & L & L & 0.01/0.80& 7.8E+05/1.4E+04 & \vline~        G12.914+0.493 & S & L & 0.28/0.87 & 3.9E+04/1.3E+04	\\
G8.127+0.255 & L & L &  	0.26 &   	4.3E+04 & \vline~	G12.02-0.03 & S & - & 0.45/0.68 & 1.9E+04/1.3E+04	\\
G8.138+0.246 & L & L &  	0.54 &   	9.7E+03 & \vline~ G11.902-0.100 & S & L & 0.18/0.47 & 4.7E+04/8.1E+04	\\
G8.13+0.22 & L & L &  	 	0.57 &   	3.7E+04 & \vline~ G11.903-0.140 & S & L & 0.86/2.27 & 5.3E+03/2.0E+03	\\
G5.948-1.125 & S & L &  	0.18 &   	1.4E+04 & \vline~ G11.861-0.183 & S & - & 0.16/0.43 & 4.2E+04/1.6E+04	\\
G5.975-1.146 & L & L &  	0.13 &   	5.0E+04 & \vline~ G11.93-0.14 & D & L & 0.34/0.90 & 2.4E+04/9.2E+03	\\
G5.971-1.158 & L & L &  	0.22 &   	2.0E+04 & \vline~ G11.942-0.157 & D & L & 0.37/0.98 & 1.9E+04/7.2E+03	\\
G5.97-1.17 & L & L &  	 	0.40 &   	3.6E+04 & \vline~ G12.200-0.003 & S & - &   	0.95 & 	 	1.2E+04	\\
G10.10+0.73 & S & - & 0.03/1.20 & 2.9E+05/7.1E+03 & \vline~ 	G11.956-0.177 & S & L & 0.14/0.38 & 5.6E+04/2.1E+04	\\
G9.62+0.19 & S & - &  	 	0.46 &  	2.6E+04 & \vline~ G12.112-0.125 & S & - &  	1.71 & 	1.6E+03	\\
G8.68-0.36 & D & L &  	 	0.92 & 	 	2.9E+04 & \vline~	G12.20-0.09 & S & L &  	 	2.48 & 	4.0E+03	\\
G8.686-0.366 & D & L &  	 0.63 &   	2.2E+04 & \vline~ 	G11.942-0.256 & S & - & 0.33/0.73 & 3.4E+04/1.6E+04	\\
G8.644-0.395 & S & L &  	 0.30 &   	1.6E+04 & \vline~ 	G12.18-0.12 & D & L &  	 	1.44 & 	 	3.0E+03	\\
G8.713-0.364 & S & L &  	 0.76 &   	3.2E+03 & \vline~ 	G12.216-0.119 & D & L &  	 0.84 &	 	2.8E+04	\\
G8.735-0.362 & S & L &  	 0.75 &   	5.8E+03 & \vline~	G11.99-0.27 & S & - & 0.36/0.79 & 1.4E+04/6.4E+03	\\
G8.724-0.401a & I & L &  	 0.23 &   	1.1E+04 & \vline~	G12.43-0.05 & S & - &  	 	1.7 &  	 	3.5E+03	\\
G8.724-0.401b & I & L &  	 0.40 &   	1.4E+04 & \vline~	G12.68-0.18 & S & - & 1.16/2.91 & 5.7E+03/2.3E+03	\\
G8.718-0.410 & I & L &  	 0.09 &   	2.9E+05 & \vline~ 	G11.94-0.62B & D & M  &  	 0.57 &  	2.3E+04	\\
G9.966-0.020 & L & L & 0.12 & 1.8E+05 & \vline~ G11.93-0.61 & D & M  &  	 0.59 &	 	2.7E+04	\\
G9.99-0.03 & D & L & 0.54 & 1.4E+04 & \vline~	G12.722-0.218 & S & - & 0.57/1.43 & 1.7E+04/6.6E+03	\\
G10.001-0.033 & D & L & 0.26 & 3.5E+04 & \vline~ 	G12.855-0.226 & S & L &  	 Ind & 	 	Ind	\\
G10.47+0.02 & S & - & 1.77/3.18 & 1.1E+04/6.3E+03 & \vline~  	G12.885-0.222 & L & L & 0.17/0.56 & 8.2E+04/2.5E+04	\\
G10.44-0.01 & S & - & 0.72/1.32 & 1.1E+04/5.8E+03 & \vline~ 	G12.892-0.226 & L & L & 0.20/0.65 & 3.5E+04/1.1E+04	\\
G10.287-0.110 & L & M  & 0.24/1.97 & 5.0E+04/6.0E+03 & \vline~  	G12.90-0.25B & L & L & 0.58/1.88 & 8.8E+03/2.7E+03	\\
G10.284-0.126 & L & M  &  	0.21 &   	7.3E+04 & \vline~ 	G12.859-0.272 & S & L & 0.47/1.53 & 2.4E+04/7.4E+03	\\
G10.288-0.127 & L & M  & 0.18/1.46 & 4.1E+04/5.0E+03 & \vline~	G13.87+0.28 & S & - &  	 	0.87 & 	 	1.4E+04	\\
G10.29-0.14 & L & M  &  	0.34 &   	5.3E+04 & \vline~ 	G12.90-0.26 & L & L & 0.85/2.74 & 1.6E+04/4.8E+03	\\
G10.343-0.142 & D & M  & 0.12/0.96 & 2.7E+05/3.2E+04 & \vline~  	G12.878-0.226 & S & L & 0.22/0.73 & 2.6E+04/8.1E+03	\\
\hline
\end{tabular}
\end{center}
\end{table*}

\clearpage
\begin{table*}
\begin{center}
\contcaption{ }
\begin{tabular}{@{}lcccclccccc@{}}
       \hline
       \hline

Source Name~$^a$& M~$^b$ & C~$^c$  & Radius~$^d$  & $n_{H_2}$~$^d$ &Source Name~$^a$ & M~$^b$ & C~$^c$ & Radius~$^d$ & $n_{H_2}$~$^d$  \\
       & & & parsecs~ &cm$^{-3}$~ & & &   & parsecs~ &cm$^{-3}$~\\
(1)&(2)&(3)&(4)&(5) &(1)&(2)&(3)&(4)&(5)\\
\hline		
G12.897-0.281 & S & L & 0.11/0.37 & 1.9E+05/6.0E+04 & \vline~ G23.319-0.298 & S & L & 0.40/1.04 & 1.6E+04/5.9E+03 \\
G12.914-0.280 & S & L & 0.13/0.42 & 1.2E+05/3.6E+04 & \vline~	G23.754+0.095 & S & L & 0.60/0.80 & 1.6E+04/1.2E+04	\\
G12.938-0.272 & S & L & 0.10/0.33 & 2.2E+05/6.8E+04 & \vline~	G24.792+0.099 & D & L & 0.71/0.95 & 3.6E+04/2.7E+04	\\
G11.49-1.48 & S & - & 0.26/3.67 & 2.5E+04/1.7E+03 & \vline~	G24.78+0.08 & D & L & 1.16/1.55 & 2.7E+04/2.0E+04	\\
G14.60+0.01 & S & - & 0.53/2.60 & 8.5E+03/1.7E+03 & \vline~	G24.84+0.08 & S & L & 0.45/0.57 & 5.2E+04/4.1E+04	\\
G10.84-2.59 & S & - &  	 0.34 &  	3.3E+04 & \vline~	G24.850+0.082 & D & L & 0.25/0.31 & 1.5E+05/1.2E+05	\\
G15.022-0.618 & S & T  & 0.17/1.00 & 1.4E+05/2.4E+04 & \vline~	G24.919+0.088 & S & L & 0.92/1.16 & 1.5E+04/1.2E+04	\\
G14.987-0.670 & L & T  & 0.14/0.82 & 2.2E+05/3.8E+04 & \vline~ 	G25.70+0.04 & S & - &  	 	1.32 &  	 	8.4E+03	\\
G15.027-0.651 & S & T  & 0.30/1.73 & 8.8E+04/1.5E+04 & \vline~	G25.802-0.159 & S & L & 0.45/0.80 & 2.6E+04/1.5E+04	\\
G15.054-0.641 & S & T  & 0.10/0.60 & 2.5E+05/4.2E+04 & \vline~	G25.83-0.18 & S & L & 0.80/1.43 & 2.3E+04/1.3E+04	\\
G14.983-0.687 & L & T  & 0.33/1.93 & 5.2E+04/9.0E+03 & \vline~	G28.14-0.00 & S & - & 0.60/0.85 & 1.0E+04/7.2E+03	\\
G15.012-0.671 & I & T  & 0.41/2.41 & 8.5E+04/1.5E+04 & \vline~	G28.231+0.367 & S & L & 0.45/0.67 & 1.3E+04/8.5E+03	\\
G15.03-0.67 & I & T  & 0.57/3.34 & 6.7E+04/1.1E+04 & \vline~	G28.287+0.010 & S & L & 0.65/0.96 & 4.9E+03/3.3E+03	\\
G14.99-0.70 & I & T  & 0.25/1.45 & 9.7E+04/1.7E+04 & \vline~	G28.20-0.04 & A & L & 1.02/1.52 & 1.7E+04/1.2E+04	\\
G15.009-0.688 & I & T  & 0.15/0.86 & 4.1E+05/7.1E+04 & \vline~	G28.198-0.063 & A & L & 0.31/0.46 & 2.2E+04/1.5E+04	\\
G15.005-0.695 & I & T  & 0.14/0.81 & 1.7E+05/2.9E+04 & \vline~	G29.193-0.073 & S & L & 0.27/0.40 & 4.3E+04/2.9E+04	\\
G15.079-0.663 & S & T  & 0.25/1.47 & 3.9E+04/6.6E+03 & \vline~ 	G28.28-0.35 & S & L &  	 	0.44 &  	 	3.5E+04	\\
G15.016-0.702 & I & T  & 0.13/0.76 & 2.3E+05/4.0E+04 & \vline~	G28.31-0.38 & S & L & 0.47/0.93 & 1.9E+04/9.5E+03	\\
G15.029-0.703 & I & T  & 0.61/3.54 & 1.2E+04/2.0E+03 & \vline~	G29.888+0.001 & A & M  & 0.12/0.18 & 6.2E+05/4.3E+05	\\
G15.089-0.673 & I & T  & 0.26/1.50 & 9.4E+04/1.6E+04 & \vline~	G29.889-0.006 & A & M  & 0.36/0.52 & 5.8E+04/4.0E+04	\\
G15.028-0.710 & I & T  &  	 	Ind &  	 	Ind & \vline~	G29.918-0.014 & S & M  & 0.16/0.23 & 1.8E+05/1.2E+05	\\
G15.098-0.681 & S & T  & 0.18/1.05 & 7.4E+04/1.3E+04 & \vline~	G29.86-0.04 & I & M  & 0.49/0.64 & 2.5E+04/1.9E+04	\\
G16.580-0.079 & S & L & 0.49/1.29 & 6.1E+03/2.3E+03 & \vline~	G29.861-0.053 & I & M  & 0.38/0.55 & 3.3E+04/2.3E+04	\\
G16.58-0.05 & S & L & 0.63/1.66 & 1.7E+04/6.6E+03 & \vline~	G29.853-0.062 & I & M  & 0.41/0.60 & 3.1E+04/2.1E+04	\\
G18.087-0.292 & L & M  &  	0.12 &  	1.5E+04 & \vline~	G29.96-0.02B & S & M  & 1.16/1.69 & 1.5E+04/1.1E+04	\\
G18.095-0.299 & L & M  &  	0.21 &   	2.4E+04 & \vline~	G29.912-0.045 & I & M  & 1.06/1.54 & 7.4E+03/5.1E+03	\\
G18.105-0.304 & L & M  &  	0.20 &   	2.9E+04 & \vline~	G29.930-0.040 & I & M  & 0.15/0.21 & 3.6E+05/2.5E+05	\\
G18.15-0.28 & D & M  &  	0.38 &   	2.2E+04 & \vline~	G29.9687-0.033 & S & M  & 0.40/0.58 & 2.1E+04/1.5E+04	\\
G18.142-0.297 & D & M  &  	0.35 &   	1.2E+03 & \vline~	G29.937-0.054 & I & M  & 0.34/0.49 & 7.9E+04/5.4E+04	\\
G18.165-0.293 & D & M  &  	0.14 &   	5.0E+04 & \vline~	G29.945-0.059 & I & M  & 0.90/1.30 & 8.2E+03/5.7E+03	\\
G18.112-0.321 & S & M  &  	0.12 &   	5.7E+04 & \vline~	G29.978-0.050 & S & M  & 0.80/1.16 & 9.9E+03/6.8E+03	\\
G18.177-0.296 & D & M  &  	0.18 &   	7.5E+04 & \vline~	G30.533-0.023 & S & - & 0.16/0.63 & 8.7E+04/2.3E+04	\\
G18.30-0.39 & S & - &  	 	0.53 &   	2.4E+04 & \vline~	G30.855+0.149 & S & L & 0.64/0.72 & 1.8E+04/1.6E+04	\\
G19.61-0.1 & S & - & 0.42/1.25 & 2.1E+04/6.9E+03 & \vline~	G30.89+0.16 & S & L & 0.47/0.52 & 2.6E+04/2.3E+04	\\
G19.607-0.234 & S & - & 0.49/1.74 & 1.0E+05/2.9E+04 & \vline~	G30.894+0.140 & S & L & 0.53/0.59 & 2.6E+04/2.3E+04	\\
G19.70-0.27A & S & - & 0.30/1.07 & 3.7E+04/1.0E+04 & \vline~	G30.908+0.137 & S & L & 0.08/0.09 & 1.1E+06/9.8E+05	\\
G16.871-2.154 & D & T  &  	 	Ind &  	 	Ind & \vline~	G30.869+0.116 & S & L & 0.73/0.81 & 2.4E+04/2.2E+04	\\
G16.86-2.15 & D & T  & 0.45/3.85 & 3.9E+04/4.6E+03 & \vline~	G30.59-0.04 & S & - & 0.37/1.45 & 3.9E+04/1.0E+04	\\
G16.883-2.188 & S & L & 0.09/0.74 & 1.5E+05/1.8E+04 & \vline~	G30.924+0.092 & S & L & 0.46/0.52 & 1.8E+04/1.6E+04	\\
G21.87+0.01 & S & - & 0.14/1.02 & 9.9E+04/1.3E+04 & \vline~	G30.760-0.027 & I & M  & 0.48/0.73 & 3.6E+04/2.4E+04	\\
G22.36+0.07B & S & - & 0.53/1.12 & 3.1E+04/1.5E+04 & \vline~	G31.41+0.30 & S & - &  	 	1.28 &  	 	3.3E+04	\\
G22.35+0.06 & S & - & 0.48/1.03 & 3.5E+04/1.6E+04 & \vline~	G30.682-0.072 & I & M  & 0.60/0.91 & 2.8E+04/1.8E+04	\\
G23.71+0.17 & S & L &  	 	0.61 &  	4.2E+04 & \vline~	G30.78-0.02 & S & M  &   	1.21 &  	8.2E+03	\\
G23.689+0.159 & S & L &  	 0.23 &  	8.7E+04 & \vline~	G30.705-0.065 & I & M  & 0.98/1.48 & 2.2E+04/1.5E+04	\\
G24.450+0.489 & D & M  & 0.14/0.25 & 1.4E+05/8.3E+04 & \vline~	G30.76-0.05 & I & M  & 0.91/1.37 & 7.5E+03/4.9E+03	\\
G24.47+0.49 & D & M  & 0.73/1.25 & 2.3E+04/1.3E+04 & \vline~	G30.769-0.048 & I & M  & 0.43/0.65 & 4.3E+04/2.8E+04	\\
G25.65+1.04 & S & - &  	 	0.49 &   	4.1E+04 & \vline~	G30.740-0.060 & I & M  & 0.73/1.11 & 2.2E+04/1.4E+04	\\
G23.949+0.163 & I & L &  	 0.44 &   	1.6E+04 & \vline~	G30.749-0.048 & S & M  & 0.42/0.64 & 7.3E+04/4.8E+04	\\
G23.96+0.15 & I & L &  	 	0.67 &  	1.3E+04 & \vline~	G31.094+0.111 & S & L & 0.03/0.30 & 1.2E+06/1.1E+05	\\
G23.281-0.201 & S & L & 0.24/0.63 & 3.4E+04/1.3E+04 & \vline~	G30.716-0.082 & I & M  & 0.58/0.88 & 5.3E+04/3.5E+04	\\
G23.268-0.210 & S & L & 0.38/0.99 & 1.4E+04/5.3E+03 & \vline~	G30.729-0.078 & I & M  & 0.49/0.74 & 2.9E+04/1.9E+04	\\
G24.012+0.173 & S & L & 	0.16 &   	4.9E+04 & \vline~	G31.06+0.09 & S & L & 0.07/0.80 & 9.2E+04/8.3E+03	\\
G23.976+0.150 & L & L &	 	0.19 &   	4.2E+04 & \vline~	G30.81-0.05 & S & M  & 1.22/1.44 & 3.0E+04/2.6E+04	\\
G23.960+0.137 & S & L &	 	0.05 &   	1.5E+06 & \vline~	G31.065+0.051 & S & L & 0.07/0.75 & 1.3E+05/1.2E+04	\\
G23.987+0.148 & L & L &	 	0.23 &   	2.7E+04 & \vline~	G31.119+0.029 & S & L & 0.06/0.65 & 1.2E+05/1.1E+04	\\
G23.25-0.24 & S & L & 0.21/0.55 & 5.2E+04/2.0E+04 & \vline~ 	G31.28+0.06 & D & M  &  	 	1.35 &  	5.8E+03	\\
G24.016+0.150 & S & L &	 	0.09 &   	3.5E+05 & \vline~ 	G31.256+0.061 & D & M  &  	 0.46 &  	2.3E+04	\\
G23.268-0.257 & S & L & 0.55/1.45 & 3.0E+04/1.1E+04 & \vline~	G31.40-0.26 & D & T  &  	 	0.92 &  	1.5E+04	\\
G23.43-0.18 & S & L & 0.80/1.28 & 2.0E+04/1.3E+04 & \vline~	G31.388-0.266 & D & T  &  	 	0.27 &  	2.7E+04	\\
G23.409-0.228 & A & L & 0.31/0.50 & 7.9E+04/4.9E+04 & \vline~	G32.80+0.19 & S & - &  	 	1.78 &  	 	2.0E+04	\\
G23.420-0.235 & A & L & 0.37/0.59 & 5.0E+04/3.1E+04 & \vline~  	G33.13-0.09 & S & - &  	 	0.82 &  	 	1.3E+04	\\
\hline
\end{tabular}
\end{center}
\end{table*}
\clearpage

\begin{table*}
\begin{center}
\contcaption{ }
\begin{tabular}{@{}lcccclccccc@{}}
       \hline
       \hline

Source Name~$^a$& M~$^b$ & C~$^c$  & Radius~$^d$  & $n_{H_2}$~$^d$ &Source Name~$^a$ & M~$^b$ & C~$^c$ & Radius~$^d$ & $n_{H_2}$~$^d$  \\
       & & & parsecs~ &cm$^{-3}$~ & & &   & parsecs~ &cm$^{-3}$~\\						
(1)&(2)&(3)&(4)&(5) &(1)&(2)&(3)&(4)&(5)\\
\hline		
G33.92+0.11 & S & - &  	 	1.35 &  	 1.2E+04 & \vline~ G49.456-0.354 &I &T & 	 0.59& 	 	3.9E+04\\
G34.256+0.155 &A &M & 	 	1.10& 	 	3.4E+04&\vline~	G49.482-0.355 &I &T & 	 	0.42& 	 	3.6E+04	\\
G34.24+0.13 &A &M & 	 	0.26& 	 	3.5E+04&\vline~	G49.472-0.366 &I &T & 	 	0.46& 	 	8.9E+04	\\
G35.02+0.35 &S &-&0.40/1.45&2.6E+04/7.2E+03&\vline~	G49.533-0.338 &I &T & 	 	0.09& 	 	6.4E+05	\\
G35.57+0.07 &A &M & 	 	1.15& 	 	7.4E+03&\vline~	G49.513-0.356 &I &T & 	 	0.21& 	 	1.7E+05	\\
G35.586+0.061 &A &M & 	 	0.38& 	 	6.3E+04&\vline~	G49.528-0.348 &I &T & 	 	0.28& 	 	1.1E+05	\\
G35.575+0.048 &S &L& 	 	0.39& 	 	3.7E+04&\vline~	G49.49-0.37 &I &T & 	 	0.85& 	 	1.1E+05	\\
G35.575+0.010 &S &L& 	 	0.68& 	 	1.2E+04&\vline~	G49.488-0.385 &I &T & 	 	1.17& 	 	9.9E+04	\\
G35.574+0.007 &S &L& 	 	0.51& 	 	2.5E+04&\vline~	G49.481-0.401 &I &T & 	 	0.59& 	 	6.6E+04	\\
G35.58-0.03 &L& L& 	 	1.58& 	 	6.6E+03&\vline~	G49.465-0.417 &I &T & 	 	0.09& 	 	6.2E+05	\\
G35.585-0.026 &L &L& 	 	0.44& 	 	1.8E+04&\vline~	G49.462-0.423 &I &T & 	 	0.31& 	 	4.2E+04	\\
G35.564-0.037 &L &L& 	 	0.21& 	 	1.3E+05&\vline~	G49.474-0.420 &I &T & 	 	0.34& 	 	4.6E+04	\\
G35.601-0.018 &S &L& 	 	0.72& 	 	4.6E+03&\vline~	G49.48-0.419 &I &T & 	 	0.10& 	 	4.6E+05	\\
G35.05-0.52 &S &-& 	 	1.00& 	 	6.2E+03&\vline~	G49.536-0.395 &I& T & 	 	0.15& 	 	2.3E+05	\\
G37.475-0.106 &S &-& 	 	0.52& 	 9.9E+03&\vline~	G49.494-0.420 &I &T & 	 	0.19& 	 	1.8E+05	\\
G37.55-0.11 &S &-& 	 	1.14& 	 	7.9E+03&\vline~	G49.508-0.409 &I &T & 	 	0.25& 	 	6.0E+04	\\
G49.459-0.317 &I &T & 	 	0.09& 	 	6.5E+05&\vline~	G49.505-0.421 &I &T & 	 	0.48& 	 	2.4E+04	\\
G49.458-0.330 &I &T & 	 	0.51& 	 	1.7E+04&\vline~	G59.794+0.076 &D &T &0.19/0.44&5.9E+04/2.5E+04	\\
G49.465-0.324 &I &T & 	 	0.30& 	 	5.5E+04&\vline~	G59.78+0.06 &D &T &0.41/0.95&3.3E+04/1.4E+04	\\
\hline
\end{tabular}
\end{center}
\begin{flushleft}

$^a$ Sources are listed in right ascension order as per Table \ref{main}.\\
$^b$ Column 2 lists the morphological classification of the sources as per the designations discussed in $\S\ref{sampleimages}$. `S' represents singular cores, `D' for double, `A' for adjacent cores, `L' and `I' for cores spatially aligned in a linear or irregular fashion respectively. `U'  used is assigned to those sources for which a morphology is unknown. See $\S\ref{sampleimages}$ for further explanation.\\
$^c$ Column 3, lists the clustering type of the source as discussed in $\S\ref{sampleimages}$. `L' denotes a low clustering association, while `M' and `T' indicate medium and tight clustering associations respectively.\\
$^d$ For those sources for which a distance ambiguity exists, two values are reported for the radius and the density ($n_{H_{2}}$). The near distance precedes the far.

\end{flushleft}
\end{table*}

\clearpage
\section[]{Analysis and Discussion}

   The aim of this millimetre continuum survey, was to study known regions of massive star formation (MSF), previously identified by the presence of methanol masers and/or radio continuum emission, and to test a proposed evolutionary sequence for massive stars. Interestingly, this survey revealed many sources which are devoid of conventional tracers of MSF (e.g. masers and \uchii~ regions), detected solely from their millimetre continuum emission. Previous evidence of millimetre only cores has been reported by \citet{hunter98, garay04, faundez04}.  For simplicity, we have dubbed the millimetre cores detected in this survey as `mm-only' cores.  The millimetre continuum sources detected in this survey comprise four distinct classes:

\begin{enumerate}
\item[*] Class MM: sources with millimetre continuum emission, but without methanol maser sites or \uchii~ regions (`mm-only' sources). 
\item[*] Class M:  millimetre sources with methanol masers and devoid of radio continuum emission.
\item[*] Class MR: millimetre sources with methanol masers and radio continuum emission.
\item[*] Class R:  millimetre sources with radio continuum emission but without methanol masers.
\end{enumerate}

   The following analysis has been performed assuming that all sources with a distance ambiguity are at the near distance. See $\S\ref{distances}$ for further explanation. We do not expect this assumption to significantly alter our results, and as confirmation, we present, in $\S\ref{histograms}$, the range and mean values for each parameter for the entire source sample of 392 sources, assuming the near distance to 197 sources; as well as for the 195 sources with well established distances (hereafter referred to as `no-ambig sources').

\subsection{Correlation of Mass versus Distance from Tracer}

\begin{figure}[h]
         \includegraphics[width=7.5cm, height=7.5cm]{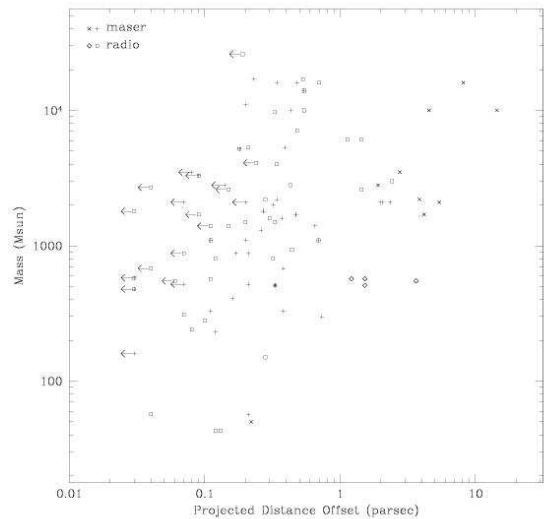}
	 \caption{Relation between the distance offset of a tracer from the peak emission of the millimetre source to the mass of the source. Maser sources are depicted with a `plus' symbol, radio continuum sources are depicted with a `box' symbol. Maser and radio sources devoid of millimetre continuum emission have been oriented 45 degrees, such that the masers are indicated by `cross' symbols, whilst the \uchii~ regions are indicated by `diamonds'. Sources with a zero offset are displayed on the graph as upper limits, determined assuming a $3''$ pointing accuracy for the SEST. \label{offsets}} 
\end{figure}

  We have examined those tracers (methanol maser and \uchii~ regions) coincident with millimetre emission which have well known distances (i.e. no distance ambiguity), to determine if there is any correlation between the mass of the sources and the projected distance offsets of the associated tracer/s from the peak millimetre emission. This relationship is shown in Fig. \ref{offsets}. Only those tracers  whose positions have been determined from interferometric observations (accurate within 1 arcsec) have been examined ($\sim100$).

   The methanol maser is depicted as a `plus' symbol, whilst the radio continuum source is depicted as a `box' symbol. Those tracers devoid of millimetre emission, (listed in Table \ref{no_mm}) with no distance ambiguity, have been included on this plot. For consistency throughout this paper, these objects are identified by the same symbols for the  maser (`plus') and \uchii~ regions (`box'). However, in order to indicate that these objects are different from those with millimetre emission, the symbols have been rotated 45-degrees such that the methanol maser is depicted as a `cross' and the \uchii~ region as a `diamond'. The offset of these sources has been determined relative to the closest millimetre source in the SIMBA map.

  Figure \ref{offsets} shows that the more massive millimetre cores tend to have associated methanol maser sites and \uchii~ regions which are further away than lower mass cores. This may simply reflect the smaller region of influence around lower mass sources.

\subsection{Parameter Analysis}

   Three physical parameters have been derived from this study, the mass, radius, and the   H$_{2}$ number density ($n_{H_{2}}$)  as listed in Tables \ref{main} and \ref{main2}.

   We examine each of the three parameters, as well as their distances, for each of the four classes of sources (MM, M, MR, R), in order to ascertain whether these populations are similar or not. Histograms for each of the four parameters for each of the sources classes, as well as for the whole sample, have been drawn and are presented in Fig. \ref{hist}. Class comparative cumulative distributions plots of the mass and the  H$_{2}$ number density ($n_{H_{2}}$) have also been drawn and are presented in Fig. \ref{cumul}, while the results of the Kolmogorov-Smirnov testing of these two parameters as well as the radius are presented in Table \ref{kstest}.

\subsubsection{Histograms of Parameter Distributions \label{histograms}}

   The histograms of the mass distributions (Fig. \ref{hist}, upper) show that the mm-only sources are less massive than class M, MR, and R sources (i.e. those sources displaying methanol maser and/ or radio continuum emission). The mass of the 392  sources ranges from 0.5~{$\times 10^{1}$}\mstar~ to 3.7~{$\times 10^{4}$}\mstar~ (both extremes are mm-only sources) with a mean mass of 1.5~{$\times 10^{3}$}\mstar. % The mean mass of the combined sample of methanol maser and \uchii~ sources is  2.5~{$\times 10^{3}$}\mstar~ with 25\% of the distribution between 500  and 1600\mstar, while the average mass of the mm-only sample is 0.9~{$\times 10^{3}$}\mstar~ with 25\% of the distribution between 125 and 400\mstar~ (see Fig. \ref{hist} upper, for distributions of each class). %25\% of the distribution between 125 and 400\mstar.
The median mass of the combined sample of the methanol maser and \uchii~ sources is 1.0~{$\times 10^{3}$}\mstar~, with 25\% of the distribution between 0.5 and 1.6~{$\times 10^{3}$}\mstar, while the median mass of the mm-only sample is 2.8~{$\times 10^{2}$}\mstar~ with 25\% of the distribution between 1.3 and 4~{$\times 10^{2}$}\mstar.

    Comparatively, the 195 no-ambig sources have masses that range from 1.8~{$\times 10^{1}$}\mstar~ to 3.7~{$\times 10^{4}$}\mstar, with a mean mass of 1.9~{$\times 10^{3}$}\mstar. The mean mass of the no-ambig combined maser and radio sample (M + MR + R) is 3.2~{$\times 10^{3}$}\mstar, while for the no-ambig mm-only sources it is 1.1~{$\times 10^{3}$}\mstar.

   The histogram of the source radii (Fig. \ref{hist}, middle) shows that the mm-only cores, as well as the methanol maser cores, generally have smaller radii than those sources with a radio continuum source (i.e. MR and R). With one exception (G0.26+0.01), the mm-only cores have radii less than 2.0~pc, with the majority ($\sim94\%$) having radii $<1.0$~pc. The radii of all the sources (392) ranges from 0.01 to 2.5~pc, with a mean radius of 0.5~pc. The mean radius of the combined maser and radio sample is 0.7~pc. The mm-only sample has a mean radius of 0.4~pc. 

  Comparatively, the 195 no-ambig sources have radii ranging from 0.05 to 2.5~pc, with a mean radius of 0.6~pc. The radius of the no-ambig maser and radio combined sample is 0.8~pc. The no-ambig mm-only sources range from 0.05 to 1.7~pc and average 0.5~pc in radius. \citet{faundez04} quote an average radius of $0.4\pm 0.2$~pc from their 1.2 mm continuum emission survey of \IRAS~ selected sources. 

  According to the H$_{2}$ number density ($n_{H_{2}}$) histogram (Fig. \ref{hist}, lower), there does not appear to be a correlation between the class of source and its density. The value of $n_{H_{2}}$ for the entire sample ranges from 1.4~$\times~ 10^3$~cm$^{-3}$, to 1.9~$\times~10^6$~cm$^{-3}$, with an average of 8.7~$\times~10^4$~cm$^{-3}$. The mm-only cores range from 1.6~$\times~ 10^3$~cm$^{-3}$, to 1.5~$\times~10^6$~cm$^{-3}$, with a mean value of 9.1~$\times~10^4$~cm$^{-3}$. This compares with the work of \citet{faundez04}, who obtained an average density of 2.1~$\times~10^5$~cm$^{-3}$ for their sample of \IRAS~ sources.

  The range of the H$_{2}$ number density ($n_{H_{2}}$) for the no-ambig sources does not vary, although the mean value of the entire no-ambig sample is 8.4~$\times~10^4$~cm$^{-3}$, whilst for the no-ambig combined maser and radio sample the mean value is 8.0~$\times~10^4$~cm$^{-3}$ and for the mm-only cores the mean value is 7.5~$\times~10^4$~cm$^{-3}$.  

   The distance histogram (Fig. \ref{hist}, lowest) does not appear to show any correlation between the type of source and its distance from us. The distances range from 0.4~kpc to 16.7~kpc with a mean distance of 5~kpc. The mode distance is 6~kpc. There is no variation in the distance results for the mm-only sample nor for the no-ambig sample.

\begin{figure*}
   \begin{minipage}{1.0\textwidth}
         \includegraphics[width=7.5cm, height=7.5cm]{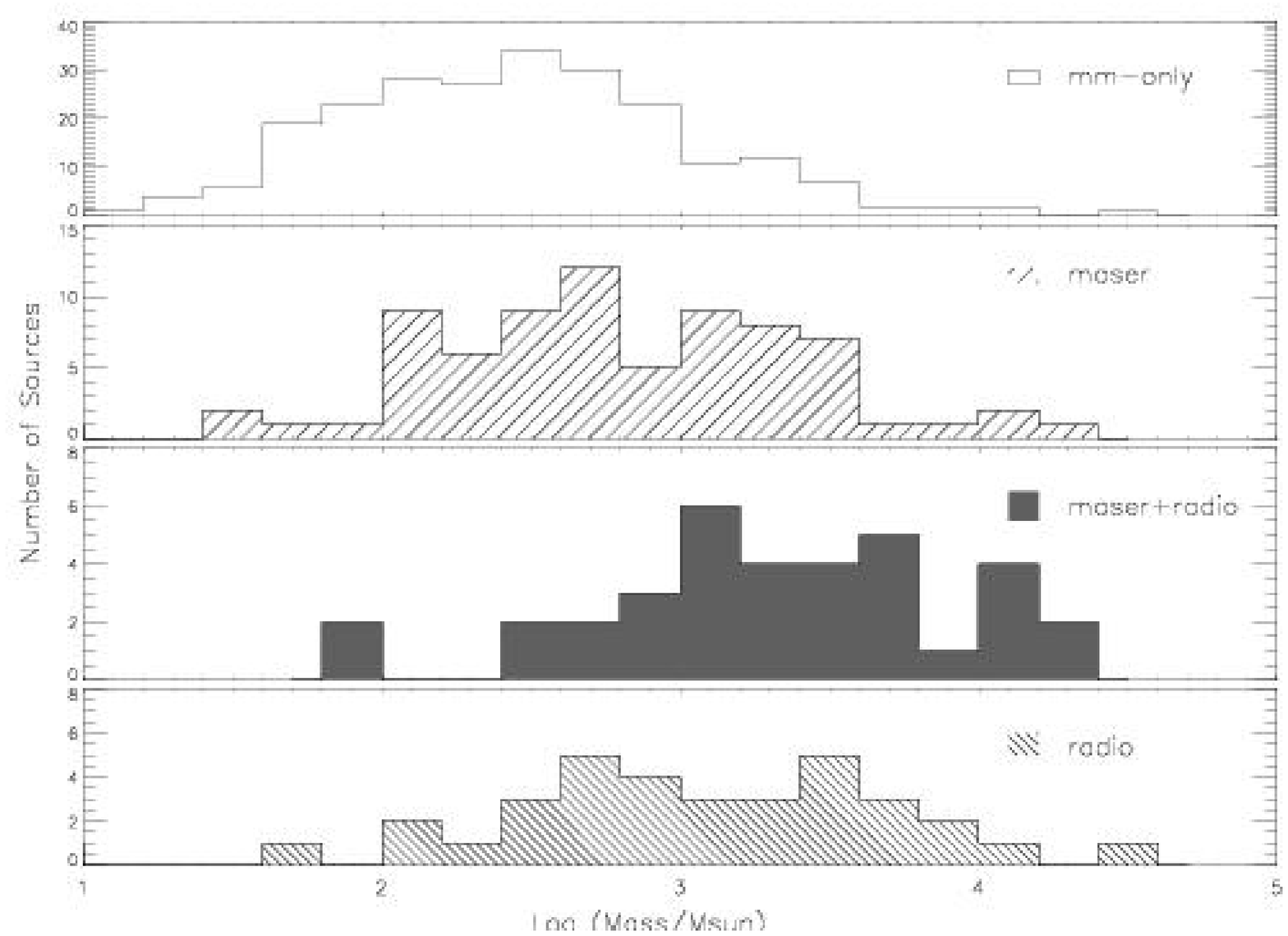}
	 \includegraphics[width=7.5cm, height=7.5cm]{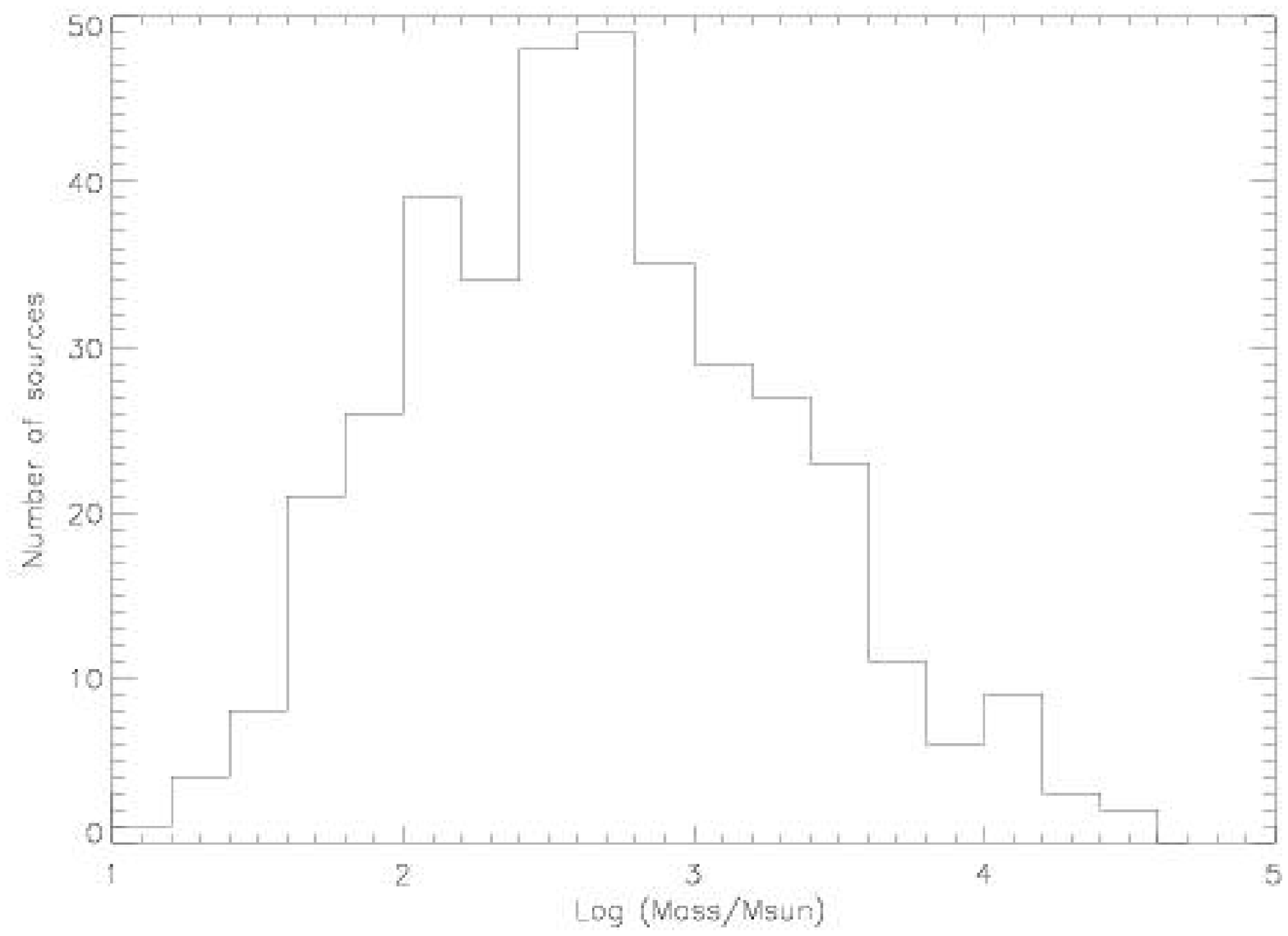}
  \end{minipage}
  \begin{minipage}{1.0\textwidth}
        \includegraphics[width=7.5cm, height=7.5cm]{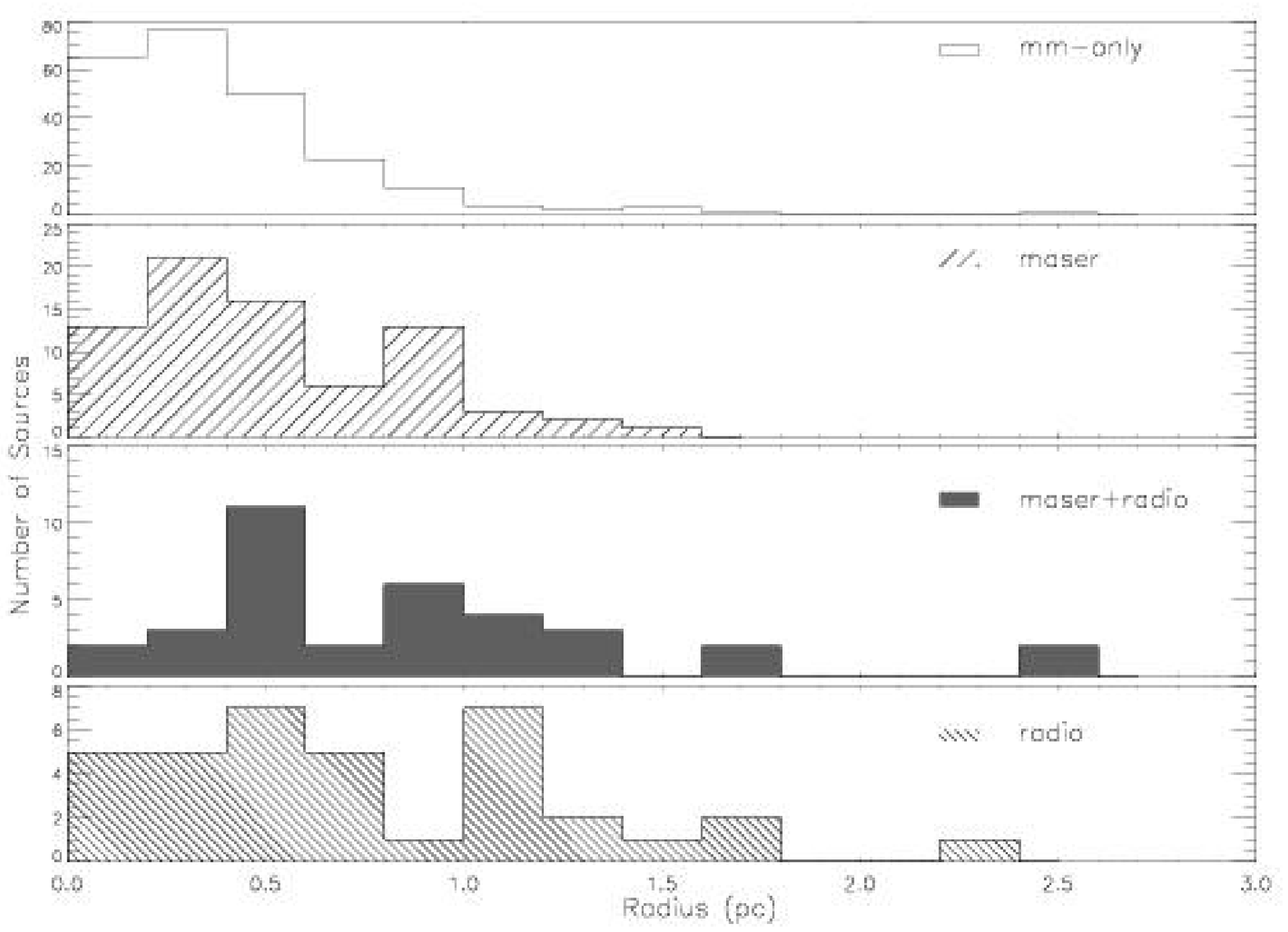}
	 \includegraphics[width=7.5cm, height=7.5cm]{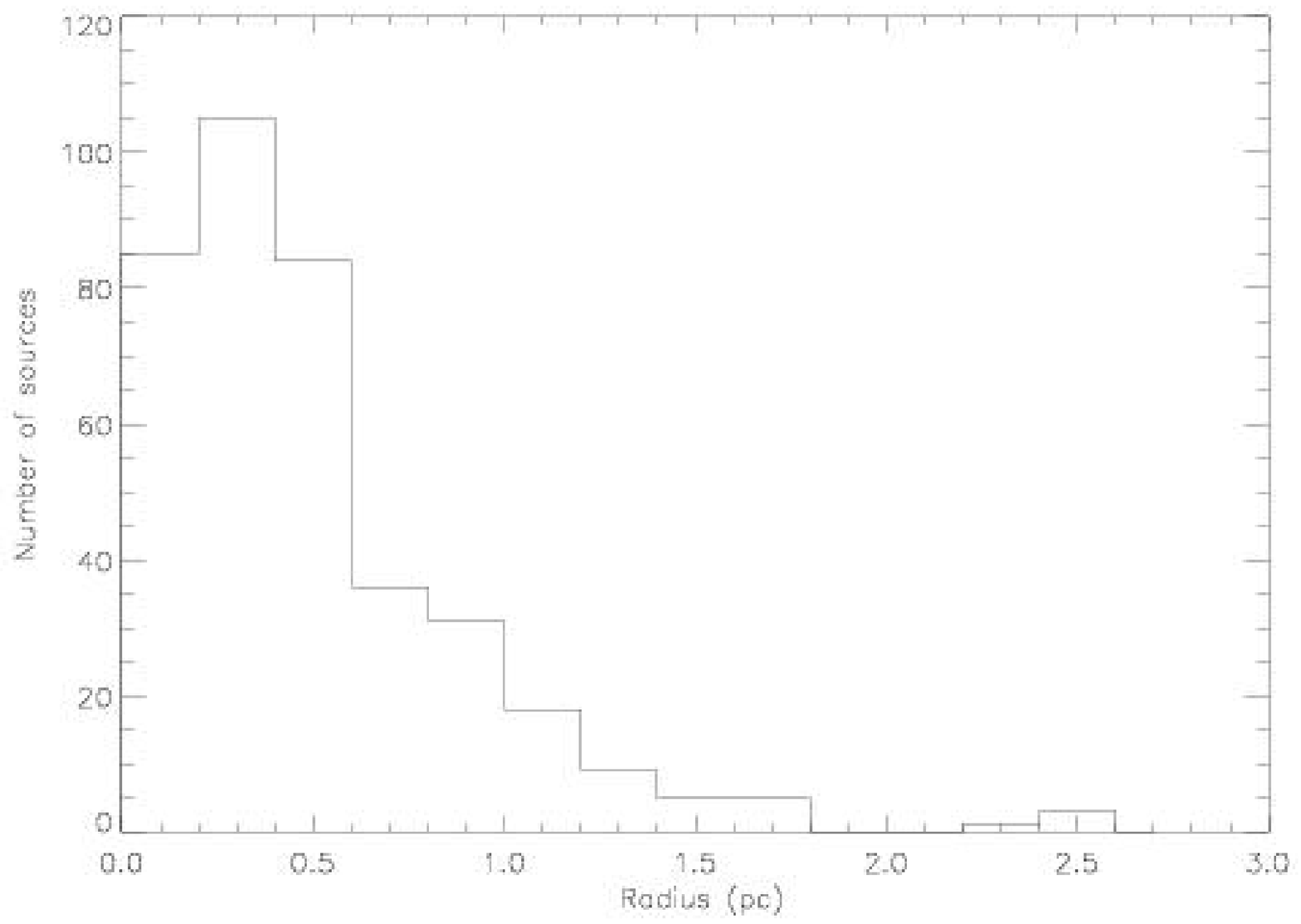}
  \end{minipage}
   \begin{minipage}{1.0\textwidth}
         \includegraphics[width=7.5cm, height=7.5cm]{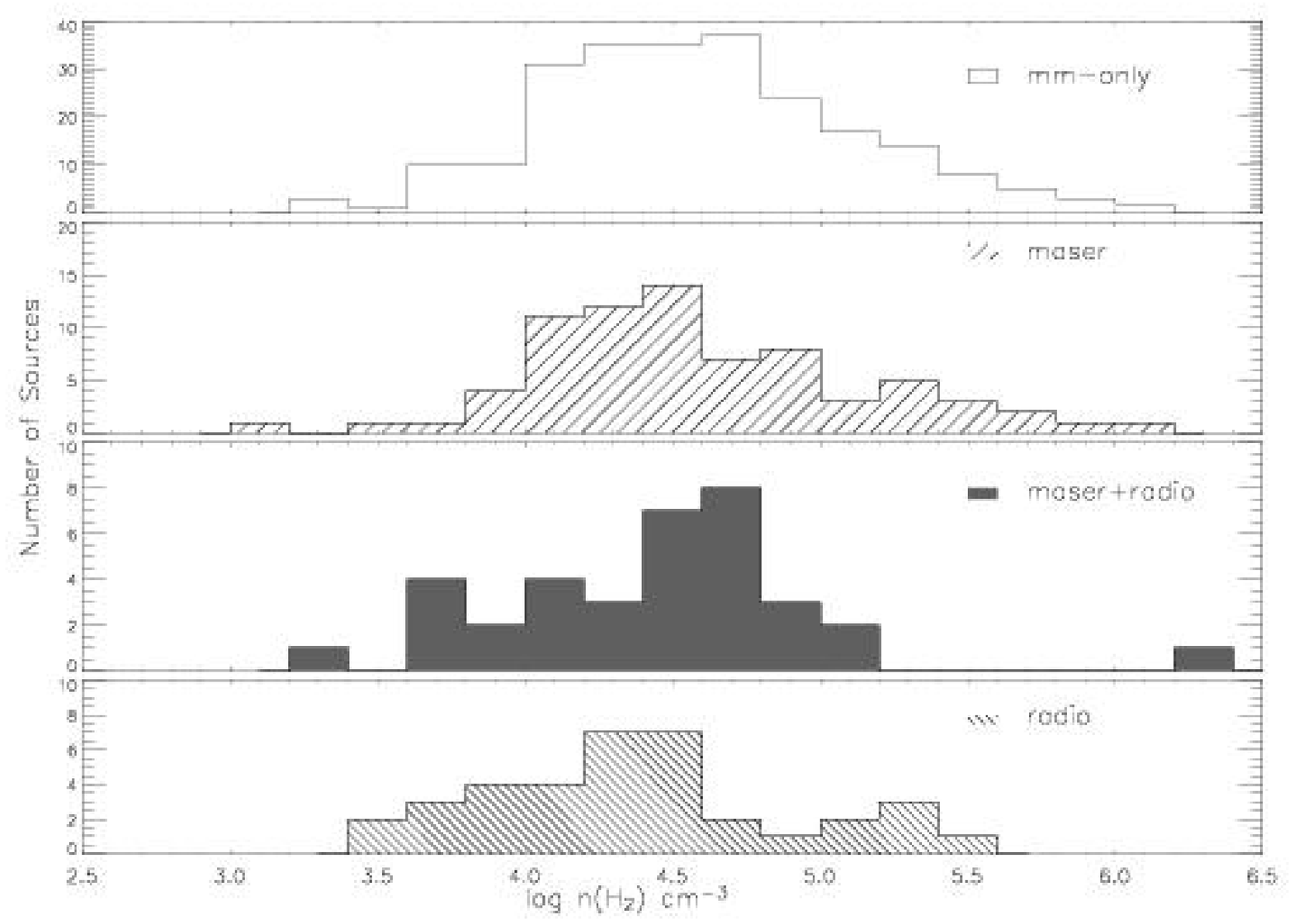}
	 \includegraphics[width=7.5cm, height=7.5cm]{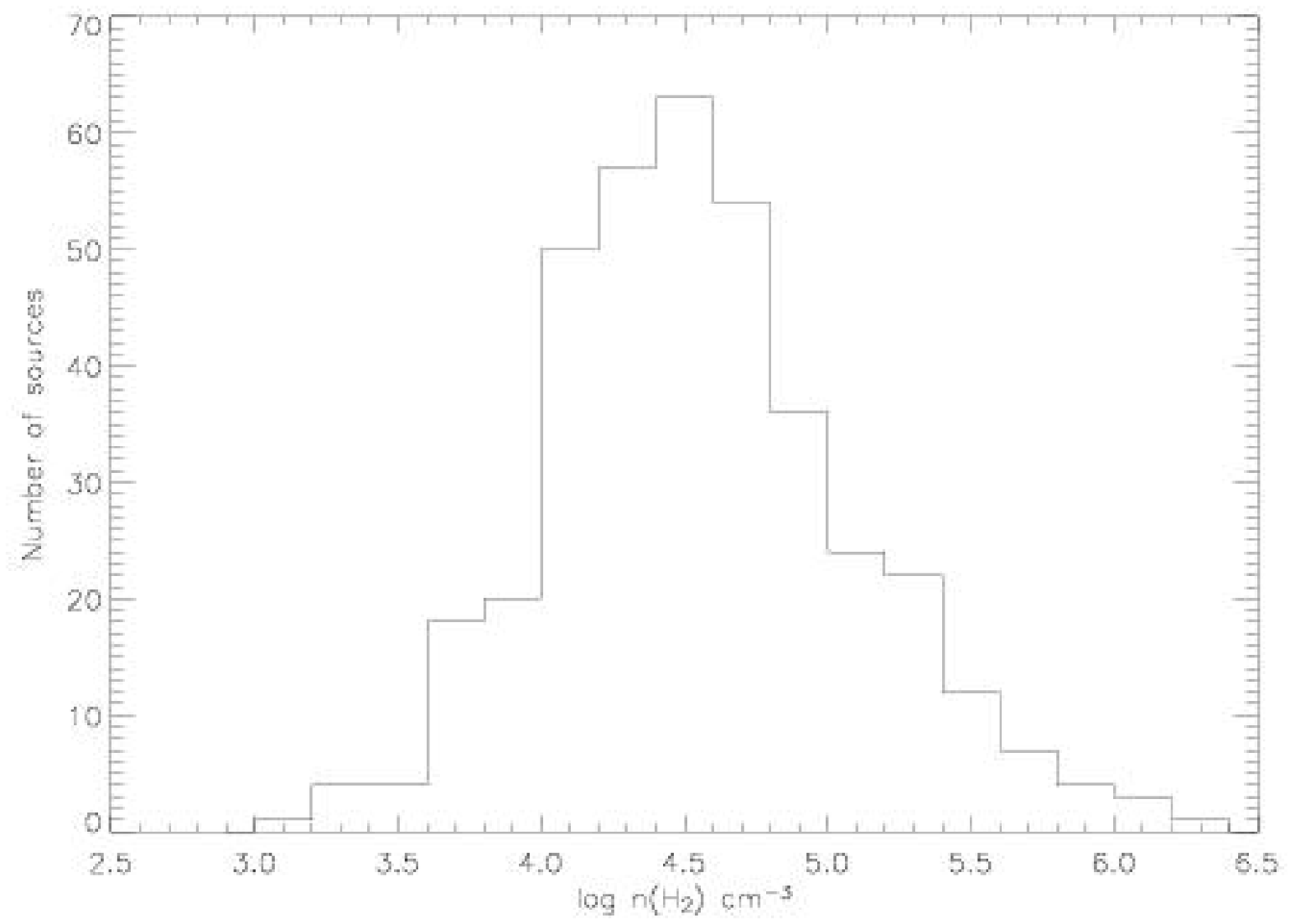}
  \end{minipage}
\caption{Distributions of the mass, distance, size and number density for all 392 sources detected in this survey (assuming the near distance to 197 sources when the distance is ambiguous). {\it Top} Histograms of mass. {\it left:} Distribution of the four classes of source (mm-only + maser + maser+radio sources + radio). {\it right:} Distribution for all sources. {\it Middle:} Radius distributions {\it left:} Distribution of the four classes. {\it right:} Distribution for all sources. {\it Bottom:} H$_{2}$ number density ($n_{H_{2}}$) distributions. {\it left:} Distribution of the four classes of source. {\it right:} Distribution of all sources.}
\label{hist}
\end{figure*}
\addtocounter{figure}{-1}

\begin{figure*}
  \begin{minipage}{1.0\textwidth}
        \includegraphics[width=7.5cm, height=7.5cm]{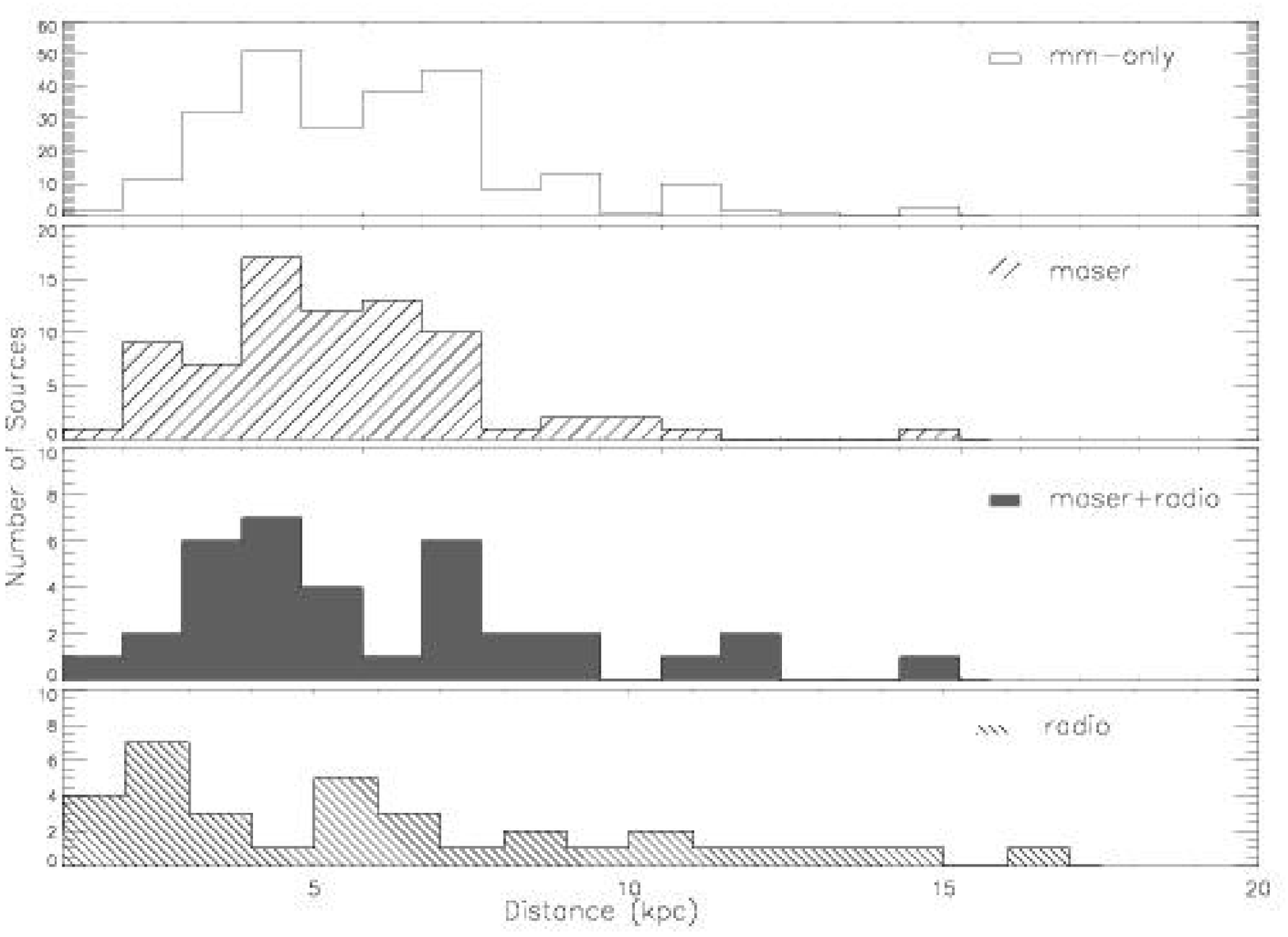}
	 \includegraphics[width=7.5cm, height=7.5cm]{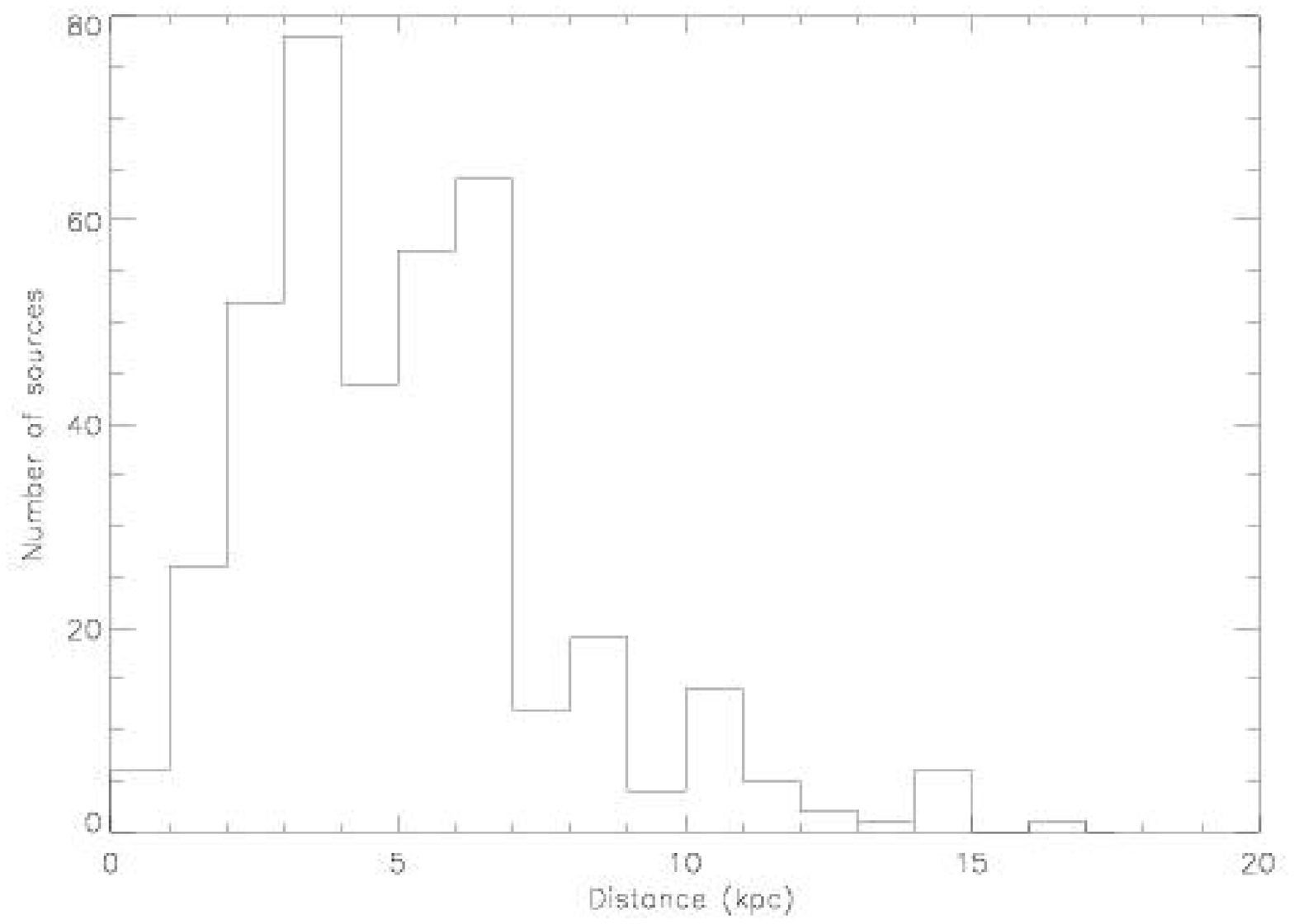}
  \end{minipage}
   \begin{minipage}{1.0\textwidth}
  \end{minipage}
  \begin{minipage}{1.0\textwidth}
  \end{minipage}
\caption{{\it - continued} Top: Histograms of distance. {\it left:} Distribution of the four classes of source. {\it right:} Distribution for all sources.}
\end{figure*}

\subsubsection{Kolmogorov-Smirnov Testing of the Cumulative Distributions}

   The Kolmogorov-Smirnov (hereafter, KS) test was also performed to test the hypothesis that the mm-only cores are drawn from the same parent population as the other three classes of source (M, MR, and R). KS tests were performed for the mass, the H$_{2}$ number density, and the radius within each of the four source classes, with the results shown in Table \ref{kstest}.

   Generally, in order to conclude that two distributions are not drawn from the same sample, the KS-probabilities must be small, $\le0.01$. The null hypothesis, that the samples are drawn from the same population, can only clearly be rejected when comparing the mass of the mm-only sample to the mass of the samples associated with methanol masers and/or radio continuum sources. This is also the case when comparing the radius of the mm-only sample to the radius of samples associated with a radio continuum source (i.e. MR and R). In the case of the H$_2$ number density, no difference between the samples can be discerned from the data.

   As a calibration measure, the KS-test was applied to the class M and R sources. The results, indicate that the likelihood of these sources being from the same parent population is 6.9\%. This result is to be expected if the two objects (maser and \uchii) often occur coincident as in the case of the class MR objects, and if the \uchii~ region succeeds the maser in the evolution of massive stars.

   The cumulative plots of mass, number density and radius, upon which the KS-tests are based, are presented in Figure \ref{cumul}. These illustrate that the mm-only sources are less massive than the other samples, as indicated by the rejection of the null hypothesis in the KS-test. The cumulative plot for the $H_2$ number density shows little variation between the four classes. The cumulative plot for the radius shows that the mm-only sources have smaller radii than the other three samples, as well as corroborating the results from the KS-test, that the mm-only sources are distinctly different from those sources associated with a radio continuum source (i.e. MR and R).

    The cumulative plots confirm the synopsis presented in the histogram plots (Fig. \ref{hist}) as discussed in $\S\ref{histograms}$.

\begin{table}
\begin{center}
\caption{Results from the KS-test of mass, density ($n_{H_{2}}$), and radius for all four classes of object. Column 1, indicates the parameter being tested. Columns 2 and 3 lists the classes of source being tested. Column 4 gives the resultant KS probability that the objects in columns 2 and 3 are from the same parent population. If this probability is $<$0.01 it is generally concluded that the samples are not drawn from the same population. \label{kstest}}
\begin{tabular}{@{}llll@{}}
       \hline
       \hline

Correlation & Source Class &  vs Source Class? & KS-prob\\
(1) & (2) & (3) & (4)\\
\hline
MASS & (MM) mm-only & (M) masers & 2.4E-03\\ % 2.9e-2
&&  (MR) maser+radio & 4.3E-10\\ % 1.2e-08
&&  (R) radio  & 2.6E-05\\  % 7.3e-6
&&&\\
& (M) maser  & (R) radio & 6.9E-02\\ %1.3e-01 
\hline
%&&&\\
H$_{2}$ NUMBER  & (MM) mm-only & (M) masers &  8.9E-01\\  % 7.2e-01
DENSITY && (MR) maser+radio & 2.5E-01 \\ % 2.2e-01
($n_{H_{2}}$)&& (R) radio & 5.3E-02\\ %2.9e-02
&&&\\
& (M) maser & (R) radio & 2.1E-01\\ %6.9E-02
\hline
RADIUS  & (MM) mm-only & (M) masers & 4.2E-02\\ 
&&  (MR) maser+radio & 2.0E-07\\
&&  (R) radio  & 3.7E-04\\   
&&&\\
& (M) maser  & (R) radio & 3.1E-02\\ 
\hline
\end{tabular}
\end{center}
\end{table}

\begin{figure*}
   \begin{minipage}{1.0\textwidth}
          \includegraphics[width=16cm, height=12cm]{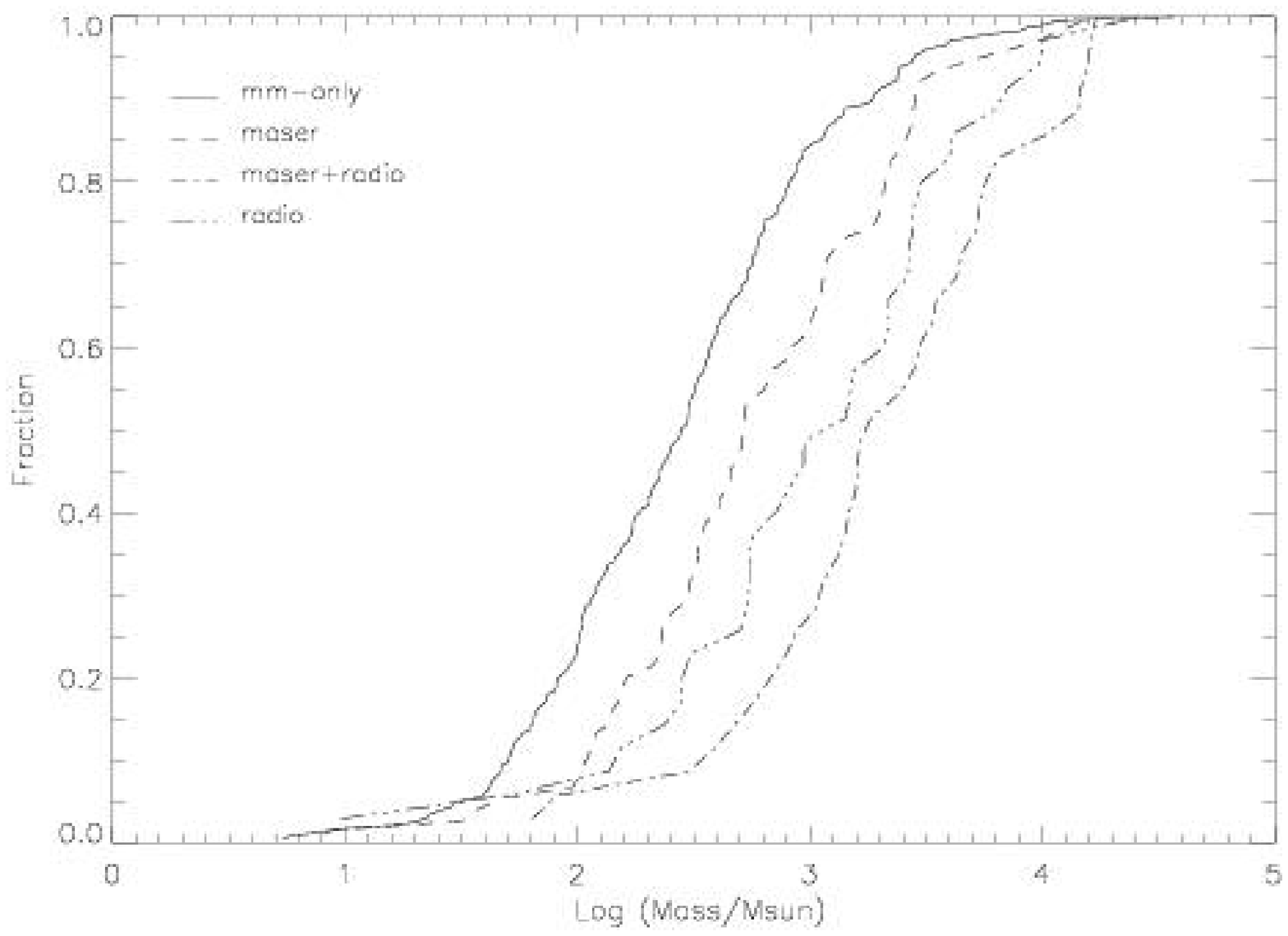}
  \end{minipage}
  \begin{minipage}{1.0\textwidth}
         \includegraphics[width=16cm, height=12cm]{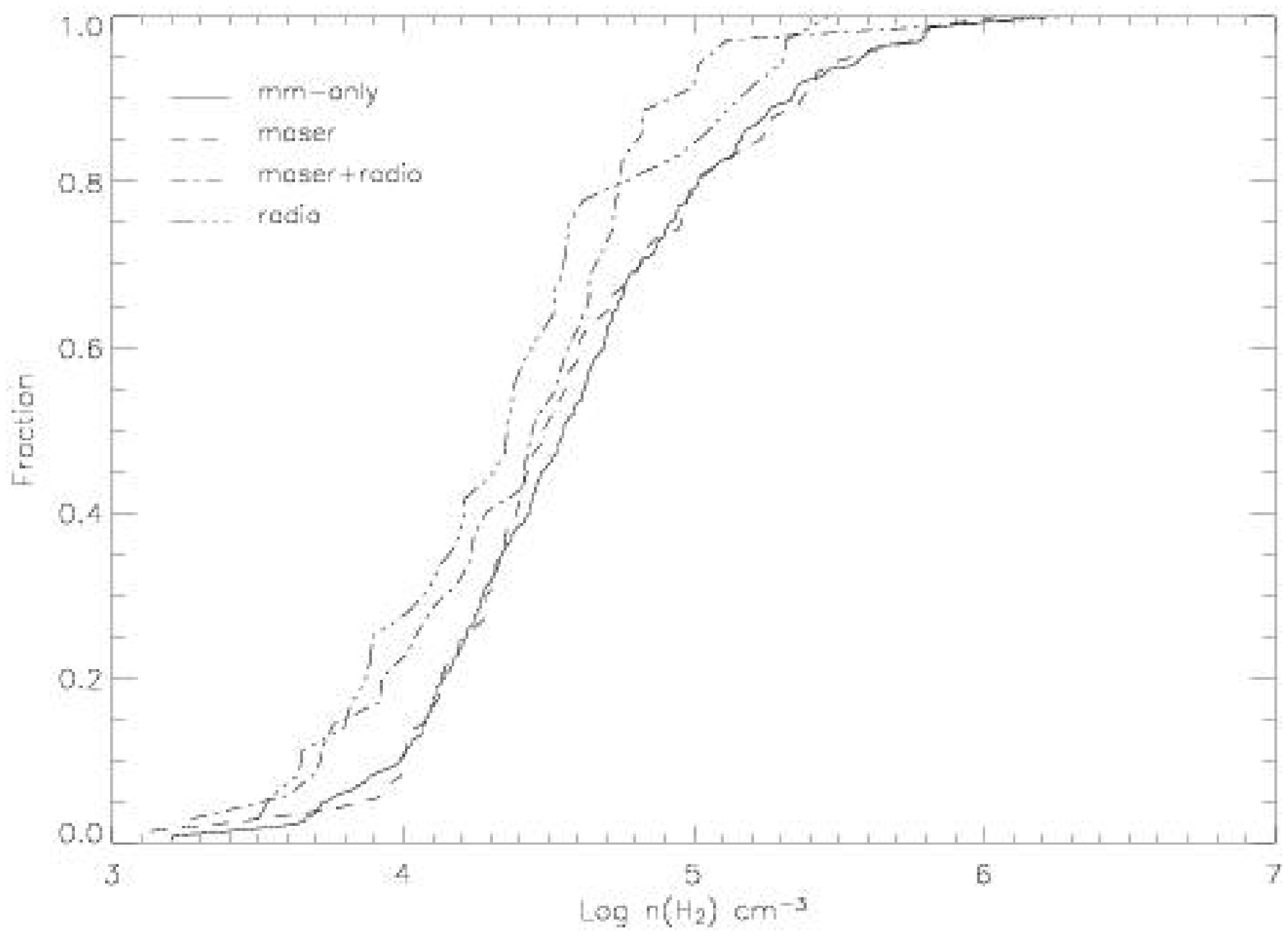}
  \end{minipage}
  \caption{Cumulative Distributions for mass (top) and H$_{2}$ number density (bottom)}
  \label{cumul}
\end{figure*}
\addtocounter{figure}{-1}

\begin{figure*}
  \begin{minipage}{1.0\textwidth}
        \includegraphics[width=16cm, height=12cm]{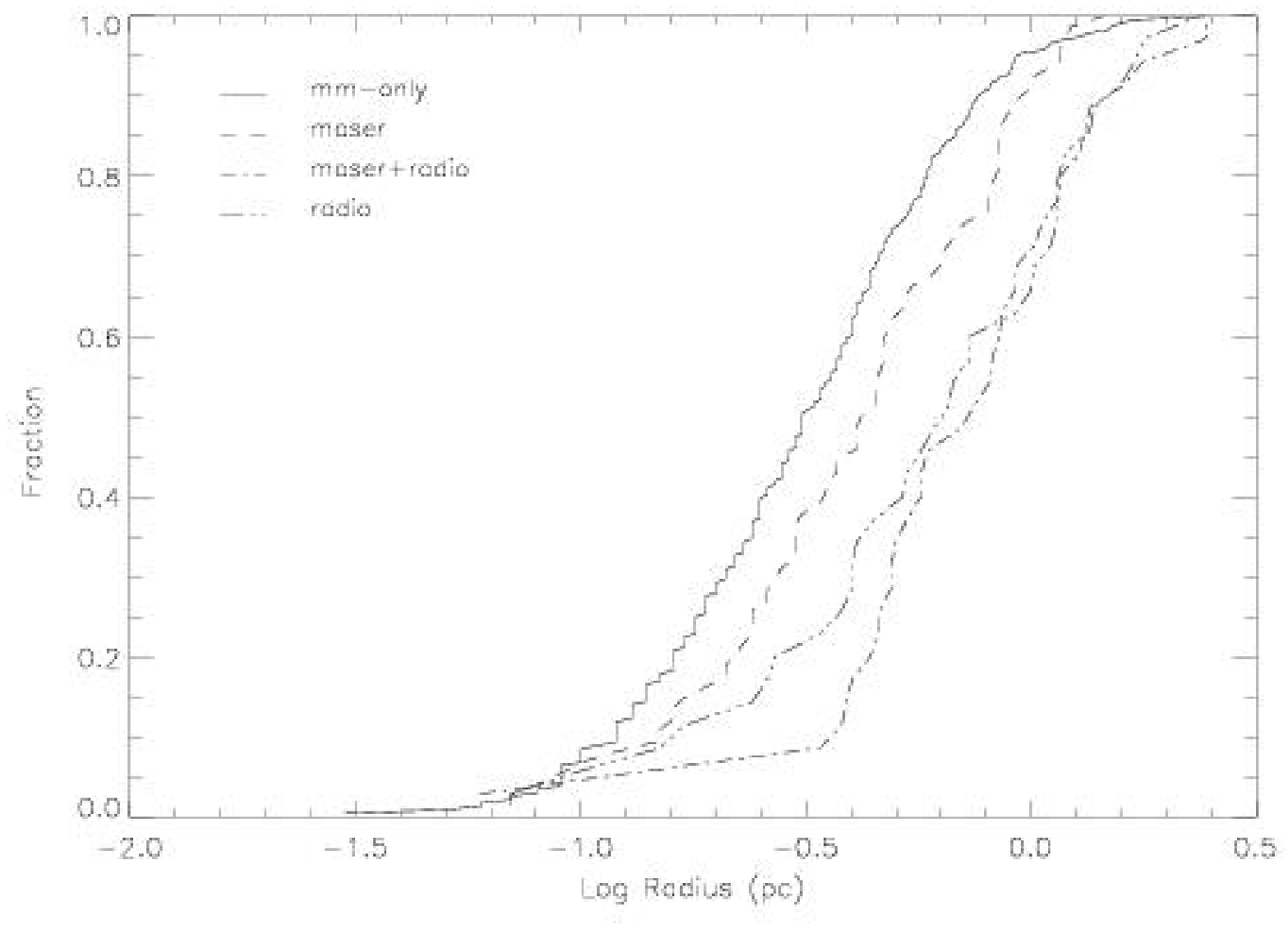}
  \end{minipage}
   \begin{minipage}{1.0\textwidth}
  \end{minipage}
  \begin{minipage}{1.0\textwidth}
  \end{minipage}
\caption{{\it - continued} Cumulative Distribution for the radius}
\end{figure*}

\subsection{Parameter Correlations}

   We compare the mass, radius, H$_{2}$ number density, and the distance of the sources in Table \ref{correlations}, to examine whether there are correlations between any of these parameters. This analysis is performed on the entire source list (assuming the near distance to 197 sources).

\subsubsection{Mass versus Radius}

   The comparison between the mass and the radius is presented in Figure \ref{correlations} (top). It is clear that these two parameters are related, with the more massive cores having larger radii,  with a best fit of M~$\propto$~R$^{2.2}$ as discussed further in $\S\ref{discussion}$. This fit has a correlation coefficient of 0.8.% (r = 1.6 m^0.46)

   A plot displaying the mass-radius correlation for each source type (MM, M, MR and R) has also been generated and can be found in Fig. \ref{correlations} (top-right). This plot shows that the mm-only cores tend to dominate the low-mass, low-radius end of the spectrum, but are just as prolific as class M, MR and R at the high-mass, high-radius end. These plots corroborate the results from the mass and radius histograms (Fig. \ref{hist}) and cumulative plots (Fig. \ref{cumul}), confirming that the mm-only cores can be less massive with smaller radii than sources with masers and/or \uchii~ regions (class M, MR, and R).

\subsubsection{Mass versus H$_{2}$ Number Density}

   The comparison between the mass and the number density is presented in Figure \ref{correlations} (middle). There is a much weaker inverse correlation apparent between these two quantities, with M~$\propto$~$n_{H_{2}}$$^{-1.1}$ i.e., the more massive a core, the lower its average density. The correlation coefficient for this fit is low at 0.2.

   There is also no clear distinction between the different categories of source.

\subsubsection{H$_{2}$ Number Density  versus Radius}

   The relation between the number density and radius  is shown in Fig. \ref{correlations} (lower). Like the mass-radius plot, there is a clear correlation between these two quantities. Denser cores have the tendency to be smaller, on average, with $n_{H_{2}}$~$\propto$~R$^{-2}$. This fit has a correlation coefficient of 0.7.

   The mm-only cores have a tendency to be both smaller, and have higher densities than the other three categories of source comprising the masers and radio continuum sources.

\subsubsection{Mass versus Distance}

   A comparison between the mass and the distance of the sources is presented in Fig. \ref{correlations} (bottom). No relation is apparent as would be expected. 

  The sensitivity limit of our observations, in terms of the lower mass detection limit has been added to the graph as a dotted line.

\begin{figure*}
 \begin{minipage}{1.0\textwidth}
         \includegraphics[width=7.5cm, height=7.5cm]{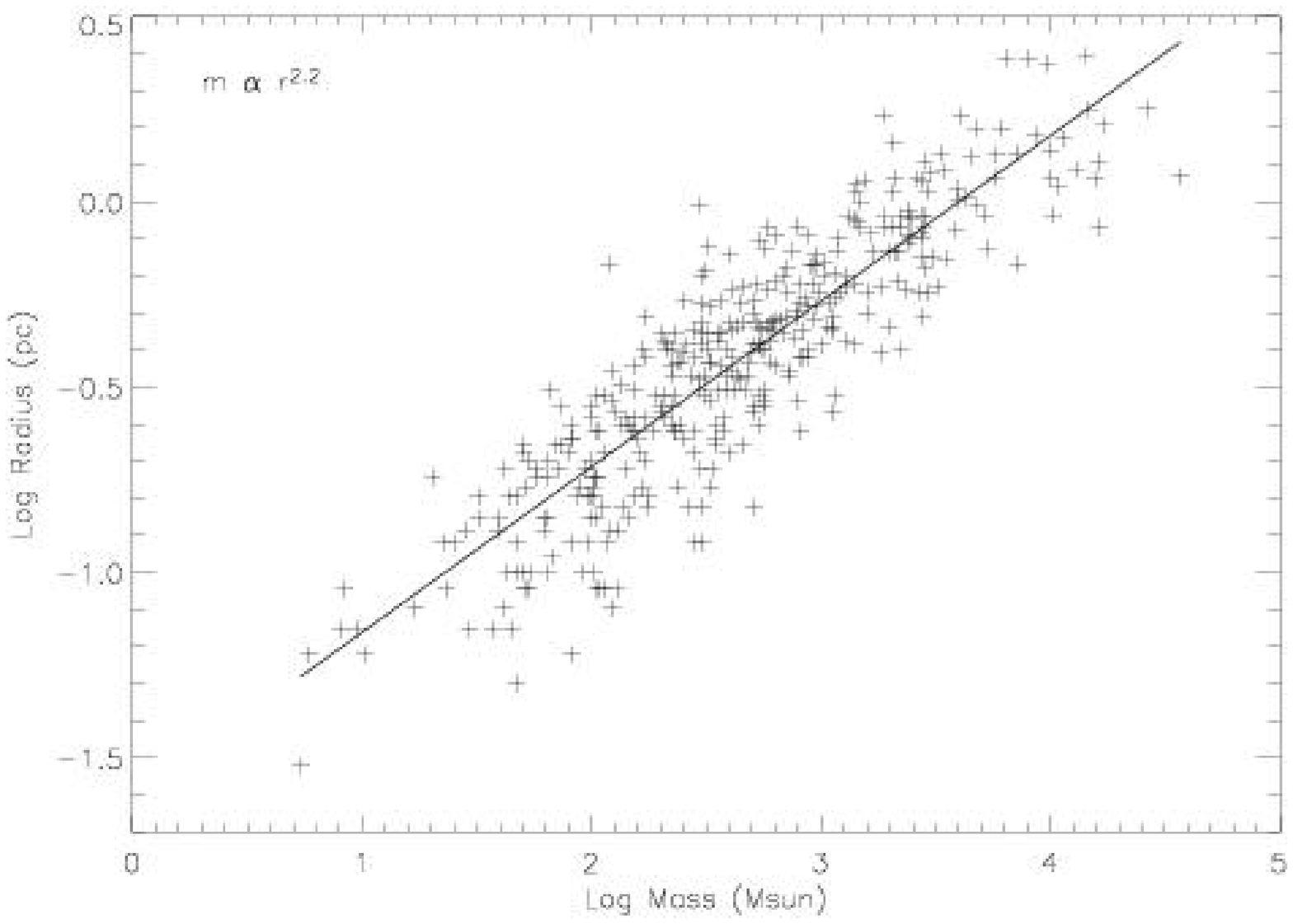}
	 \includegraphics[width=7.5cm, height=7.5cm]{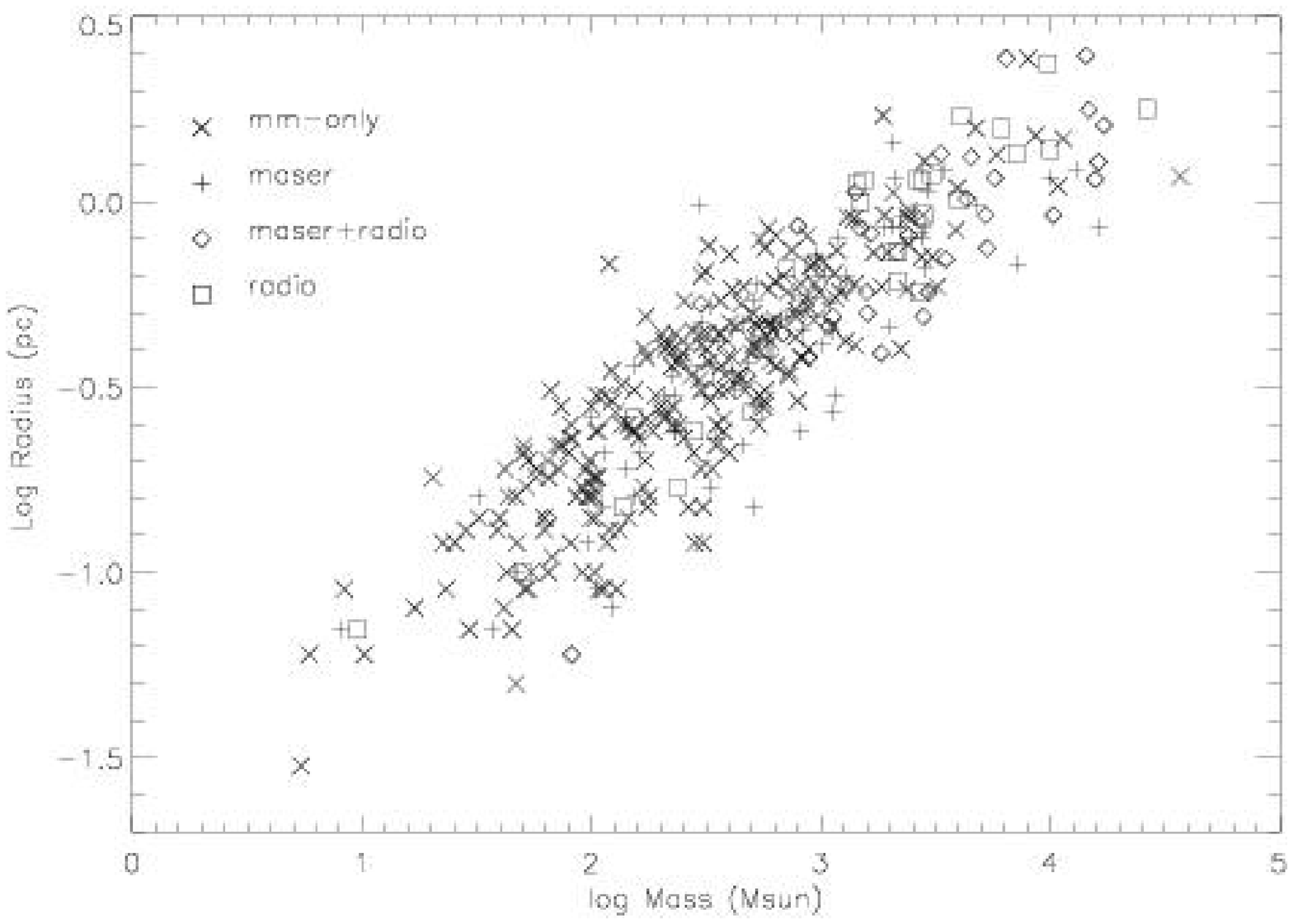} 
 \end{minipage}
 \begin{minipage}{1.0\textwidth}
	 \includegraphics[width=7.5cm, height=7.5cm]{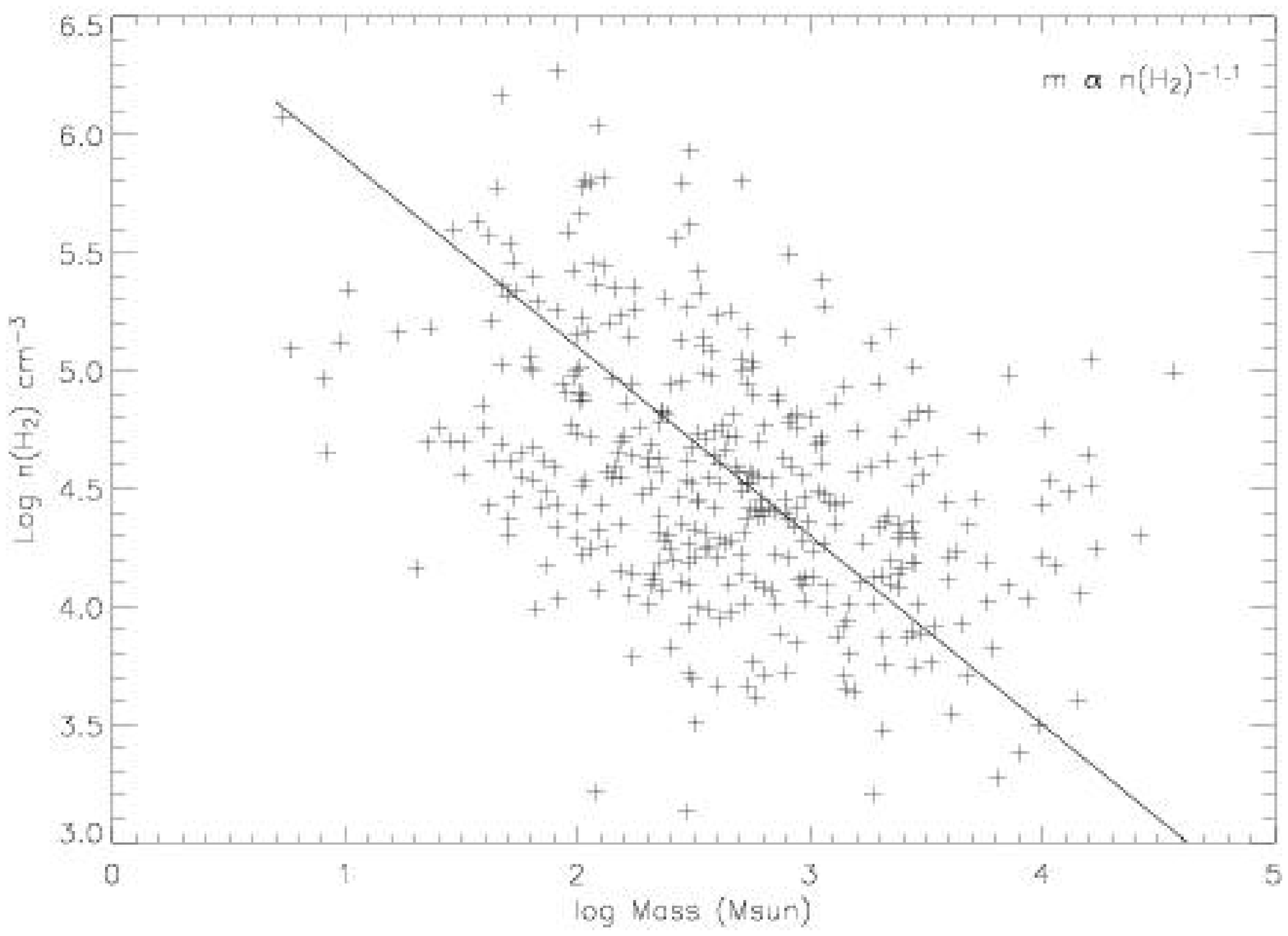}
          \includegraphics[width=7.5cm, height=7.5cm]{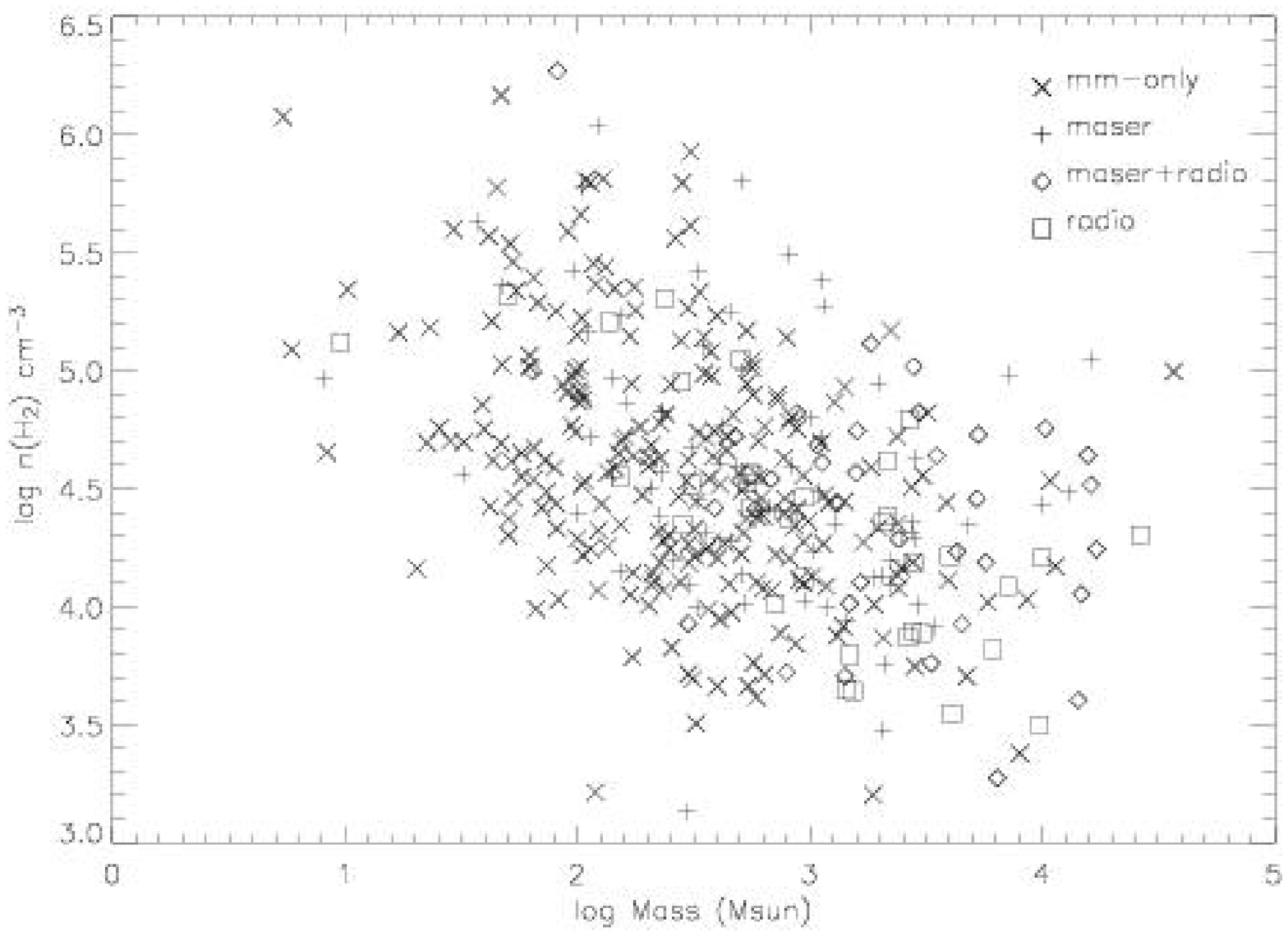}
 \end{minipage}
 \begin{minipage}{1.0\textwidth}
          \includegraphics[width=7.5cm, height=7.5cm]{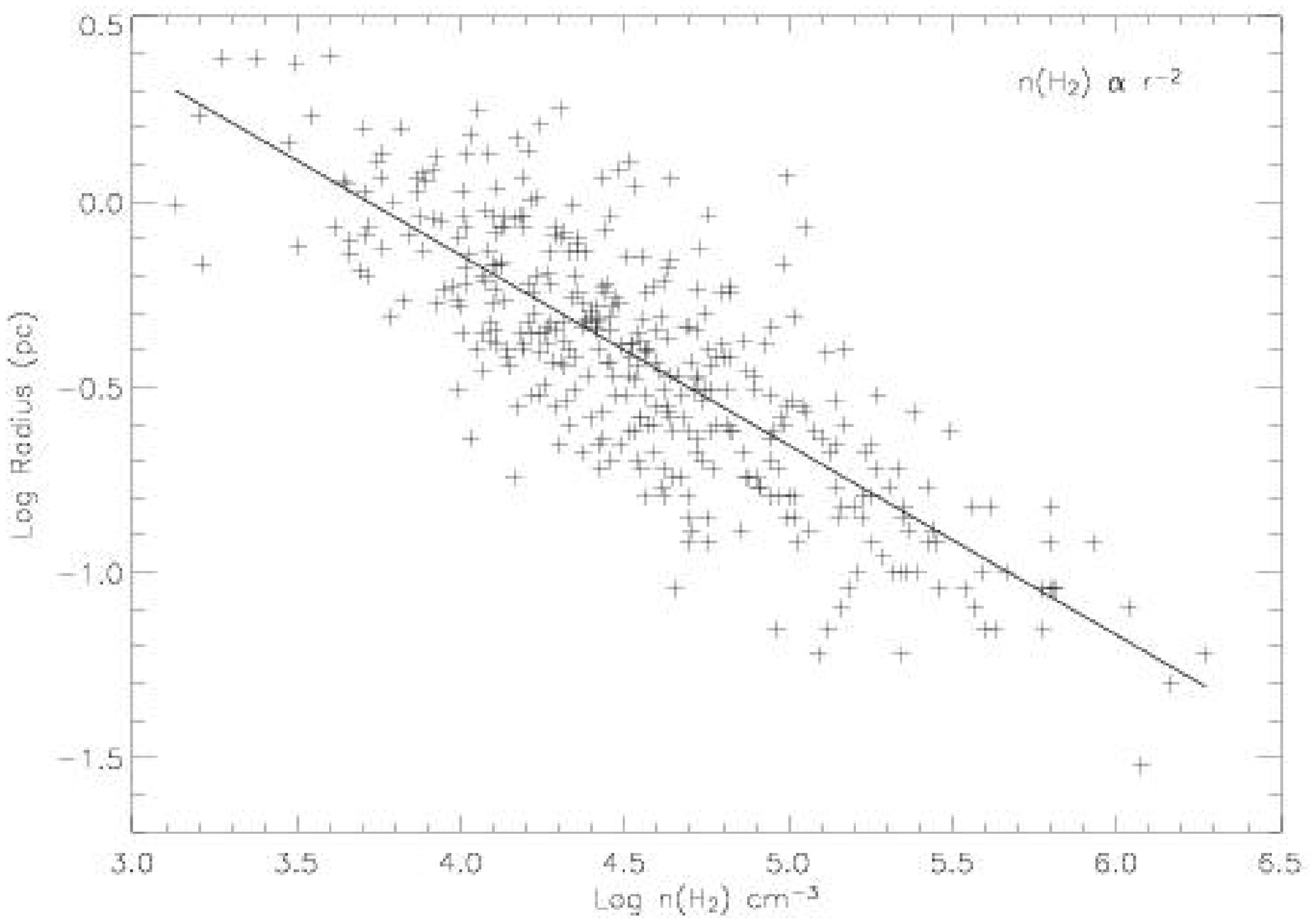}
	 \includegraphics[width=7.5cm, height=7.5cm]{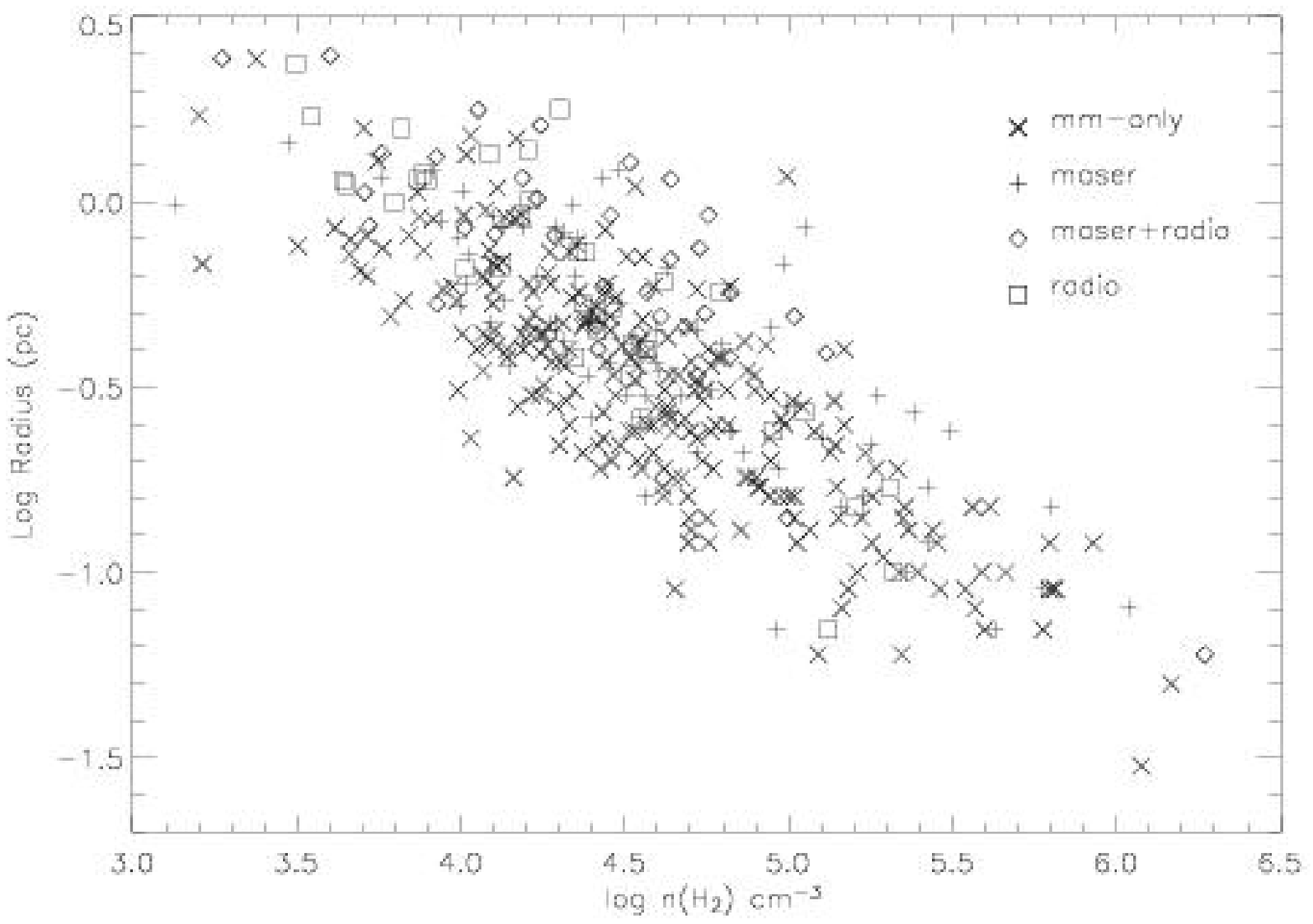}
 \end{minipage}
 \caption{Correlation plots of mass, H$_{2}$ number density ($n_{H_{2}}$), radius and distance. The plots on the left correspond to the full source sample (assuming the near distance to 197 sources). The relation best describing the two parameters is printed on the plots. The plots on the right indicate the individual class distributions with a symbol key presented on the plots. {\it Top:} correlation plot of the mass and radius of the sources. {\it Middle:} Plot of mass and H$_2$ number density ($n_{H_{2}}$). {\it Bottom:} Plot of H$_2$ number density ($n_{H_{2}}$) and radius.
\label{correlations}}
\end{figure*}
\addtocounter{figure}{-1}

\begin{figure*}
  \begin{minipage}{1.0\textwidth}
         \includegraphics[width=7.5cm, height=7.5cm]{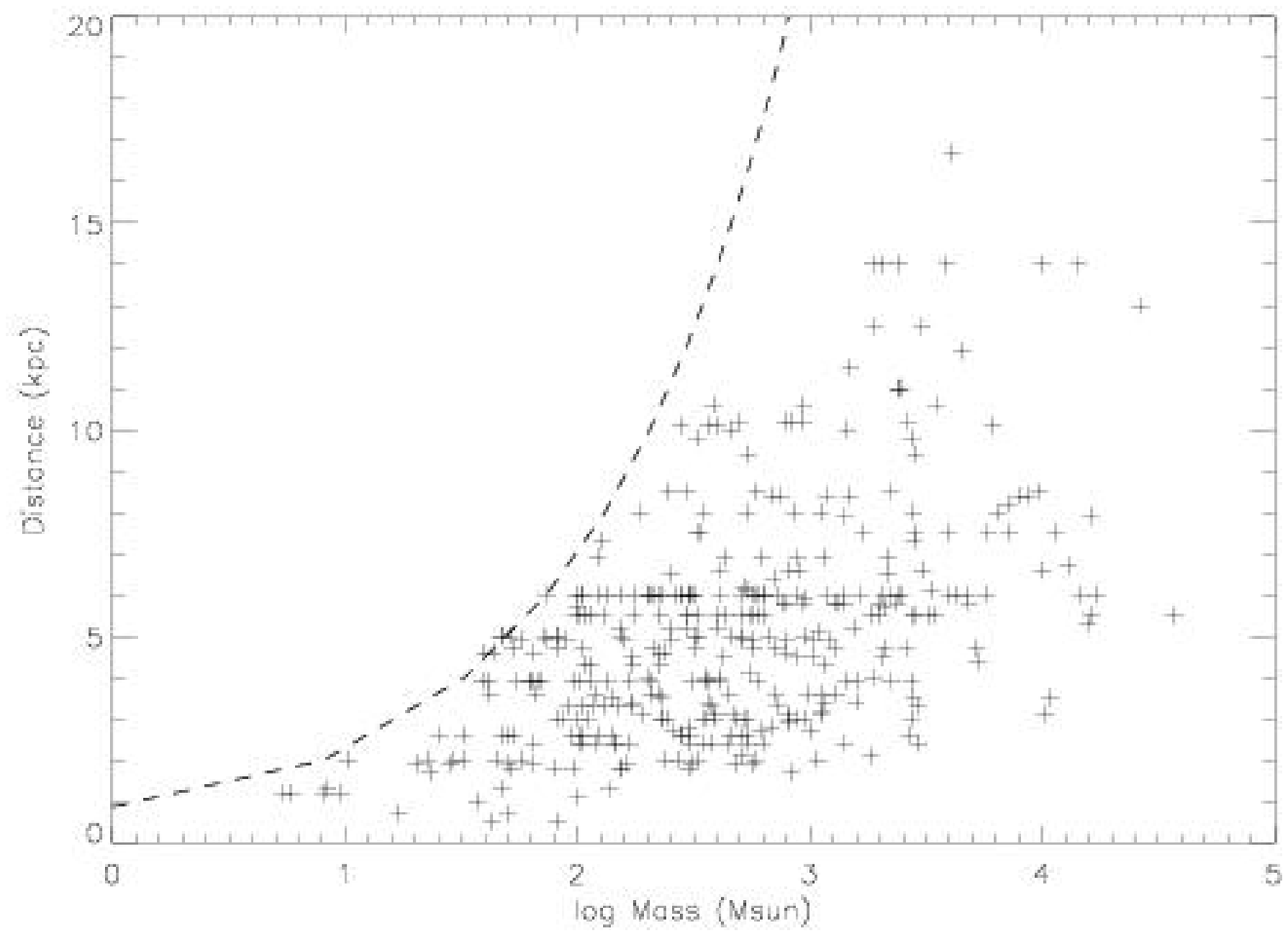}
	 \includegraphics[width=7.5cm, height=7.5cm]{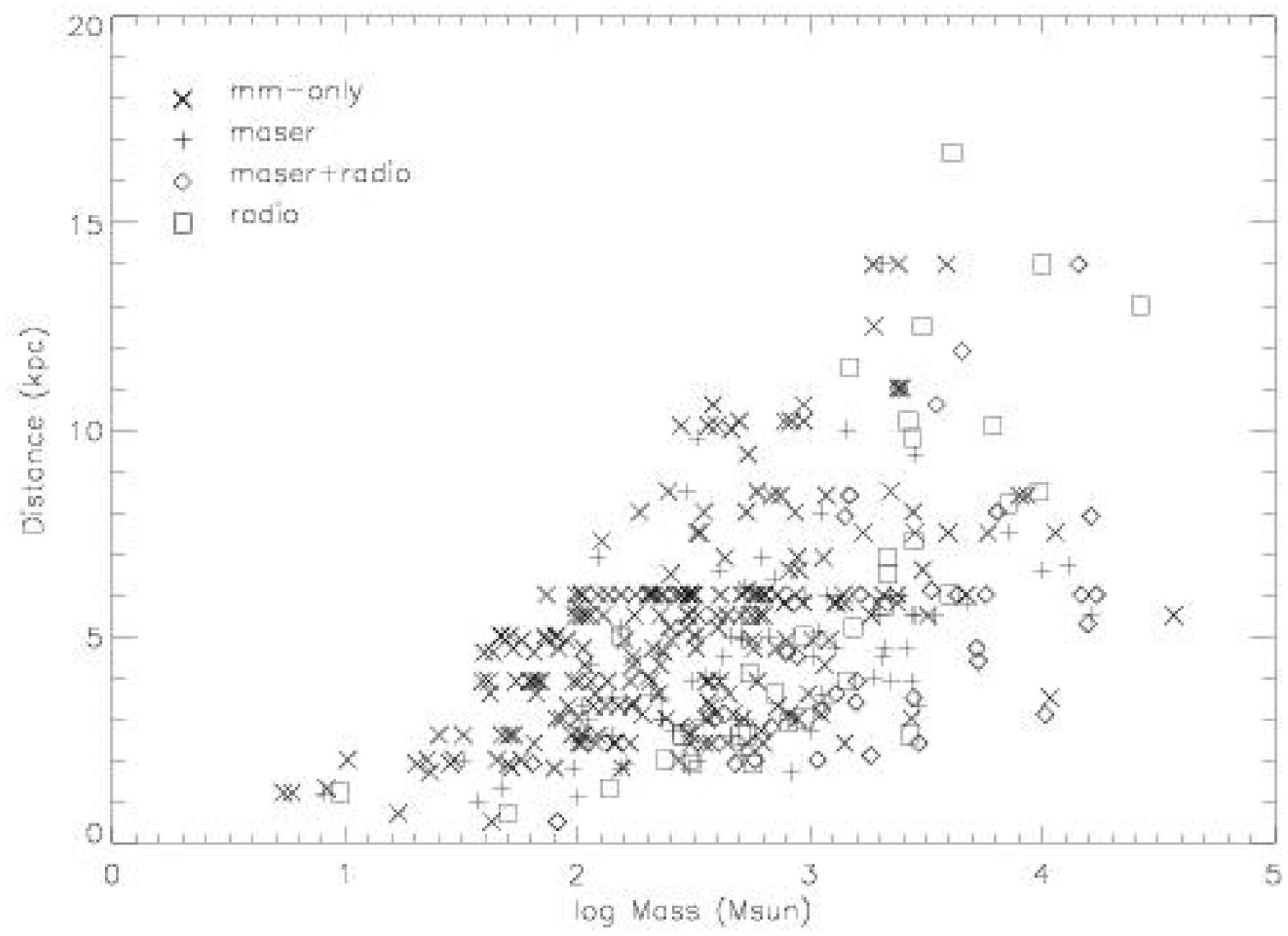}
  \end{minipage}
  \caption{{\it - continued} Plot of the mass with distance. The dotted line indicates the 3$\sigma$ mass-sensitivity limit of our survey at 50 mJy.}
\end{figure*}

\subsection{Discussion}\label{discussion}

   The physical attributes of mass, radius and $H_{2}$ number density ($n_{H_{2}}$) of each of the four classes of source detected in this survey (MM, R, MR, and R) have been examined to determine whether there are any correlations between each of these parameters, and whether those correlations are source class dependent.

   Examination of the mid infrared \MSX~ images for the  SIMBA fields has revealed many sources devoid of mid-IR emission and also many that are associated with mid-IR dark clouds. At a lower limit, we estimate that 30\% of the 404 sources detected in this survey have no mid-IR emission, implying that they are cold sources, 90\% of which are mm-only cores. Conversely to this, it is not possible to conclusively state that the remaining 70\% of sources {\it do} have mid-IR emission due to an excess in emission and confusion in many of the infrared images examined.

   The histograms (Fig. \ref{hist}) and cumulative distributions (Fig. \ref{cumul})  show that the mm-only cores are smaller in radius and are also less massive (0.4~pc and 0.9~{$\times 10^{3}$}\mstar~ on average) than those cores harbouring a methanol maser and/or an \uchii~ region (0.7~pc and 2.5~{$\times 10^{3}$}\mstar~ on average). Analysis of the sources with no distance ambiguity confirms the robustness of this result, and the small influence of assuming the near distance for those sources with a distance ambiguity.

    The mm-only cores display no evidence of ongoing massive star formation: they are not associated with any compact radio emission signifying the presence of \uchii~ regions, nor do they display evidence of H$_{2}$O or CH$_{3}$OH maser emission. At least 45\% of these mm-only cores do not display any mid infrared emission, as detected by the \MSX~ satellite.

  These results lead to two hypotheses about the nature of the mm-only cores.  It is possible that they are a precursor to the methanol maser, hence younger, less massive and smaller. Accordingly, these mm-only cores may represent an earlier evolutionary stage to that of cores associated with methanol masers and/or \uchii~ regions. That is, they may in fact represent the earliest stage in the formation of massive stars, prior to the onset of methanol maser emission.

   However, we can not rule out the possibility that the mm-only cores are simply intermediate mass cores that will give birth to intermediate mass stars (hypothesis 2). Undermining this hypothesis however, is the fact that even the least massive mm-only cores have masses which will support stars in excess of 10~\mstar~ forming. However, if this hypothesis is true, it suggests that the maximum mass of a star in a cluster is related to the mass of the molecular core from which it is formed.

  The mm-only cores may in fact provide examples to support both hypotheses. That is, the smaller, less massive mm-only cores may contain intermediate mass stars in the process of formation, while the larger more massive mm-only cores represent an earlier evolutionary state of massive star formation prior to the onset of methanol maser emission.

  Assuming the same temperature (20~K) for all of the cores in the sample introduces an element of uncertainty in the final mass estimates derived. If, for example, we have underestimated the temperature of the radio continuum sources, which typically have temperatures of the order 40~K (c.f. \citealp{faundez04}), then we have overestimated the mass of the cores by a factor of 2.3 (see Table \ref{scalef}). In this instance the mass ranges of each of the classes of source (MM, M, MR and R) would be comparable. If this were so, then the mm-only cores are colder cores, which are as massive as the methanol maser cores. The mm-only cores are then more likely to represent an earlier stage in the star formation process than those cores with methanol masers or \uchii~ regions. Spectral energy distribution modelling in subsequent papers will allow further investigation of the mm-only cores, in particular allowing better estimate of the temperature and hence the mass.

   The visual extinction ($A_v$) and surface density ($\Sigma$) of 392 sources have been estimated. The visual extinction of our sample varies from 10 to 500~mag with an average of 80~mag, implying a high degree of embedding. The surface density ($\Sigma$) varies from 0.2 to 18.0~kg~m$^{-2}$ (both extremes are mm-only cores), with an average of 2.8~kg~m$^{-2}$. Care must be taken in comparing surface densities among source classes, since this parameter depend on how well the source is resolved.

   It is clear (c.f. Fig. \ref{correlations}) that the mass and radius of a source are highly correlated, with the more massive cores having larger radii. The H$_2$ number density and the radius also appear to have a strong relation where the denser cores have smaller radii on average. There is a weak correlation between the mass and the H$_2$ number density.

   The relation between the mass,  mean H$_{2}$ number density ($n_{H_{2}}$), and the radius of the cores is best described by the following two equations:  M~$\propto$~R$^{2.2}$ and  $n_{H_{2}}$~$\propto$~R$^{-2}$, where M, R, and $n_{H_{2}}$ are the mass, radius and  H$_{2}$ number density respectively. Note that these relations are not inconsistent with M~$\propto$~$n_{H_{2}}$~R$^3$, despite the substitution of the fits in this formula giving M~$\propto$~R, rather then the fitted value of M~$\propto$~R$^{2.2}$. This is because this formula applies for a specific core, while the relations that we have obtained are for the average properties of the entire sample.

   This survey also revealed 20 maser sites and 9 \uchii~ regions, which are devoid of millimetre continuum emission (c.f. Table \ref{no_mm}). These sources were not specifically targeted in this survey, yet their coordinates fall within the fields mapped by SIMBA.
   
   The positions of 27 of these tracers have been determined via interferometry and hence are accurate to within  $1''$. The methanol masers are not weaker on average than those methanol masers for which millimetre emission is detected. These 27 maser and radio continuum sources also occur at distances equivalent to those maser and \uchii~ regions detected with millimetre continuum emission. Assuming a maximum distance of 16.3~kpc for a particular maser or radio continuum source, then the 3$\sigma$ upper limit at 1.2~mm would imply a mass of 600\mstar. Figure \ref{correlations} (lower)  indicates that a particular maser or \uchii~ region at this distance would not be detected in the survey with this mass.

   Since the majority of sources have masses greater than this, it prompts us to ask why these methanol maser and \uchii~ regions are devoid of millimetre continuum emission? Does the lack of millimetre emission suggest that these objects have  characteristics dissimilar to those sites associated with massive star formation regions?  Does the methanol maser and \uchii~ region exist in the later stages of massive star formation, i.e. after the cold core phase, which is no longer detected at millimetre wavelengths? Can massive star formation occur without the presence of these ubiquitous tracers? Or are these objects simply too far away with masses too small to be detected by the sensitivity limit of the SEST?  At this time we can not draw a secure conclusion, and further study of these sources is warranted.

\section[]{Conclusions}

   We have undertaken a millimetre continuum survey of 131 regions of massive star formation, traced by the presence of methanol maser and/or radio continuum emission, using the SIMBA instrument on the SEST. 404 sources are detected in this survey.

   Millimetre continuum emission is detected toward all of the methanol maser and \uchii~ regions targeted (129). The millimetre continuum emission is offset from the two NM-\IRAS~ positions however, but this may be attributed to the low resolution offered by \IRAS~ in pinpointing the peak emission of the central core.

   For 20 maser sites, and 9 \uchii~ regions, within the surveyed fields, millimetre continuum emission is not detected. Further follow-up work of these objects is required in order to ascertain their nature and the reason for the lack of millimetre emission. It is not clear whether these are simply more distant objects where continuum emission falls below the detection limit.

   Also detected in this survey are sources that have no methanol maser or radio continuum emission indicative of MSF. At a lower limit estimate, 45\% of these cores are also devoid of  mid-IR \MSX~ emission. The majority of these `mm-only' cores are separate from, and generally offset from, the targeted tracers in the same field.  

   In total, 253 new mm-only cores have been discovered. Many of the fields contain multiple sources. It is therefore likely that the mm-only cores belongs to the same star formation complex as the methanol maser and \uchii~ regions targeted in the study.

   The mass, radius and H$_{2}$ number density ($n_{H_{2}}$) have been determined from the millimetre flux and  distance to the sources, assuming a temperature of 20~K.

   Analysis of the mm-only cores reveals that they are generally less massive than those sources with a maser and/or an \uchii~ region, and also have smaller radii (c.f. Fig. \ref{hist}). The results from our analysis leads to two hypotheses about the nature of the mm-only cores. One possibility is that the mm-only core may be a precursor to the methanol maser in the evolutionary sequence of massive stars, the other is that the mm-only cores will simply give birth to intermediate mass stars (i.e. no massive stars that will produce \hii~ regions). If so, the maximum stellar mass of a cluster is related to the mass of the core from which it forms. Alternatively, the mm-only cores may in fact represent a cross-section of sources supporting both arguments.

   We note, that if the cores which only show millimetre emission are systematically cooler than the other classes of source, the temperature assumption of 20~K will introduce a bias, and the different mass distributions we have inferred may not actually occur. In this instance, only the first hypothesis above will apply. A better determination of the temperature of the cores is needed to constrain their masses in order to examine these hypotheses further.

  The relation between the mass,  H$_{2}$ number density ($n_{H_{2}}$), and the radius of the cores is best described by the following two equations:  M~$\propto$~R$^{2.2}$ and  $n_{H_{2}}$~$\propto$~R$^{-2}$.

\section*{Acknowledgements}
   The authors would like to thank the SEST staff for their support during the observations. In particular, we thank Markus Neilbock, as well as Robert Zylka of IRAM, for their data analysis support. We also thank an anonymous referee for useful suggestions and improvements.

   The authors extend thanks to Jim Caswell for the many useful and constructive comments on all aspects of this paper, which helped to shape the final product. T.H would like to thank the Australia Telescope National Facility (ATNF), a division of the CSIRO, for their support.

  T.H would like to acknowledge the computer programming support of Chris Blake, who also contributed valuable comments on earlier versions of this paper. We thank Naomi McClure-Griffiths for her assistance in creating the map for G30.70 region; and Matthew `Dr Matty' Whiting and Steve Curran for help with \LaTeX. We also thank Stuart Lumsden for his {\it MSX} script.% T.H. thanks `Beano' for their editorial support.

   The authors also thank the Australian Research Council (ARC), and acknowledge the MRF program of the Australian Nuclear Science and Technology Organisation (ANSTO) for travel support to the SEST.

   The data were reduced using the mapping software package MOPSI, developed by Robert Zylka. This software uses the `restoring' algorithm of Emerson, Klein and Haslam (1979), the `converting' algorithm of Chris Salter (1983), and partly the NOD2 and GILDAS libraries.

    This work has made use of the image production toolkit karma, developed by R. Gooch for the CSIRO; as well as the GAIA image display and analysis tool, a derivative of the skycat catalogue, developed as part of the VLT project at ESO; in conjunction with the IDL, and MIRIAD computing packages.

%\bibliographystyle{mn2e}
%\bibliography{aa,references}
%\expandafter\ifx\csname natexlab\endcsname\relax\def\natexlab#1{#1}\fi
%need for mn2e to work properly

\appendix
\section[]{Presentation of the Images}

   Maps of the 1.2mm continuum emission detected with the SIMBA instrument on the SEST are presented here. Coordinates of the images are in J2000. For consistency throughout, the methanol maser is depicted as a `plus' symbol, while the radio continuum source (\uchii) is denoted by a `box' symbol. Class MR objects (with both a methanol maser and an \uchii~ region) will house both a `plus' and a `box' symbol. The contours are drawn over the 1.2 mm greyscale emission at 10, 30, 50, 70, 90\% of the peak emission. Peak emission in each case is presented in Table \ref{main}. For the NM-\IRAS~ sources, (G305.533+0.360 and G305.952+0.555), the \IRAS~ positions have been included on the images as a black circle asterisk (*) near the centre of the image.

\newpage

\bsp

\label{lastpage}

\end{document}